\documentclass[a4paper]{aa}
\usepackage{natbib}
\usepackage[a4paper]{hyperref}
\usepackage{amssymb}
\usepackage{graphicx}
\usepackage{lscape}
\begin{document}

\title{Ten--year optical monitoring of PKS 0735+178:
historical comparison, multiband behaviour and variability
timescales}

  \author{S. Ciprini\inst{1,2,3}, L. O. Takalo\inst{1}, G. Tosti\inst{2,3}, C. M. Raiteri\inst{4}, M. Fiorucci\inst{2}, M. Villata\inst{4}\\
  G. Nucciarelli\inst{2}, L. Lanteri\inst{4}, K. Nilsson\inst{1}, J. A. Ros\inst{5}}
    \offprints{S. Ciprini \\ \email{stefano.ciprini@utu.fi}}
    \institute{Tuorla Astronomical Observatory, University of Turku, V\"{a}is\"{a}l\"{a}ntie 20,
    21500 Piikki\"{o}, Finland
    \and
    Physics Department and Astronomical Observatory, University of Perugia, via A. Pascoli, 06123
    Perugia, Italy
    \and
    INFN Perugia Section, via A. Pascoli, 06123 Perugia, Italy
    \and
    INAF, Torino Astronomical Observatory, via Osservatorio 20, 10025
    Pino Torinese, Torino, Italy
    \and Agrupaci\'{o}n Astron\'{o}mica de Sabadell, Apartado de Correos 50,
     PO Box 50, 08200 Sabadell, Barcelona, Spain
    }
\date{Received ....; accepted ......}
\authorrunning{Ciprini et al.}
\titlerunning{Ten--year optical monitoring of the blazar PKS 0735+178}
%
%
%
%
%
%
%
%
\abstract
{}
{New data and results on the optical behaviour of the prominent
blazar PKS 0735+178 (also know as OI 158, S3 0735+17, DA 237, 1ES
0735+178, 3EG J0737+1721) are presented, through the most
continuous $BVRI$ data available in the period 1994-2004 (about
500 nights of observations). In addition, the whole historical
light curve, and a new photometric calibration of comparison stars
in the field of this source are reported.}
{Several methods for timeseries analysis of sparse data sets are
developed, adapted and applied to the reconstructed historical
light curve and to each observing season of our unpublished
optical database on PKS 0735+178. Optical spectral indexes are
calculated from the multi-band observations and studied on
long-term (years) durations as well. For the first time in this
source, variability modes, characteristic timescales and the
signal power spectrum are explored and identified over 3 decades
in time with a sufficient statistics. The novel investigation of
mid-term optical scales (days, weeks), could be also applied and
compared to blazar gamma-ray light curves that will be provided,
on the same timescales, by the forthcoming GLAST observatory.}
{In the last 10 years the optical emission of PKS 0735+178
exhibited a rather achromatic behaviour and a variability mode
resembling the shot-noise. The source was in a intermediate or low
brightness level, showing a mild flaring activity and a
superimposition/succession of rapid and slower flares, without
extraordinary and isolated outbursts but, at any rate,
characterized by one major active phase in 2001. Several mid-term
scales of variability were found, the more common falling into
duration intervals of about 27-28 days, 50-56 days and 76-79 days.
Rapid variability in the historical light curve appears to be
modulated by a general, slower and rather oscillating temporal
trend, where typical amplitudes of about 4.5, 8.5 and 11-13 years
can be identified. This spectral and temporal analysis,
accompanying our data publication, suggests the occurrence of
distinctive signatures at mid-term durations that can likely be of
transitory nature. On the other hand the possible pseudo-cyclical
or multi-component modulations at long times could be more stable,
recurrent and correlated to the bimodal radio flux behaviour and
the twisted radio structure observed by several years in this
blazar.}
{}
\maketitle
%
%
\section{Introduction}\label{par:intro}
The rapid and violent optical variability is one of the defining
properties of blazars, and variability studies are important in
understanding the physics of AGN in general. Characteristic
timescales, fluctuations modes, flares shapes and amplitudes, duty
cycles and spectral changes, correlations and temporal lags
between variations in different spectral bands, provide crucial
information on the nature, structure and location of the emission
components and on their interdependencies. In particular the so
called low/intermediate--frequency peaked BL Lac objects
(LBL/IBL), have the peak of the synchrotron emission around
infrared and optical wavelengths and commonly show large-amplitude
flares characterized by prominent flux variations in a wide range
of temporal scales. The rapid optical variations of LBL and IBL
are also systematically larger and with shorter duty cycles than
those of the high energy peaked BL Lac objects (HBL). Hence a
multi-band, possibly well sampled and extended optical monitoring
is an important and subsidiary element of the standard
multiwavelength (MW) analysis. MW observing campaigns provide,
more or less, short snapshots of the targets, lacking of
information about their mid/long-term evolution. Even if the
optical band has a narrow spectral extension, it can yield useful
information about the synchrotron emission peak and possible
disk/host-galaxy contributions. Moreover long-term (historical)
records of blazar variability are available at optical wavelengths
for several bright objects, although data collected in the past
are rather sparse. Small--size and dedicated (possibly automatic)
telescopes, in conjunction with international consortiums, have
recently increased the amount of photometric data, sometimes with
a fair continuous sampling during specific observing campaigns.
\par In this paper we present more than
10 years (Feb.1993--Feb.2004) monitoring data about the blazar PKS
0735+178 (1332 photometric points in four $BVRI$ Johnson-Cousins
filters, obtained during about 500 observing nights). Our effort
represents the best optical monitoring available for this object
regarding to continuous and long--term coverage. This optical
programme allowed to study colours and the continuum spectrum
(mainly in $VRI$ bands) and, for the first time, enabled to study
mid-term scales (days, weeks), over an extended data set. These
timescales were almost unexplored in blazars due to the
irregular/poor sampling and the low statistics in the optical
regime. Time series analysis accompanying our data publication, is
performed for both our observations and the historical light curve
(1906-2004), while a new photometric calibration of comparison
stars in the field of the source is also reported, as useful
reference for future optical observations and monitoring. In the
historical light curve there are obvious differences in data
quality, accuracy and sampling over time, that can yield biases,
noise, spurious and fakes signatures. However we remark that the
last 33 years portion (1970-2004) of the historical light curve
holds a sufficiently regular sampling to allow meaningful
statistical results on long-term intervals too. Data binning when
needed, the employment and comparison of 7 different temporal
analysis methods suitable for unevenly sampled data set, and the
calculation of the power spectrum given by the gaps, secure us to
have determined and reported only real and intrinsic time
signatures. We note finally that the main aim of our paper was to
investigate the variability behaviour on such intermediate scales
through our 10-year observations. Data published in this work were
obtained by 4 optical observatories: the Perugia University
Observatory (Italy), the INAF Torino Observatory (Italy), the
Tuorla Observatory (Finland), and the Sabadell Observatory
(Spain). Perugia, Torino and Sabadell data on PKS 0735+178 are
unpublished, while part of the Tuorla data were already published
in \citet{katajainen00}. Optical data from \citet{qian04} have
been also added to improve a few the sampling.
\par The paper is organized as follows: in Sect. \ref{par:0735optpropintro}
we review briefly the optical knowledge about PKS 0735+178, while
in Sect. \ref{par:observations} we mention the observing and data
reduction techniques. A new photometric calibration of comparison
stars in the field of the source is presented in Sect.
\ref{par:compstars}, and the $BVRI$ light curves collected during
our monitoring are showed in Sect. \ref{par:data1993-2004}. The
reconstructed historical light curve is described in Sect.
\ref{par:dataHistorical} while in Sect. \ref{par:specindexes} the
analysis of the multi--band behaviour is reported computing the
optical spectral indexes. A joint temporal analysis of our data
and the historical light curve is performed in Sect.
\ref{par:timescales}, and summary and conclusions are outlined in
Sect. \ref{par:summaryconclusions}.
%
%
\section{Optical properties of PKS 0735+178}\label{par:0735optpropintro}
%
The radio object PKS 0735+178, belonging to the Parkes catalog
(other most used names are: PKS 0735+17, S3 0735+17, OI 158, DA
237, VRO 17.07.02, PG 0735+17, RGB J0738+177, 1Jy 0735+17, RX
J0738.1+1742, 3EG J0737+1721) was identified with an optical point
source by \citet{blake70}. Afterwards it was classified as a
classical BL Lac object in \citet{carswell74}. This source is
optically bright, highly variable, and both radio \citep{kuhr81}
and X-ray selected \citep{elvis92}. PKS 0735+178 has been
extensively studied in the radio regime. The radio flux appear to
vary quite slowly with some outbursts
\citep{baath91,terasranta92,aller99,terasranta04}, but there is
not any evidence of a radio-optical correlation
\citep{clements95,hanski02} or periodicity \citep{ciaramella04}.
Early radio observations of PKS 0735+178 showed a peculiar
spectrum soon interpreted as the superposition of incoherent
synchrotron radiation emitted by distinct and homogeneous radio
components, conspiring to add up to an overall very flat shape
\citep[the source was indeed nicknamed as the ``Cosmic
Conspiracy''][]{marscher77,marscher80,cotton80}. Several moving
components and an unusual, complex morphology characterized by a
twisted jet were observed in VLBA/VLBI radio imaging \citep[see,
e.g.][]{perlman94,gabuzda94,gabuzda01,gomez01,homan02,ojha04,kellermann04}.
PKS 0735+178 has one of the most bent radio jets among AGN
observed at milliarcsecond (mas) scales. A bimodal scenario in
which periods of enhanced activity with ejection of superluminal
components are followed by epochs of low activity with a highly
twisted jet geometry was suggested \citep{gomez01,agudo06}. PKS
0735+178 is also a X-ray and gamma-ray (EGRET, 3EG J0737+1721)
emitting blazar \citep[see, e.g. ]{kubo98,hartman99}. The spectral
energy distribution (SED) evinced PKS 0735+178 as a low or
intermediate--energy peaked BL Lac object (LBL/IBL), where the
IR-UV synchrotron continuum dominates the total observed power.
The very low X-ray variability with respect to the high optical-IR
variations, supported the idea that X-rays are produced by inverse
Compton mechanism in some mas radio components
\citep{bregman84,madejski88}. The $\gamma$-ray flux of this blazar
appeared likewise no strongly variable \citep{nolan03}.
\par The optical spectrum of PKS 0735+178 shows an absorption line due
to an intervening system at 3980 $\mathrm{\AA}$ that, if
identified with Mg-II, provides a lower redshift limit of
$z>0.424$ \citep{carswell74,burbidge87,falomo00,rector01}. A
strong Lyman-alpha absorption line has also been detected by the
IUE satellite at the same redshift \citep{bregman81}. This
absorption was not identified in deep optical imaging, even if a
very faint emission was detected about 3.0-3.5'' NE/E (projected
distance 22-25 kpc at $z=0.424$) from the object
\citep{falomo00,pursimo02}. The host galaxy of PKS 0735+178
remains unresolved in optical imaging
\citep{scarpa00,falomo00,pursimo02}, but the source has two well
resolved companion galaxies. The galaxy at 7'' NW, declared
distorted by interaction with PKS 0735+178 \citep{hutchings88},
does not show marks of interaction in more recent and higher
resolution images, and a redshift of $z=0.645$ was obtained for it
\citep{stickel93,scarpa00,falomo00}. In addition a brighter galaxy
located at 8.1'' SE cannot be the absorber due to the great
projected distance from our blazar. A third, very faint, elongated
structure at 3-3.5'' NE was detected as well
\citep{pursimo99,falomo00}, and this could be related to the
intervening absorption at $z = 0.424$. The lower limit for the
redshift $z > 0.5$ obtained assuming typical properties for the
host \citep{falomo00} is consistent with the limit derived from
intervening absorption.
%
%
\begin{figure}[t!!]
\begin{center}
\begin{tabular}{c}
{\resizebox{8.5cm}{!}{\includegraphics{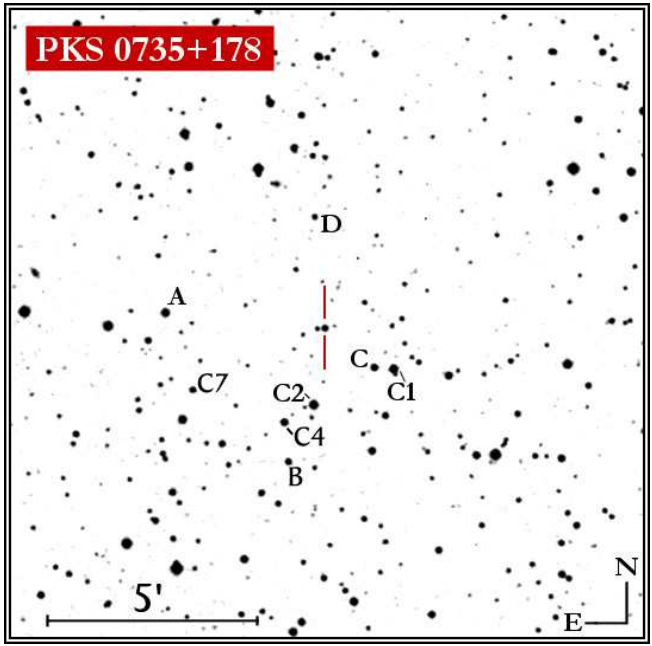}}}\\
\end{tabular}
\end{center}
\vskip -0.5 true cm \caption{Finding chart with the optical
comparison stars identified in a field of 15'$\times$15' centered
on PKS 0735+178 (within the double bar). The field is elaborated
from a frame of the Digitized Sky Survey. The new photometric
$VRI$ calibration of the stars C1, C, D, C2, C4, C7, and A is
reported in Table \ref{tab:ourcompstars}. A, C, and D stars belong
also to the photometric sequence calibrated by \citet{smith85}, to
the photometric sequence published in  \citet[][ here with a B
sequence too]{wing73}, and the sequence appeared in \citet[][ A
and D stars only]{veron75}. Stars C2, C4, C7 belong also to the
photometric sequence calibrated by \citet[][ named there
1,2,3]{mcgimsey76}, while for the C1 star the calibration is
totally new.} \label{fig:starfield}
\end{figure}
%
%
\par When combined with our data the historical optical curve of PKS
0735+178, starting from JD 2417233, i.e. Jan. 22, 1906
\citep{fan97}, spans almost 100 years. Optical variations are
often of larger amplitudes than the infrared one \citep{fan00}.
Correlations between the spectral index and the optical brightness
were observed \citep{sitko91,lin98}, alongside a spectral
flattening (blueing) with the source brightening
\citep{brown89,lin98}. On the other hand this type of correlation
appear to be weak \citep{gu06} or opposite \citep[i.e. spectral
steepening, reddening; ][]{ghosh00} in other multiband
observations. The largest optical variations registered are on the
order of 3-4 mag \citep{pollock79,fan97} at long (years) ranges.
Some intra-day (IDV) and inter-day variations until 0.5 mag were
reported in the optical history of this blazar
\citep{xie92,fan97,massaro95,zhang04}. Over the period 1995-1997,
the optical IDV and microvariations were rare and with a small
amplitude \citep{bai99}, while no clear evidence was found in more
recent observations \citep{sagar04}. Several possible recurrent
and pure periodical components were claimed, with values of 1.2,
4.8 years \citep{smith88,webb88,smith95}; 14.2, 28.7 years
\citep{fan97}; 8.6, 13.8, 19.8, 37.8 years \citep{qian04}, even if
we remark that such scales are derived by different data sets and
different epochs. The long--term analysis performed with the
Jurkevich's method on a more complete data set
\citep{fan97,qian04} postulated a main periodical component of
about 13.8-14.2 years, but our temporal analysis
(Section\ref{par:timescales}) suggest other and shorter long-term
signatures.
\par PKS 0735+178 has also a relatively high degree of optical polarization
showing very different levels covering the whole range from about
1\% up to 30\% \citep[see, e.g.
][]{mead90,takalo91,takalo92,valtaoja91,valtaoja93,tommasi01}.
Only a modest variability of this optical polarization was
observed on inter/intra night durations, that was interpreted as
owed to substructures of different polarization and variable
intensity in the jet. A preferred polarization level over few
years \citep{tommasi01} could indicate quiescence and stability in
the underlying jet structure.
%
\begin{table*}[t!]
\caption{A new optical photometric calibrations for field of PKS
0735+178, with $VR_{c}I_{c}$ Johnson-Cousins sequences of
comparison stars C1, C, D, C2, C4, C7 and A (see Fig.
\ref{fig:starfield}). Data are obtained at the Perugia University
Observatory and are adopted in this work. The previous available
comparison sequences were published in
\citet[][]{smith85,wing73,veron75,mcgimsey76, smith85}. Magnitudes
of common stars are roughly in agreement within the uncertainties.
The $VR_{c}I_{c}$ sequence reported in this table, joint with the
$U,B$ values reported by \citet{smith85}, is suggested for future
optical observations and monitoring.} \label{tab:ourcompstars}
\vspace{-0.5cm}\centering {}
\par
\begin{tabular}{lccccc}
\vspace{2mm} \\
\hline \hline %
\scriptsize{{$~~~~~~~~~~~~~~~~~~~~~~~~~$PHOTOMETRIC SEQUENCES FOR
PKS 0735+178 COMPARISON STARS} $~~~~~~~~~~~~~~~~~~~~~~~~~~~~$  }
\end{tabular}
\begin{tabular}{lccccc}
\vspace{-4mm} \\
\hline \hline
%
\vspace{-2mm} \\
 Star   &    R.A.   &    Dec.   &    $V$    &    $R_{c}$    &    $I_{c}$    \\
    &   {\footnotesize (J2000.0)}   &    {\footnotesize (J2000.0)}  &    {\footnotesize [mag]}  &    {\footnotesize [mag]}  &    {\footnotesize [mag]}     \vspace{1mm} \\
 \hline \hline
\vspace{-3mm} \\
\textbf{C1}........................................    &   07 38 00.5  &   +17 41 19.9     &    13.26 $\pm$ 0.04   &    12.89 $\pm$ 0.04   &    12.57 $\pm$ 0.04 \\
 \hline
\textbf{C}..........................................    &   07 38 02.4  &   +17 41 22.2     & 14.45 $\pm$ 0.04   &  13.85 $\pm$ 0.04   &    13.32 $\pm$ 0.04\\
\hline
\textbf{D}..........................................    &   07 38 08.3  &   +17 44 59.7     &    15.90 $\pm$ 0.05   &   15.49 $\pm$ 0.05    &    15.12 $\pm$ 0.06 \\
\hline
 \textbf{C2}........................................    &   07 38 08.5  &   +17 40 29.2     &    13.31 $\pm$ 0.04   &    12.79 $\pm$ 0.04   &    12.32 $\pm$ 0.04  \\
\hline
\textbf{C4}........................................     &   07 38 11.6  &   +17 40 04.4     &    14.17 $\pm$ 0.05   &    13.80 $\pm$ 0.04   &    13.48 $\pm$ 0.04  \\
\hline
\textbf{C7}........................................     &   07 38 20.7  &   +17 40 51.2     &    15.01 $\pm$ 0.06   &    14.70 $\pm$ 0.06   &    14.37 $\pm$ 0.05  \\
\hline
\textbf{A}..........................................    &   07 38 23.4  &   +17 42 43.0     &    13.40 $\pm$ 0.05   &    13.10 $\pm$ 0.05   &    12.82 $\pm$ 0.05  \\
\hline \hline
\end{tabular}
%
\end{table*}
%
%
%
%
%
\begin{figure*}[t!!]
\centering
\resizebox{\hsize}{!}{\includegraphics[angle=0]{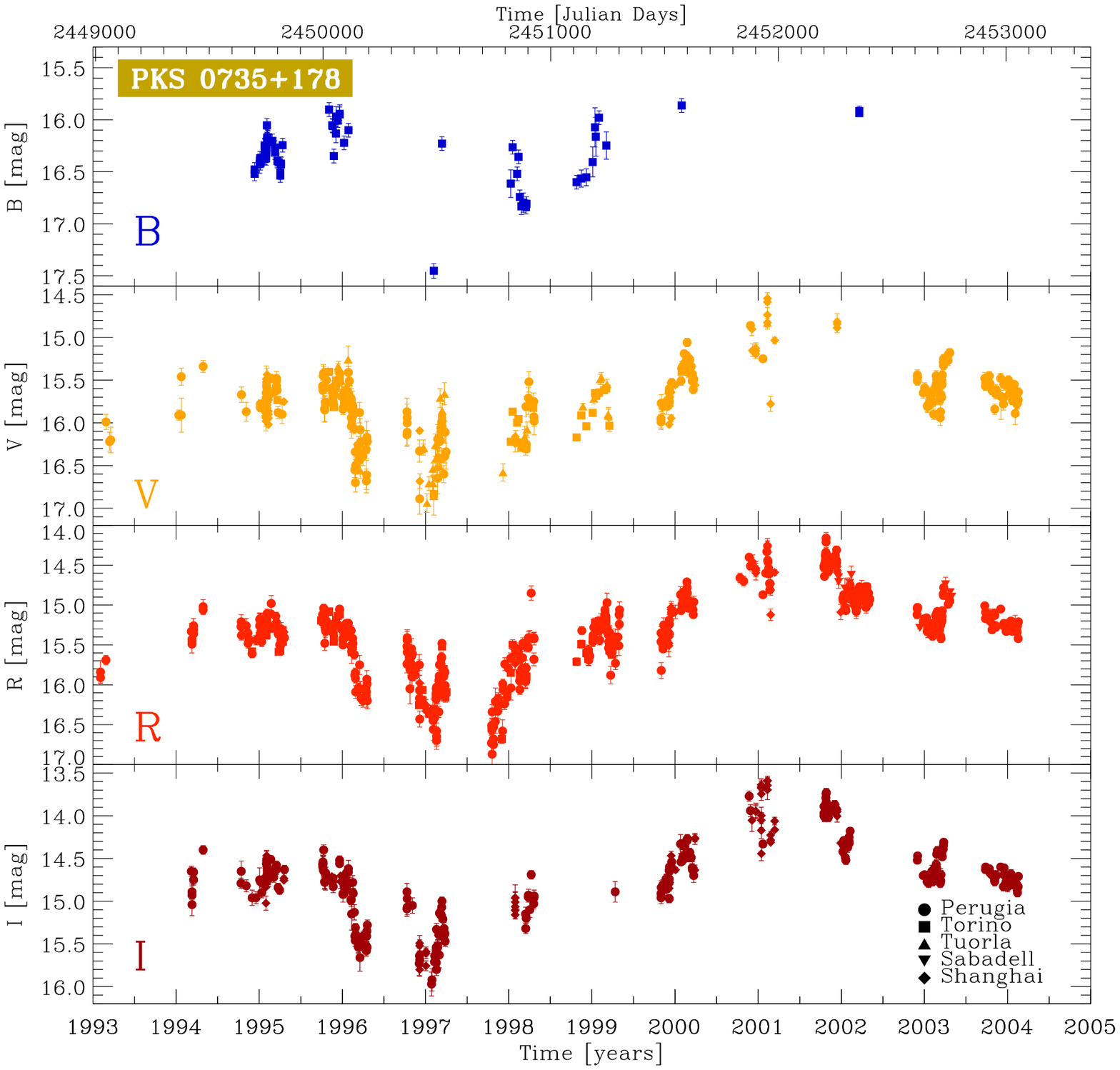}}
\vspace{-0.5 cm} \caption{$BVRI$ magnitude light curves of PKS
0735+178 from 1993 to beginning of 2004. Data cames from our
10-year observing monitoring. Published observations from Shanghai
Observatory \citep[][]{qian04} are added in order to improve the
sampling. Data sets of different observatories are in agreement
within the uncertainties.} \label{fig:clucedatinostri}
\end{figure*}
%
%
\section{Observations and data reduction}\label{par:observations}
%
Photometric observations were carried out with four telescopes.
The Newtonian f/5, 0.4 m, Automatic Imaging Telescope (AIT) of the
Perugia University
Observatory\footnote{\texttt{http://astro.fisica.unipg.it}}, Italy
(451 meters above sea level, a.s.l.), a robotic telescope equipped
with a $192 \times 165$ pixels CCD array, thermoelectrically
cooled with Peltier elements \citep{tosti96}. The REOSC f/10, 1.05
m, astrometric reflector of the Torino
Observatory\footnote{\texttt{http://www.to.astro.it}}, Italy (622
meters a.s.l.), mounting a $1242 \times 1152$ pixel CCD array,
cooled with liquid nitrogen and giving an image scale of $0.467''$
per pixel. The Dall-Kirkham f/8.45, 1.03 m reflector of the Turku
University Tuorla
Observatory\footnote{\texttt{http://www.astro.utu.fi}}, Finland
(60 meter a.s.l.), equipped with a $1530 \times 1020$ pixel CCD
camera, thermoelectrically cooled. The Newtonian 0.5m telescope of
the Sabadell
Observatory\footnote{\texttt{http://www.astrosabadell.org}}, Spain
used in two interchangeable configurations (Newton at f/4, and
Cassegrain-Relay at f/15), and equipped with a $512 \times 512$
FLI CM-9 CCD array. The Perugia and Torino telescopes were
provided with standard $BV$ (Johnson) and $R_{c}I_{c}$ (Cousins)
filters \citep{bessell79,fiorucci03,bessell05}. The Tuorla 1m
telescope, and the Sabadell telescope were equipped with V and
$R_{c}$ filters.
\par
\begin{figure}[t!] \centering
\resizebox{9.3cm}{!}{\hspace{-0.7cm}%
\includegraphics[angle=0]{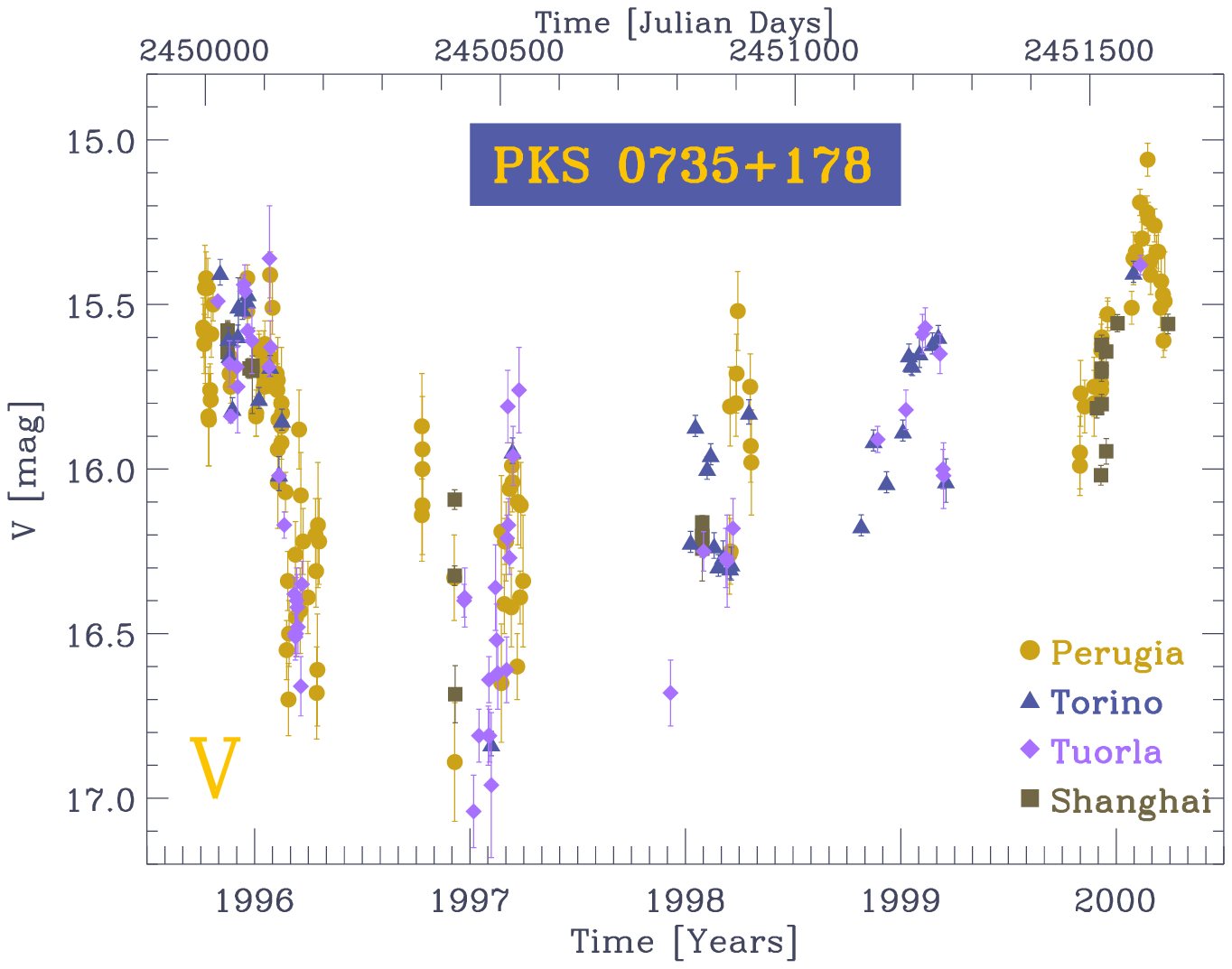}}
\vspace{-0.2cm} \caption{A portion of the V-mag light curve where
is possible to see in more detail data from the different
observatories.}
\label{fig:diff-obs-detail}
\end{figure}

 All the observatories took CCD frames and performed a first automatic
data reduction using standard methods, to correct each raw image
(for dark and bias background signals where needed) and to flat
fielding, to recognize the field stars, and to derive instrumental
magnitudes via aperture photometry (Perugia and Tuorla) or
circular Gaussian fitting (Torino). The single frames are then
inspected to evaluate the quality of the image, the reliability of
the data, and to search for spurious interferences. Comparison
among data obtained with these different telescopes on the same
night reveals a good agreement, and no detectable offset is found.
The matching with data taken in simultaneous epochs at Shanghai
Observatory \citep{qian04} showed a good agreement too (see Fig.
\ref{fig:diff-obs-detail}). The precision level in the light curve
of PKS 0735 assembled in this way is enough for a variability
analysis performed on intermediate and long--term timescales.
Moreover the time series analysis of shorter timescales ($<200$
days) is performed in each single observing season using $R$-band
data, that were mainly obtained by 1 telescope (see Tab.
\ref{tab:samplingprop}).
%
%
\section{Comparison stars photometry}\label{par:compstars}
Calculation of the source magnitude is easily obtained by
differential photometry with respect to comparison stars in the
same field of the object. The discussion of the adopted comparison
star sequence is crucial for analysis of optical data obtained
during blazar monitoring observations, as the photometric sequence
affects data quality and reliability. In order to obtain a
dependable photometric sequence for PKS 0735+178, we selected a
set of non--variable stars with brightness comparable to the
object and different colours (see the finding chart in
Fig.~\ref{fig:starfield}). Photometric calibrations of these stars
were derived from 13 optimal photometric nights between 1994 and
1996 at the Perugia University Observatory using Landolt
standards. The stability of the sequence for the stars C1, C, D,
C2, C4, was well tested and verified during data reduction of the
overall database (for stars A and C7 there are less measurements
because of the small FOV of the instrument). Our new photometric
sequence is presented in Table~\ref{tab:ourcompstars}, showing the
$V$ (Johnson), and $R_{c},I_{c}$ (Cousins) photometric values.
Previous calibrations of this star--field were performedin in
\citet{wing73,veron75,mcgimsey76,smith85}. In particular the
determinations of \citet{mcgimsey76} are photoelectric in the
$UBV$ Johnson system, thus well similar to our calibrations in
$B,V$ bands using a CCD detector. Discrepancies are small if the
specifications of \citet{bessell90} are respected and determined
\citep{fiorucci03}.
\par This new photometric calibration of the PKS 0735+178 field,
(suitable also for telescopes with small FOV), is slightly more
extended and accurate with respect to the past calibrations. Table
\ref{tab:ourcompstars} reports the $V,R_c$, $I_c$ values for seven
comparison stars, while the sequence for the star denoted with C1
is completely new. In each photometric night standard Landolt
stars were observed at different airmasses, and the calibration
line as a function of the airmass was constructed (neglecting the
colour corrections being always smaller than the instrumental
errors). The standard magnitudes of such stars were derived (with
error equal to the quadrature sum of the linear regression error
and the instrumental error on the single star). The typical error
for each night and each star is between 0.03 and 0.1 mag
(depending on the luminosity of the star and the atmospheric
conditions). Data shown in Table \ref{tab:ourcompstars} are the
result of weighted averages on the values of each night (weight
equal to $1/\sigma^2$), whereas the error estimation is equal to
the standard deviation weighted on the averages. This uncertain is
higher than the standard deviation on each single night, because
of some systematic errors different on each night. We chose to
report a reliable calibration with reliable errors in Table
\ref{tab:ourcompstars} with respect to more accurate but more
doubtful smoothed values. Colour transformations to comparison
stars were not applied, because from the analysis of Landolt stars
it was not possible to separate this effect from the instrumental
statistical errors given mainly by the effective limits of the
Perugia instrument and site. Then again the photometric system of
the Perugia telescope was developed to follow at best the standard
Johnson-Cousins system. Finally we note that some of the stars
listed in Table \ref{tab:ourcompstars} are quite red ($V-I$ index
ranges from +0.58 to +1.13 mag, suggesting that they have spectral
types F, G, possibly K), but these colour indexes are similar to
the colour indexes of PKS 0735+178 (average $V-I = +0.95 \pm 0.1$,
average $B-R = +0.91 \pm 0.09$) hence colour effects are
correspondent, and $B$-band data were obtained only with the
larger 1m Torino telescope. A detailed description of the
observing and data reduction procedures, filter system, software
adopted in the calibrations, and comparison with other works can
be found in \citet{fiorucci98,fiorucci96}.
%
%
%
\begin{table}[t!]
\caption[]{The number of photometric $B V R I$ data points of PKS
0735+178 obtained by each observatory in the period 1993-2004 and
published in this paper. In the bottom panel a summary of the
basic information and statistics about our data in each band.}
\label{tab:samplingprop} \vspace{-0.2cm} \centering {}
\begin{tabular}{ccc}
\vspace{-3mm} \\
\hline \hline
$~~~~~~~~~~~~~$   & \scriptsize{DATA POINTS PER OBSERVATORY} & $~~~~~~~~~~~~~$   \\
\end{tabular}
\begin{tabular}{lccccccc}
\vspace{-4mm} \\
\hline
\vspace{-3mm} \\
 Obs.& $B$  & $V$& $R$ & $I$ & Tot. & \scriptsize{Period}  \\
\hline \hline
Perugia &    0  & 226  & 490 & 281  & 997   & \scriptsize{Feb1993-Feb2004}\\
Torino  &    75 & 38   & 150 &  0   & 263   & \scriptsize{Dec1994-Apr2002}\\
Tuorla  &    0  &  55  & 0   &  0   &  55   & \scriptsize{Oct1995-Feb2001}\\
Sabadell&    0  &  0   & 17  &  0   &  17   & \scriptsize{Dec2001-Feb2004}\\
Shanghai&    0  &  115 & 52  &  138 & 305   & \scriptsize{Jan1995-Dec2001}\\
\hline
Total &      75 & 434 & 709 & 419 & 1637  \\
\hline \hline
\end{tabular}
$~$\\ $~$ \\
\begin{tabular}{ccc}
\vspace{-3mm} \\
\hline \hline
 $~~~~~~~~~~~~~~~~~~~~~$   & \scriptsize{ 1993-2004 DATA STATISTICS } & $~~~~~~~~~~~~~~~~~~$   \\
\end{tabular}
\begin{tabular}{lcccc}
\vspace{-4mm} \\
\hline
\vspace{-3mm} \\
$~$ &  $B$  & $V$& $R$ & $I$   \\
\hline \hline
Total data points &  75 & 434 & 709 & 419    \\
Start date [JD-2449000] & 698  & 45   & 21  & 420   \\
End date [JD-2449000] &   3354 & 4053 & 4053 & 4053   \\
Total period $N_{tot}$ [days] & 2657  & 4001 & 4032 & 3633    \\
Nights with data $N_{on}$ & 52  & 297  & 459 & 259    \\
$N_{on}/N_{tot}$ fraction & 0.019  & 0.074 & 0.171  & 0.071    \\
Mean num. points $\times$ night & 1.44  & 1.46 & 1.51 & 1.62    \\
Total mean gap $\Delta t$ [days]&  35.9 & 9.3 & 5.8 & 8.7   \\
Longest gap [days]&  780  & 352 & 375 & 356   \\
Average brightness [mag]& 16.319  &  15.760 & 15.301 & 14.693    \\
Max brightness [mag]& 15.863  & 14.544 & 14.16 & 13.59    \\
Min brightness [mag]& 17.453  & 16.94 & 16.87 & 15.97    \\
Variab. range $\Delta m$ [mag]& 1.59  & 2.39 & 2.71 & 2.38  \\
Absorption coeff.$^{\dag }$ [mag]& 0.152 & 0.117 & 0.094 & 0.068  \\
Data standard deviation & 0.256  &  0.368 & 0.515 & 0.453    \\
Data skewness & 1.23  & 0.386 & 0.329 & 0.155    \\
Data kurtosis &  3.791  &  1.087 & 0.019 & 0.400    \\
Max flux [mJy] & $2.21 $  & $6.1 $ & $7.3 $ & $9.9 $    \\
Min flux [mJy] & $0.51 $  & $0.67 $ & $0.60 $ & $1.1 $   \\
\hline \hline
\end{tabular}
\begin{list}{}{} \item[$\dag $] Values for the galactic extinction by NED database
\citep{schlegel98}.
\end{list}
\end{table}
%
\section{Optical light curves from 1993 to 2004}\label{par:data1993-2004}
We monitored the BL Lac object PKS 0735+178 in the four
$B,V,R_{c},I_{c}$ optical bands for more than 10 years, from
February 2, 1993 to February 17, 2004 (JD=2449021--2453053). A
total of 1332 $BVRI$ reduced and validated photometric points were
obtained over a period of 4032 days (see Fig.\
\ref{fig:clucedatinostri}). In order to obtain a more complete
light curve, data from the Shanghai Observatory
(Jan.1995-Dec.2001) are added \citet{qian04}. In the best sampled
band (the $R$--band, analyzed in detail in
Sect.\ref{par:timescales}), 709 photometric points were collected
over 12 observing seasons (see Tab.\ref{tab:samplingprop} and
Tab.\ref{tab:timescalestable}) with 459 nights in total having at
least one $R$ data point. The last and best sampled 10 observing
seasons (from the III to the XII, i.e. from October 1994 to
February 2004) have a duration spanning from 144 to 203 days
Tab.\ref{tab:timescalestable}), an average number of data points
per night equal to 1.5, an average empty gap between subsequent
observations of 3 days, and an average coverage of nights with
data respect to each season duration of about 27\%. Practically
speaking such numbers mean that data are not clustered or bunched,
and that an enough regular monitoring was performed (when
permitted by atmospheric/technical conditions). The priority of
our observing programme during these years was to perform a
constant and possibly uniform optical monitoring. Consequently,
for the first time in PKS 0735+178, this allowed to obtain data
suitable for a deep and detailed statistical analysis on
days/weeks timescales. The majority of such $R$--filter
observations were obtained by only two telescopes (over the $69\%$
obtained by the Perugia telescope, and a further 21\% by the
Torino telescope) having known inter--instrumental offsets below
0.1 mag in this band \citep[e.g. ][]{villata02,boettcher05}. Three
examples of such R-band seasonal light curves with the
accompanying time series analysis functions are reported in
Fig.\ref{fig:statisticalplotIVseason},
Fig.\ref{fig:statisticalplotVIIseason} and
Fig.\ref{fig:statisticalplotXseason}.
\par A direct visual inspection of our 10-year multiband
light curve (Fig.\ \ref{fig:clucedatinostri}), show an average
optical brightness placed at a mid or low levels, displaying rapid
variability with a moderate flaring but no extraordinary
big/isolated outburst ($R>14$ in the whole data set). Luminosity
drops/increasing of about 2 magnitudes were common on time
intervals smaller than half year. From the end of 1997, a slow
increase of the average brightness was clearly detected (e.g. the
$R$-band magnitude never dropped to values higher than 16 from
beginning of 1998), while in 2001 a clear brightening phase can be
well identified. This moderate-outburst phase can be considered
comparable to the other outbursts seen in the optical history of
PKS 0735+178 (see, Sec. \ref{par:dataHistorical}).
%
\section{The historical light curve}\label{par:dataHistorical}
\begin{figure*}[t!!] \centering
\hspace{-0.5cm}\resizebox{\hsize}{!}{\includegraphics[angle=90]{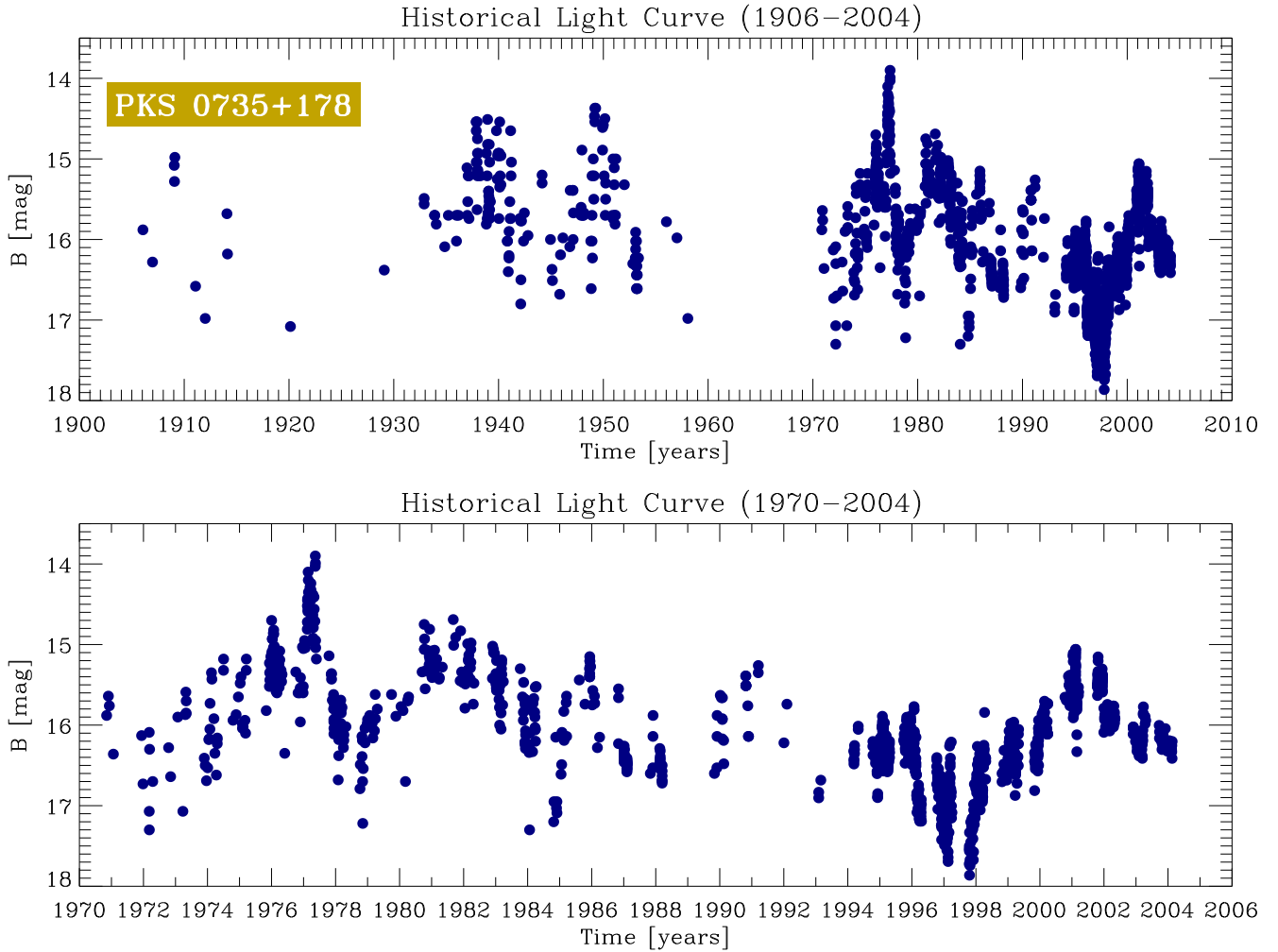}}
\vspace{-0.2cm} \caption{The historical optical light curve of PKS
0735+178 in $B$ band reconstructed by published data \citep[data
points mainly from the compilation of ][ with more few data from
other literature]{qian04} and adding our original $B$ and
$R$-derived data (see text). Error bars are not represented for
clarity. The total historical time series obtained in this way is
composed of 1725 final data points. At least five main outbursts
followed by a general humped and rather oscillatory long-term
trend can be visually recognized. This behaviour is identifiable
better in the 1970-2004 portion
(lower panel), thanks to a higher sampling (lower panel).}
\label{fig:storica}
\end{figure*}

The optical history of PKS 0735+178 together with our data (Fig.
\ref{fig:storica}) extends over almost one century (from Jan 22,
1906, JD 2417233, to Feb 17, 2004, JD 2453053).
%
%
The older points in the light curve were obtained using plates
\citep[mainly from the Landessternwarte Heidelberg-K\"{o}nigstuhl
Observatory, Germany, and the Rosemary Hill Observatory, Florida,
USA ][]{zekl81,webb88}, from which a photographic magnitude
$m_{pg}$ can be extracted and converted in the photometric $B$
magnitude following a semi--empirical correction \citep[see, e.g.
][]{lu72,kidger89}. More recent data have been obtained directly
with photoelectric or CCD instruments. The historical data
collection was taken directly from \citep{qian04} with few
additions and appending our original and derived $B$-magnitudes
\citep[the derived $B$-band data are estimated from our best
sampled $R$-mag data after 1993, using a constant colour index
with value equal to previous works
 $B-R=0.993$, ][]{fan97,qian04}. In general a prudential error
estimation (taking into account different offsets, different data
quality, systematic errors and instrumental dispersion), needs to
be figured out, before to use heterogeneous historical optical
light curves for a quantitative analysis. In this specific case
the further errors introduced by using a constant conversion index
from the $R$ and $B$ band has also to be counted on. This
estimation is difficult without all the original data sets
(plates, frames, etc.), but basing on experience a reasonable and
prudential upper limit to the errors in Fig.\ref{fig:storica}
might be considered around the value of $\pm 0.4$ mag.
\par The largest outbursts or brightening phases (mag $B\lesssim 15$)
occurred in the period Dec.1937-Feb.1941, Apr.1949-Feb.1950,
around Feb.-May1977, in the period Oct.1980-Mar.1981, and
Feb.2001-Oct.2001. The brightest outburst was observed around mid
of May 1977 (JD 2443277-78), when PKS 0735+178 reached its
historical optical maximum ($B=13.9$). A rather humped, swinging
and oscillating long-term trend appear to modulate the rapid
variability of PKS 0735+178. This aspect might suggest a cyclical
or intermittent trend, with possibly pseudo-periodic or
multi-component oscillations running on long-term ranges.
%
%
\begin{figure}[t!]
\centering
\begin{tabular}{l}
\hspace{-5mm}
{\resizebox{\hsize}{!}{\includegraphics{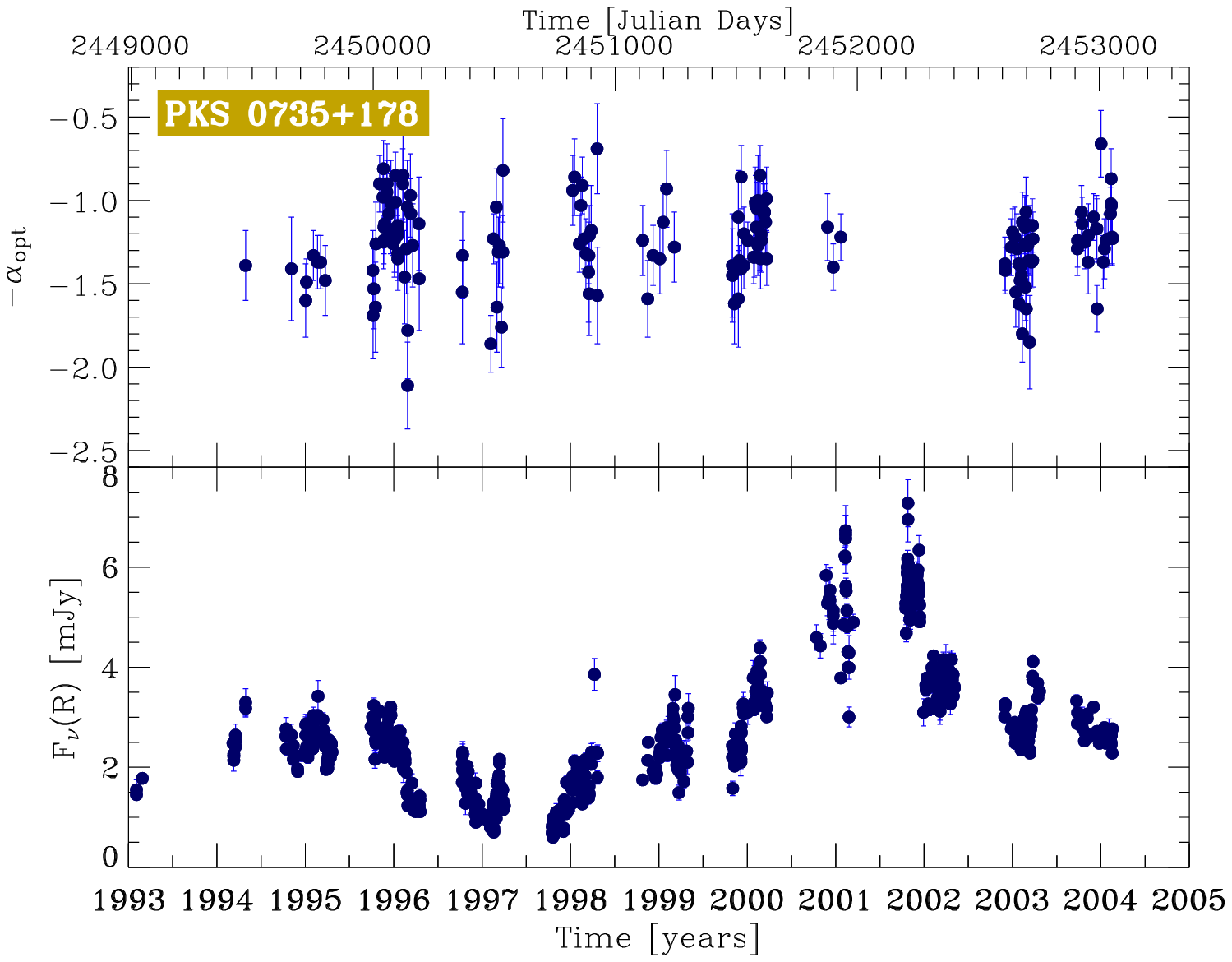}}} \vspace{-5mm}\\
\hspace{-3mm} \vspace{-5mm}
{\resizebox{8.3cm}{!}{\includegraphics{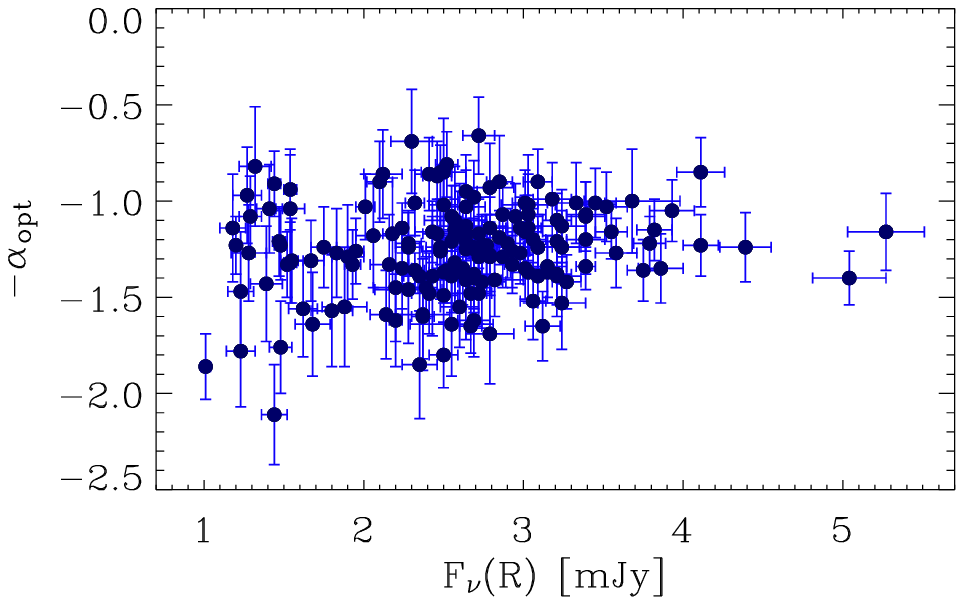}}} \\
\hspace{-3mm} \vspace{-5mm}
{\resizebox{8.3cm}{!}{\includegraphics{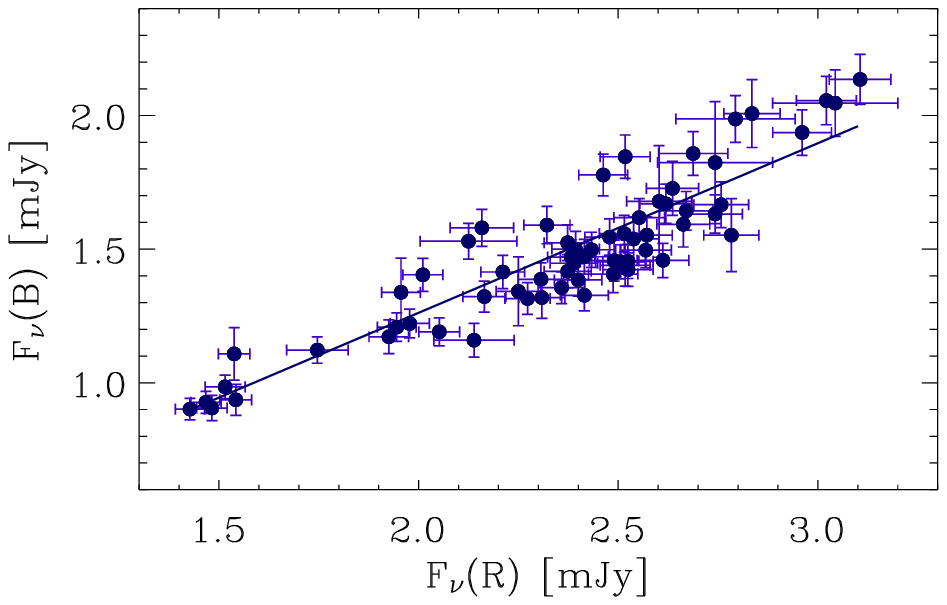}}} \\
\end{tabular}
\caption{\textit{Upper panel}: temporal behaviour of the optical
spectral index $\alpha$ and the $R$-band flux in PKS 0735+178.
This plot shows that the long--term variability is essentially
achromatic, even if flares can imply spectral changes.
\textit{Middle panel}: dependence of the optical spectral index
$\alpha$ on the flux intensity in the $R$-band. The scatter in the
data can be explained as produced by statistical fluctuations due
to instrumental and calculation errors, and by the intrinsic
scatter of $\alpha$ during the more rapid and larger flares. A
weak linear correlation is outlined. \textit{Lower panel}: the
scatter plot between the fluxes in $R$ and $B$ bands shows, as
well, a linearly correlated emission (linear correlation
coefficient $r_{B-R}=0.90\pm 0.18$ and slope $0.64 \pm 0.04$).}
\label{fig:alphavsflux}
\end{figure}
%
\section{Optical spectral indexes}\label{par:specindexes}
The continuum spectral flux distribution of blazars in the optical
range can be analyzed to distinguish properly all the emission
components that, together with synchrotron radiation, contribute
to the observed spectrum shape. Moreover optical flux variations
in blazars are frequently associated by changes in the spectral
shape. This can be revealed by analyzing the magnitude color
indexes or the flux spectral indexes. In calculating the color
indexes and the continuum spectral slopes, we selected the more
accurate multi-band data (3 filter at least) about PKS 0735+178,
obtained by a single telescope, and coupling frames with a maximum
time lag of 20 minutes (in order to reduce possibly
intrinsic/extrinsic, instrumental micro--variations). Since both
the host galaxy of PKS 0735+178 and the feature possibly
responsible for the intervening absorption at $z=0.424$ are rather
faint (see sect. \ref{par:0735optpropintro}), it is reasonable to
neglect the galaxy color interference and any thermal contribution
in the observed continuum optical spectra. The observed magnitudes
were transformed into flux densities, corrected by the Galactic
absorption \citep[derived by ][ see Tab. \ref{tab:samplingprop}
lower panel]{schlegel98} for the source (located at moderate
Galactic latitude, b=18.07). The absorption is rather small (i.e.
$B-I = 0.8$ mag), accordingly the colour correction is little in
comparison to the mean $B-I$ value of the source. Fluxes relative
to zero-magnitude values are taken from the Johnson-Cousins system
calibration presented in \citet{bessell79,bessell90} and
\citet{fiorucci03}.
\par The optical spectral energy distribution (SED), can be expressed
conveniently by a power law $\nu F_\nu \propto \nu^{-\alpha+1}$,
($\nu$ being the frequency of radiation and $\alpha$ the spectral
index). In the optical regime the degree of correlation between
$\alpha$ and the flux sheds light on the non-thermal emission
processes (e.g. synchrotron and inverse-Compton processes),
produced by a population of relativistic electrons in the jet. The
degree of correlation between the spectral index $\alpha$ and the
flux in various bands, through a least-square linear regression
was checked, and values characterized by large errors and bad
$\chi^2$ were rejected. We found that the spectral index $\alpha$
of PKS 0735+178 varies between $2.11 \pm 0.26$ to $0.66 \pm 0.20$,
with an average value of $1.25 \pm 0.15$ in the last ten years.
These values for $\alpha$ are roughly in agreement with the values
calculated previously in \citet{fiorucci04} using only the Perugia
Observatory data set.
%
\par In Fig.\ref{fig:alphavsflux} (upper panel) the temporal behaviour
of $\alpha$ is represented in comparison with the flux light curve
in the better-sampled $R$-band. The long--term variability seems
essentially achromatic, and there is no obvious correlation
between the light curve and the spectral index, whereas flares and
short term variations can imply spectral changes. This is the same
behaviour found in BL Lac \citep{villata02,villata04} and in S5
0716+71 \citep{ghisellini97,raiteri03}. Unfortunately there is
almost no spectral information during the 2001 outburst (lack of
$B$ and $V$ data), to check a spectral flattening. The few data
suggests a rather constant spectral index. In the same
Fig.\ref{fig:alphavsflux} (middle panel) the scatter plot between
$\alpha$ and the flux is reported. Data dispersion is evident, and
can be explained as statistical fluctuations due to uncertainties
and to scattering in $\alpha$ during the more rapid and larger
flares. Such linear correlation seems weak and a general spectral
flattening is not detected clearly. Uncorrelated random
fluctuations in the emitted flux might introduce a statistical
bias, due to the spectral index dependence by the flux
\citep{massaro96}, but values computed for the central frequency
(close to the $R$-band, as plotted in Fig. \ref{fig:alphavsflux}),
can be considered unbiased and representative of the brightness
state. The spectral index $\alpha$ showed consistent variations
even when the light curve has a rather small variations. On the
contrary in the lower panel of Fig. \ref{fig:alphavsflux} the
scatter plot between the fluxes in the $R$ and $B$ bands shows a
well correlated emission as expected (linear correlation
coefficient $r_{B-R}=0.90\pm 0.18$ and slope $0.64 \pm 0.04$),
without a detectable curvature.
\par During well-defined and large flares at X-ray bands
(especially observed in HBL), the X-ray spectral index versus the
flux frequently displays a characteristic loop-like pattern
\citep[see, e.g. ][]{georganopoulos98,kataoka00,ravasio04}. That
patterns outline a hysteresis cycle arising whenever the spectral
slope is completely controlled by radiative cooling processes
\citep[see, e.g. ][]{kirk98,boettcher02}. In few sources this
feature was found in the optical regime too
\citep{fiorucci04,ciprini04}. Consequently we can claim that
around and beyond the synchrotron peak frequency, the behaviour of
the LBL sources during flares in the optical band, is scaled in
frequency but possibly very similar to the behaviour of the HBL in
X-rays bands.
%
\begin{figure*}[t!!]
\centering
\begin{tabular}{l}
\hspace{-9mm}
{\resizebox{18.8cm}{!}{\includegraphics{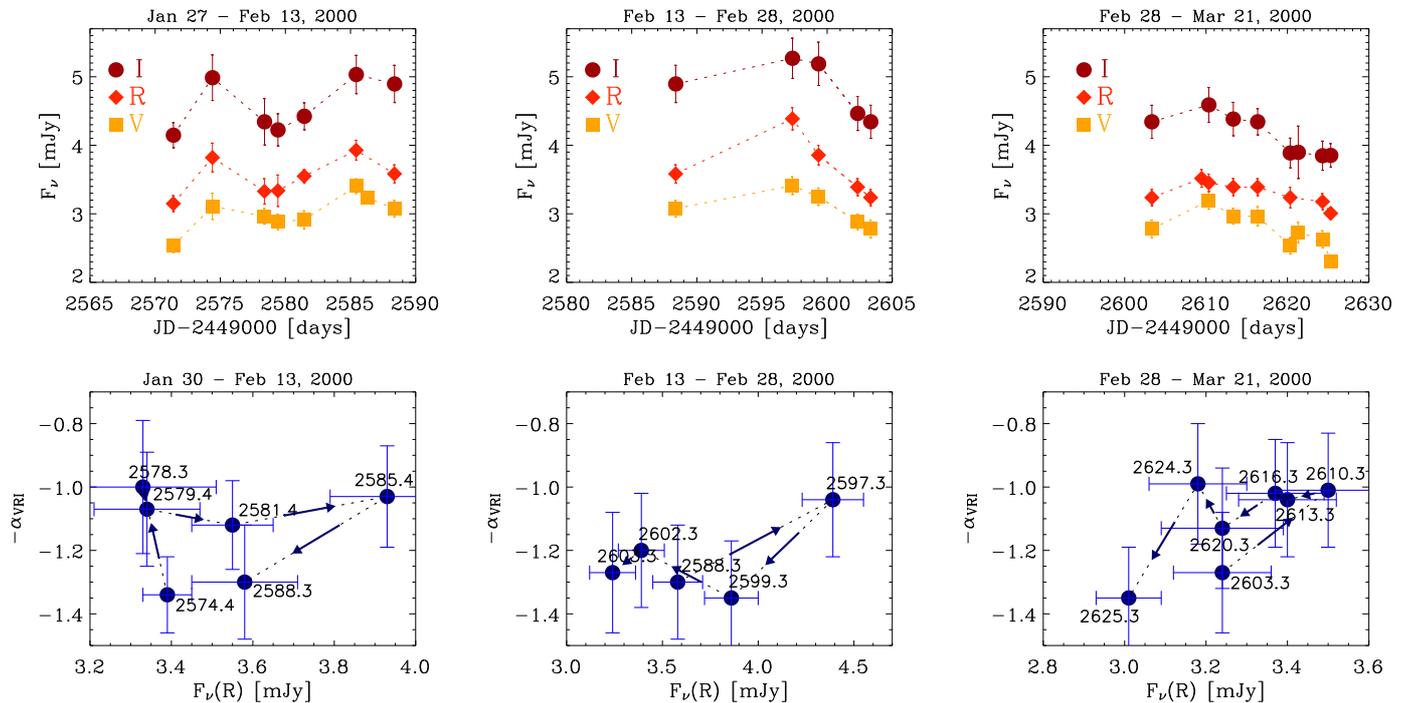}}} \\
\end{tabular}
\caption{Evolution of the continuum optical spectrum of PKS
0735+178 (spectral index $\alpha$) as a function of the flux in
$R$-band, during 3 contiguous periods of flickering variability
with moderate flaring activity observed by the same telescope
(Perugia Observatory, in period: January 29 - March 21, 2000). An
approximative loop-like behaviour of $\alpha$ is hinted in the
first (clockwise) and in the third (counterclockwise) patterns
(see text for details).} \label{fig:3loops}
\end{figure*}
%
\par In our 10-year light curve, PKS 0735+178 showed several moderate-amplitude
outbursts, wider bump of longer duration, and a general flickering
or shot-noise type of variability on mid-term scales. The
evolution of $\alpha$ as a function of the flux is erratic and did
not show evident hysteresis loops caused by non-thermal cooling.
In Fig. \ref{fig:3loops} the evolution of $\alpha$ during 3
contiguous observing periods (from January 29, to March 21, 2000)
is reported as example. A rough loop-like behaviour is hinted,
meaning that radiative cooling can dominate the optical SED also
during mild-flaring activity. Consequently variations at higher
frequency band could lead those at the lower frequency bands
during both the increasing and decreasing brightness phases,
reflecting differences in electron cooling times.
\par
The rather limited amplitude of the optical variability in the
epochs of Fig. \ref{fig:3loops}, the possible superimposition of
different emission processes in the optical band, the
under-sampling and the error propagation in the $\alpha$
calculation, can be the main reasons for the lack of well defined
loops in such $\alpha$ vs flux diagrams. Our data are not
sufficient to make a final judgement, and an improved multi-band
monitoring and a better data sampling would probably clarify  the
existence of that patterns also during mild variability in this
object.
%
\section{Temporal variability analysis}\label{par:timescales}
%
Time series analysis (evolved from both signal-processing
engineering and mathematical statistics) provides very useful
methods to study blazar variability. These methods allows to
explore and extract temporal signatures, structures and
characteristic timescales (the powerful scales of variations),
duty cycles (the fraction of time spent in an active state) and
trends, to determine the dominant fluctuation modes and the power
spectrum of the signal. Moreover time series analysis allows to
detect and study auto/cross-correlations, time lags, transient
events, periodicity and composite modulations, scaling and
coherency, oscillations, beatings and instabilities, intermittence
and drifts, dissipation, dumping, long-memory patterns and
self-similarity, resonance and relaxation processes, random and
deterministic features, linear and non-linear processes,
stationary and non-stationary activity, as well as to perform
filtering and forecast.
%
%
\begin{figure}[b!!] 
\begin{center}
\begin{tabular}{c}
\hspace{-3mm}{\resizebox{\hsize}{!}{\includegraphics{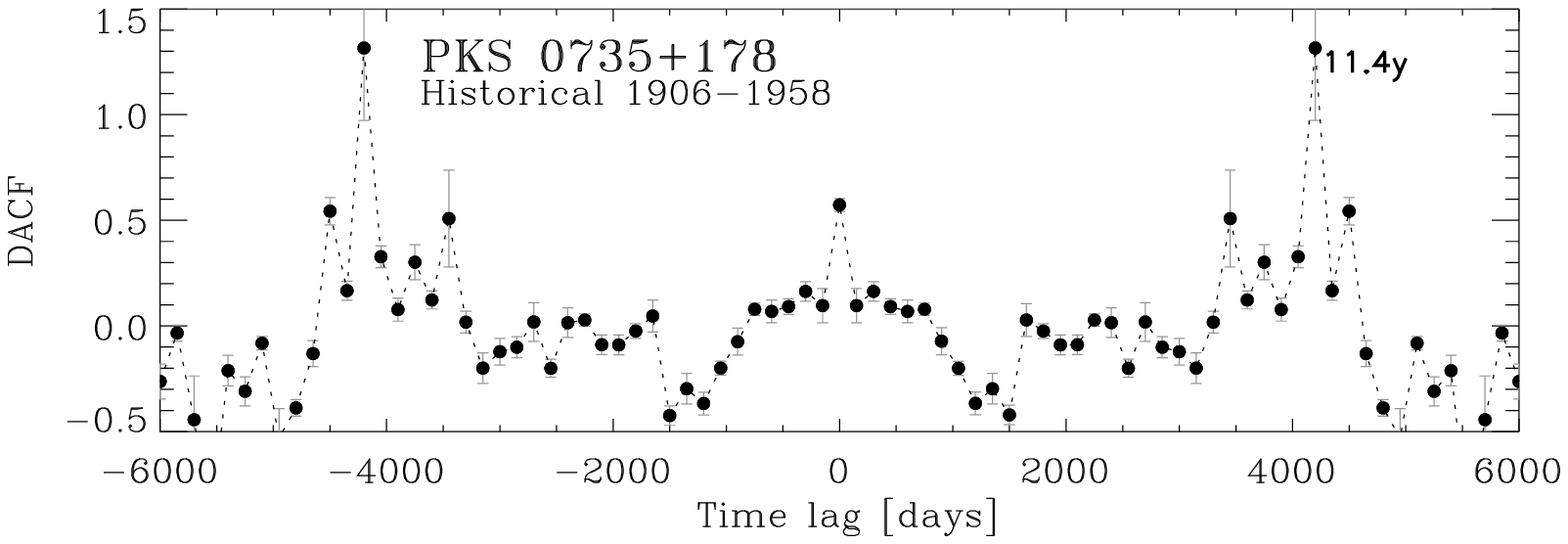}}} \\[-3mm]
\hspace{-3mm}{\resizebox{\hsize}{!}{\includegraphics{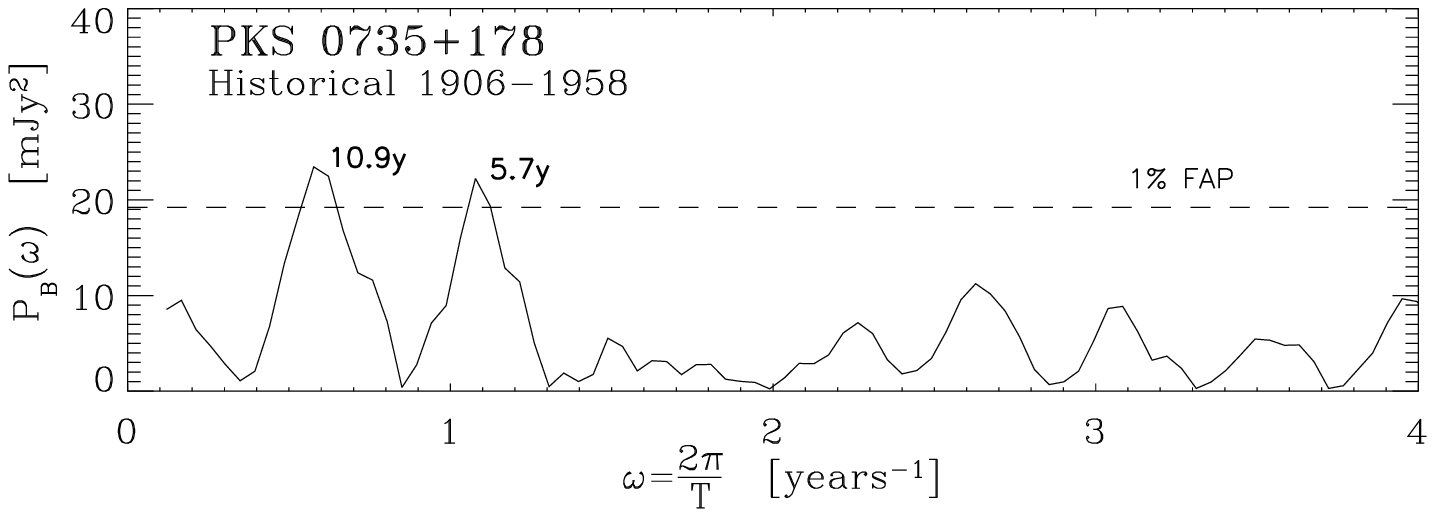}}} \\[-3mm]
\vspace{-5mm}
\end{tabular}
\end{center}
\caption{Discrete Autocorrelation Function (DACF; data bin: 7
days, DACF bin: 150 days) and Lomb-Scargle Periodogram (LSP;
dashed line indicates the threshold of false alarm probability
[FAP] fixed at 1\%) of the pre-1970 light curve. A characteristic
timescale around 11 years is hinted also by this poorly sampled
portion of the historical light curve.}
 \label{fig:statisticalplot19061958}
\end{figure}
%
The analysis of the flux evolution over time in a blazar, joint
with the multiwavelength and cross-correlation analysis, provides
crucial information on the location, size, structure and dynamics
of the emitting regions, and shed light on physical mechanisms of
particle acceleration and radiation emission.
\par In this section a quantitative analysis of the
optical variability observed in PKS 0735+178 is performed using 7
different methods in 13 different light curves. The aim of this
work is to examine in detail the optical behaviour on long-term
timescales (months/years) using the whole 1906-2004 historical
light curve, the pre-1970 portion and the best sampled 1970-2004
part, and to explore mid-term scales (days/weeks in intervals
$<200$ days) using our 10-year monitoring data set with improved
sampling. Each seasonal light curve in the best sampled $R$-band
is analyzed separately (a part of the first 2 pilot seasons, where
we obtained only few observations). For the first time a rather
comprehensive temporal analysis of the optical variability was
performed over 3 decades in time in this peculiar blazar,
investigating scales between about 2 and 30 years regarding the
historical dataset, and between few days and about 100 days in the
seasonal monitoring (the duration of season with data points spans
between 144 and 203 days).
%

%
\begin{figure*}[htp!!!]
\begin{center}
\begin{tabular}{c}
\hspace{-4mm}{\resizebox{8.8cm}{!}{\includegraphics[angle=90]{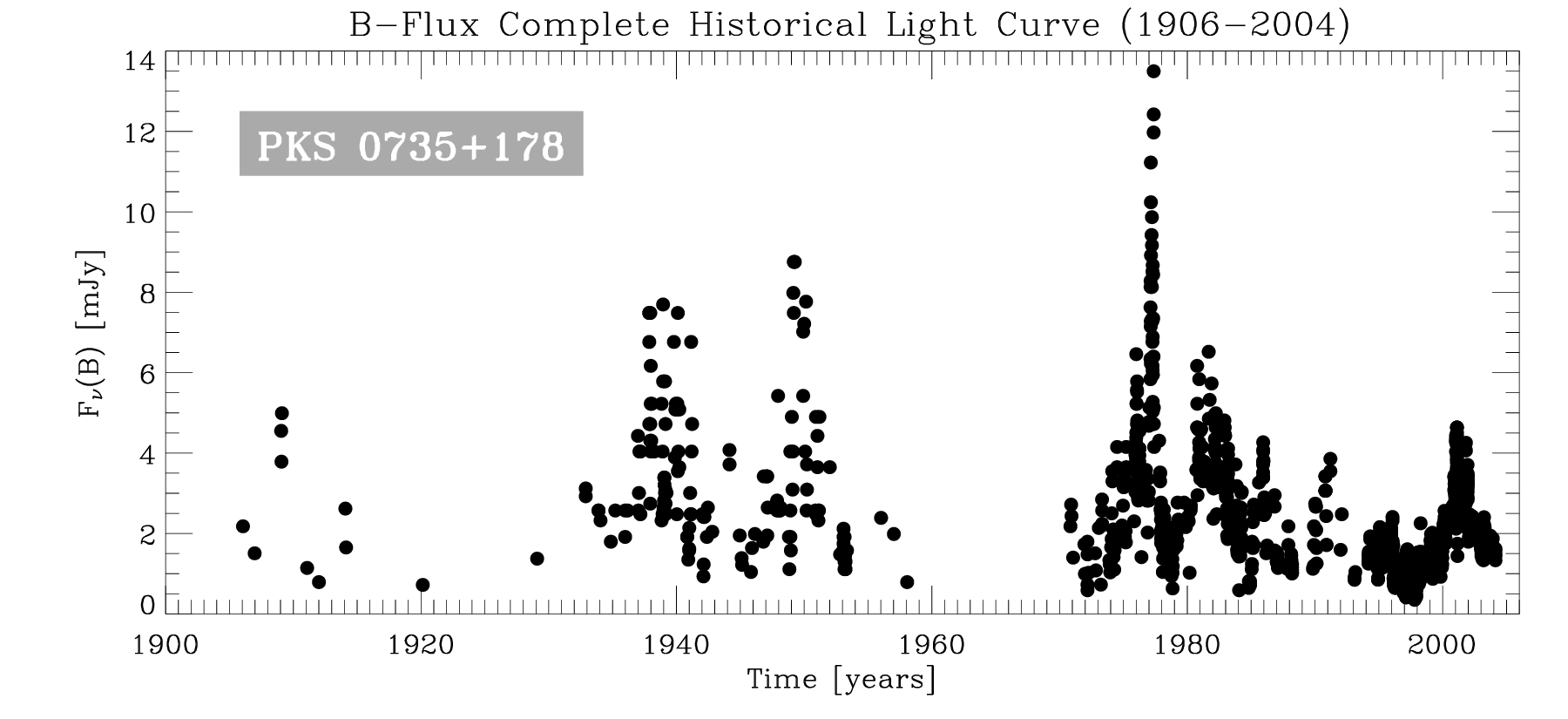}}}
\hspace{-4mm}{\resizebox{8.6cm}{!}{\includegraphics{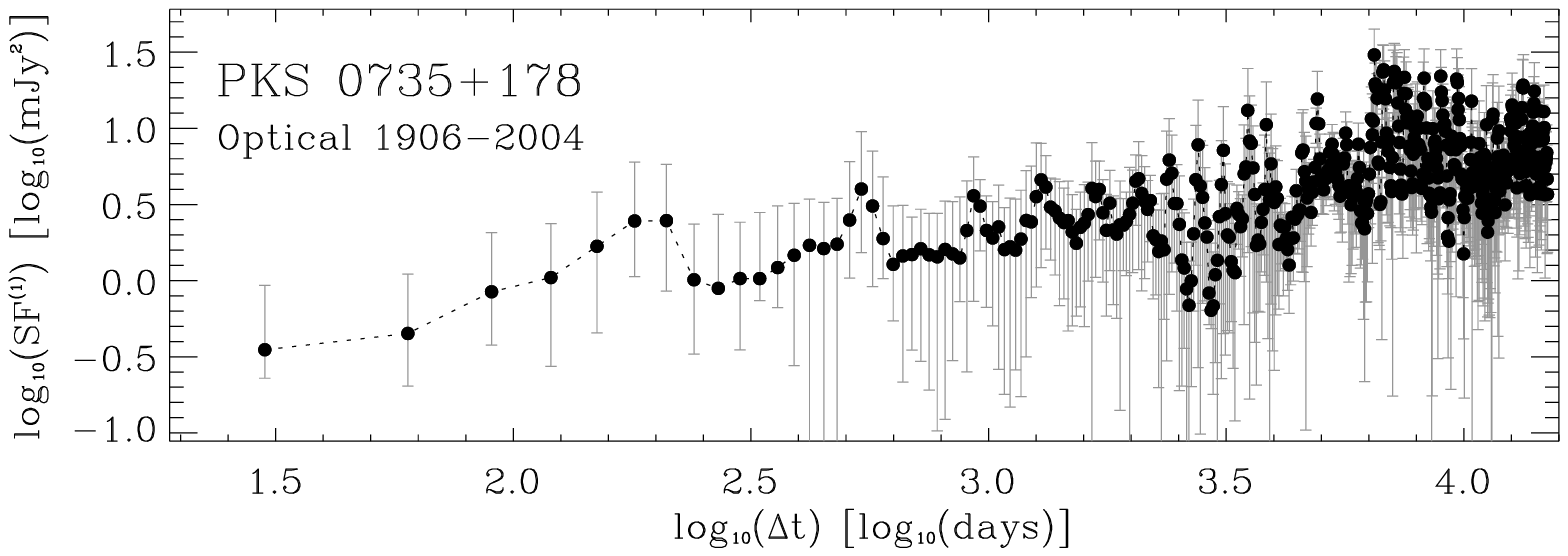}}} \\[-3mm]
\hspace{-4mm}{\resizebox{8.6cm}{!}{\includegraphics{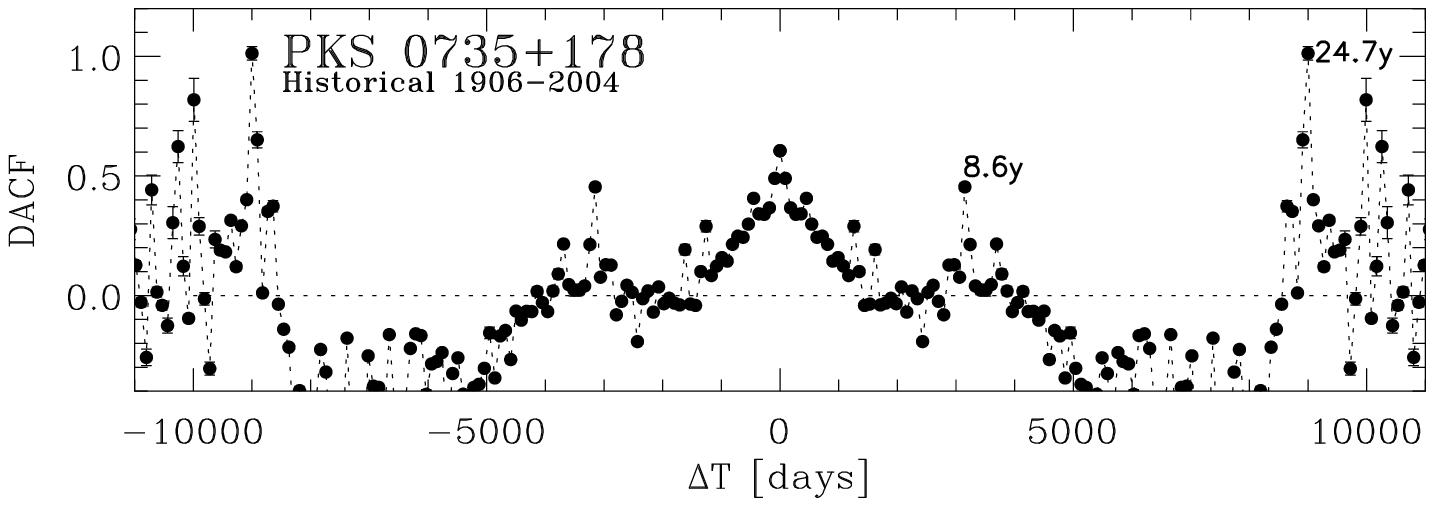}}}
\hspace{-4mm}{\resizebox{8.6cm}{!}{\includegraphics{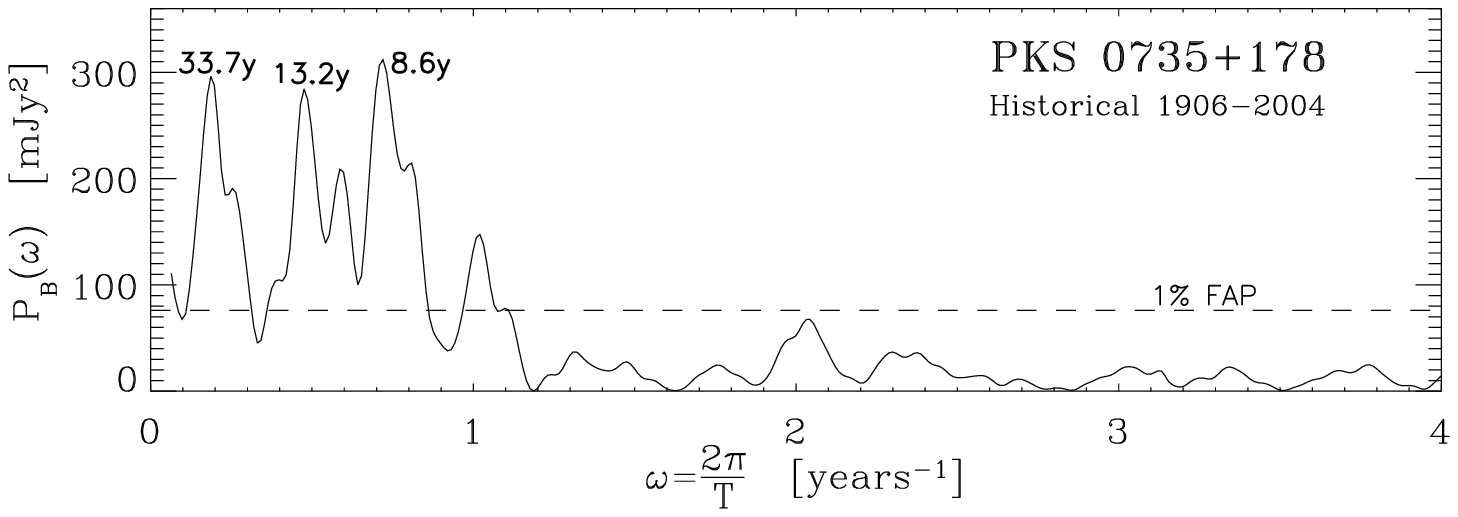}}} \\[-3mm]
\hspace{-4mm}{\resizebox{8.6cm}{!}{\includegraphics{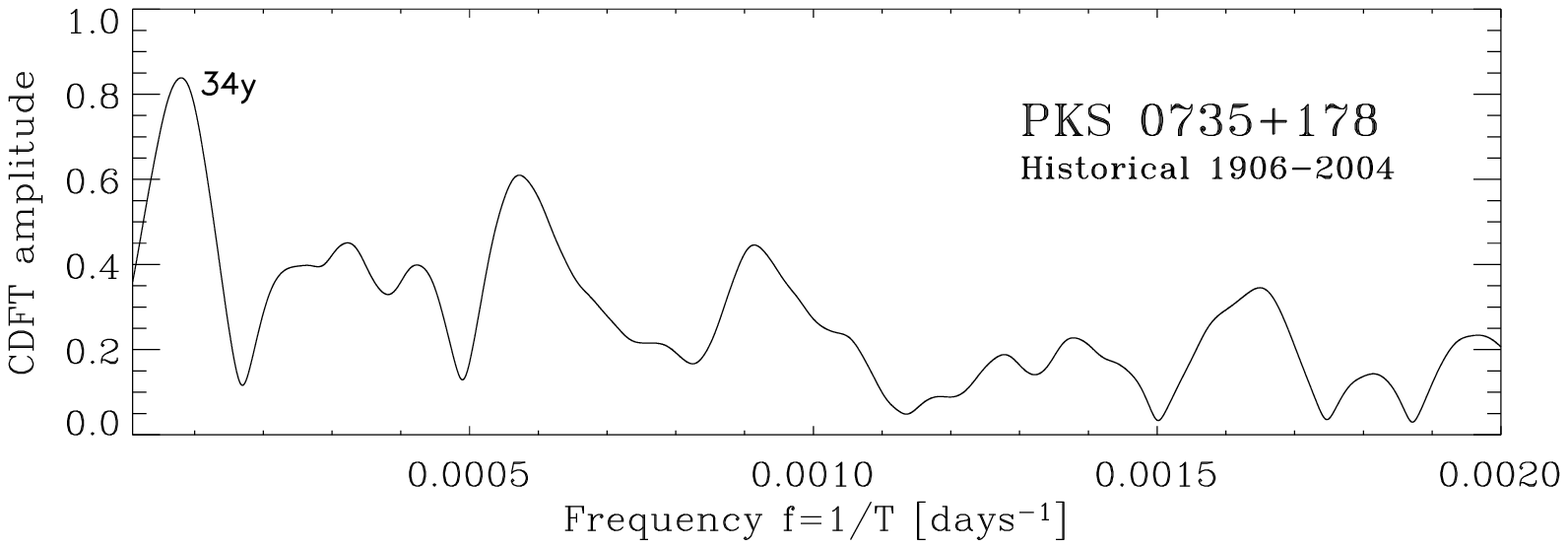}}}
\hspace{-4mm}{\resizebox{8.6cm}{!}{\includegraphics{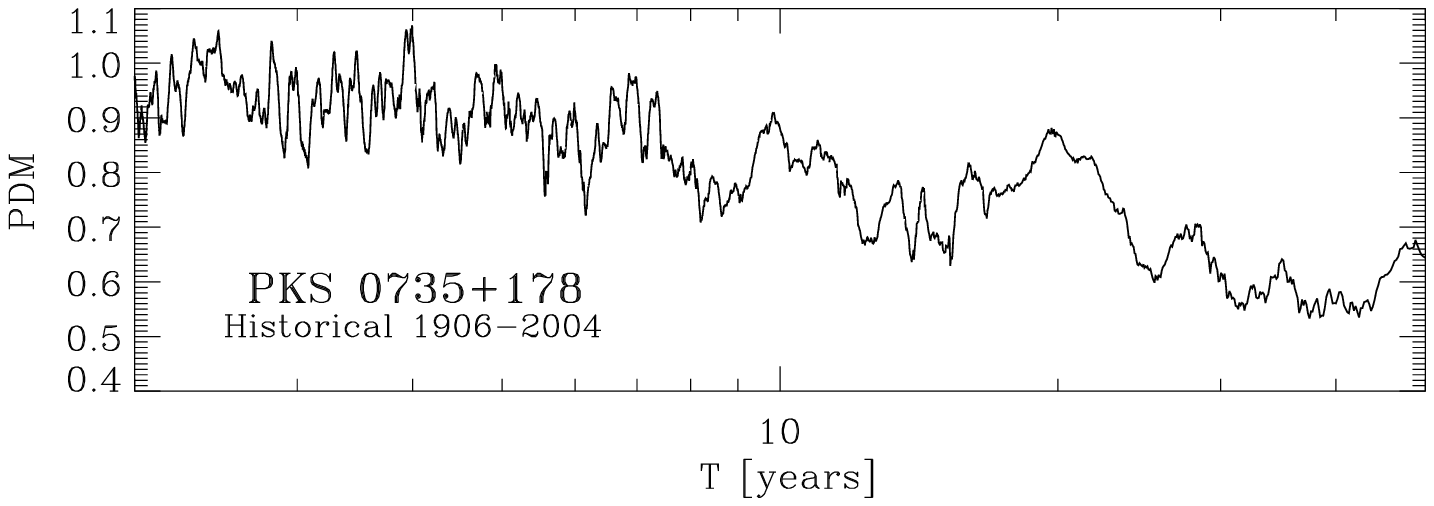}}} \\[-3mm]
\hspace{-4mm}{\resizebox{8.6cm}{!}{\includegraphics{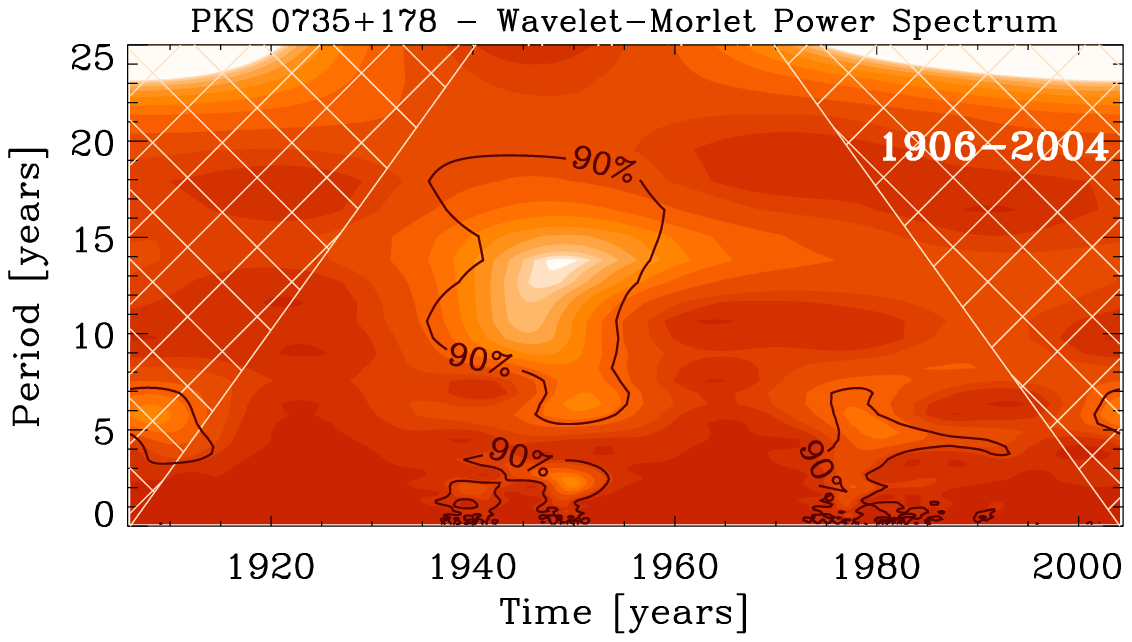}}}
\hspace{-4mm}{\resizebox{8.6cm}{!}{\includegraphics{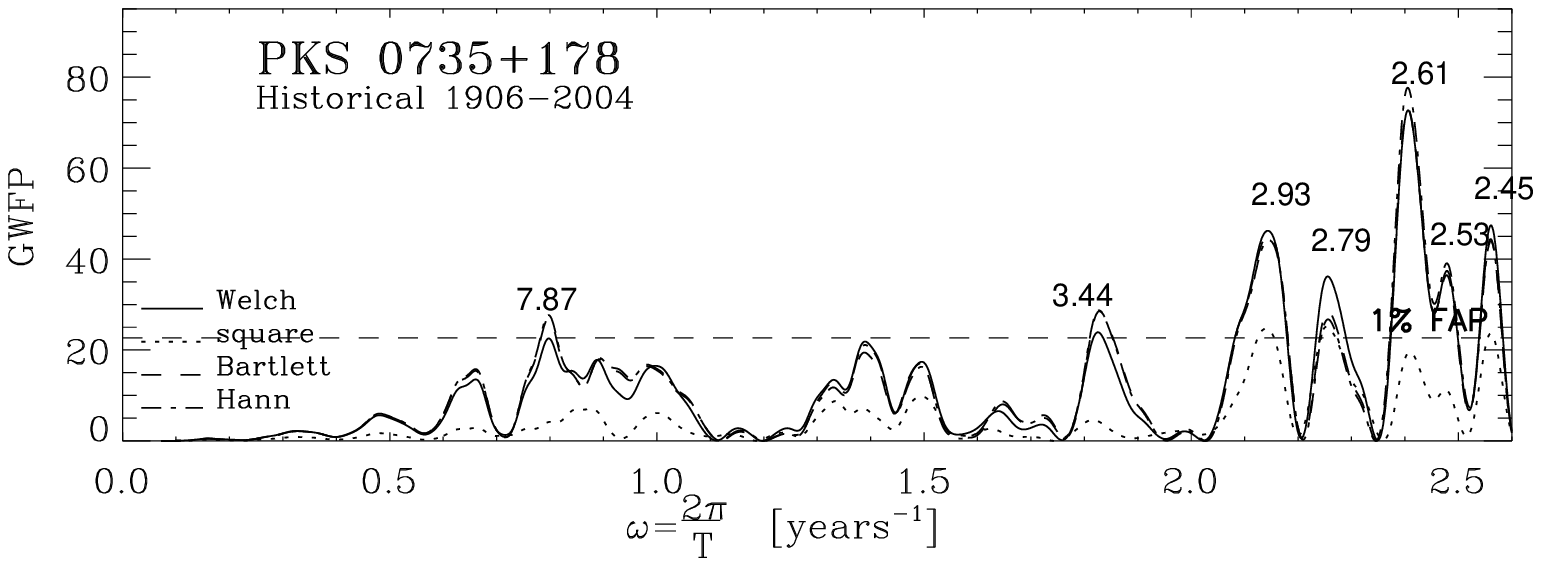}}}
\vspace{-6mm}
\end{tabular}
\end{center}
\caption{Panels from left to right and following below: the whole
historical (1906--2004) flux light curve of PKS 0735+178 in
$B$-band and plots from the related time-series analysis. First
order Structure Function (SF; data bin: 7 day, SF bin: 60 days) in
log-log representation, Discrete Auto Correlation Function (DACF;
data bin: 7 day, DACF bin: 90 days), Lomb-Scargle Periodogram
(LSP; dashed line is the 1\% false alarm probability [FAP]
threshold), ``Clean'' implementation of the Discrete Fourier
Transform (CDFT), Phase Dispersion Minimization function (PDM),
plane contour plot of the wavelet scalogram (i.e. the
two-dimensional energy density function $\left| CWT(t,T)
\right|^{2}$, CWT being the Continuous Wavelet Transform computed
using a Morlet waveform, t and T the time and period scale
respectively). In the last panel the periodogram of the synthetic
light curve constructed upon the empty gaps using different window
functions (GWFP; gap threshold 1 year). In the CWT scalogram the
power spectral density is represented by filled-colour contour
levels, while the thick black contours are the 90\% confidence
levels of true signal features against white/red noise background,
and the cross-hatched regions represent the ``cone of influence'',
where edge effects become important. The descriptions of the plots
above can be applied to the following
Fig.\ref{fig:statisticalplot19702004},
Fig.\ref{fig:statisticalplotIVseason},
Fig.\ref{fig:statisticalplotVIIseason}, and
Fig.\ref{fig:statisticalplotXseason}. Issues and results from
these diagrams are described in the text.}
\label{fig:statisticalplot19062004}
\end{figure*}
%
\begin{figure*}[htp!!!]
\begin{center}
\begin{tabular}{c}
\hspace{-4mm}{\resizebox{8.8cm}{!}{\includegraphics[angle=90]{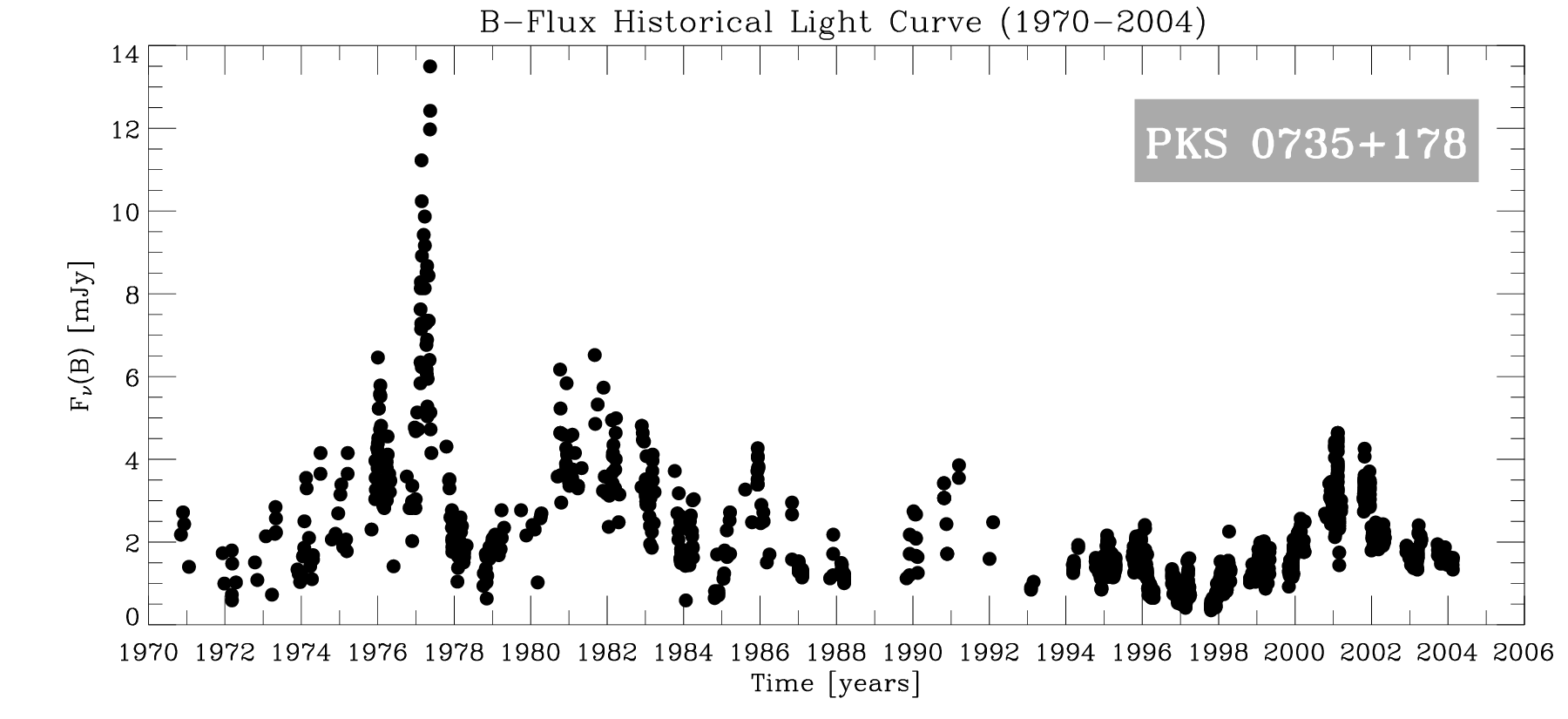}}}
\hspace{-4mm}{\resizebox{8.6cm}{!}{\includegraphics{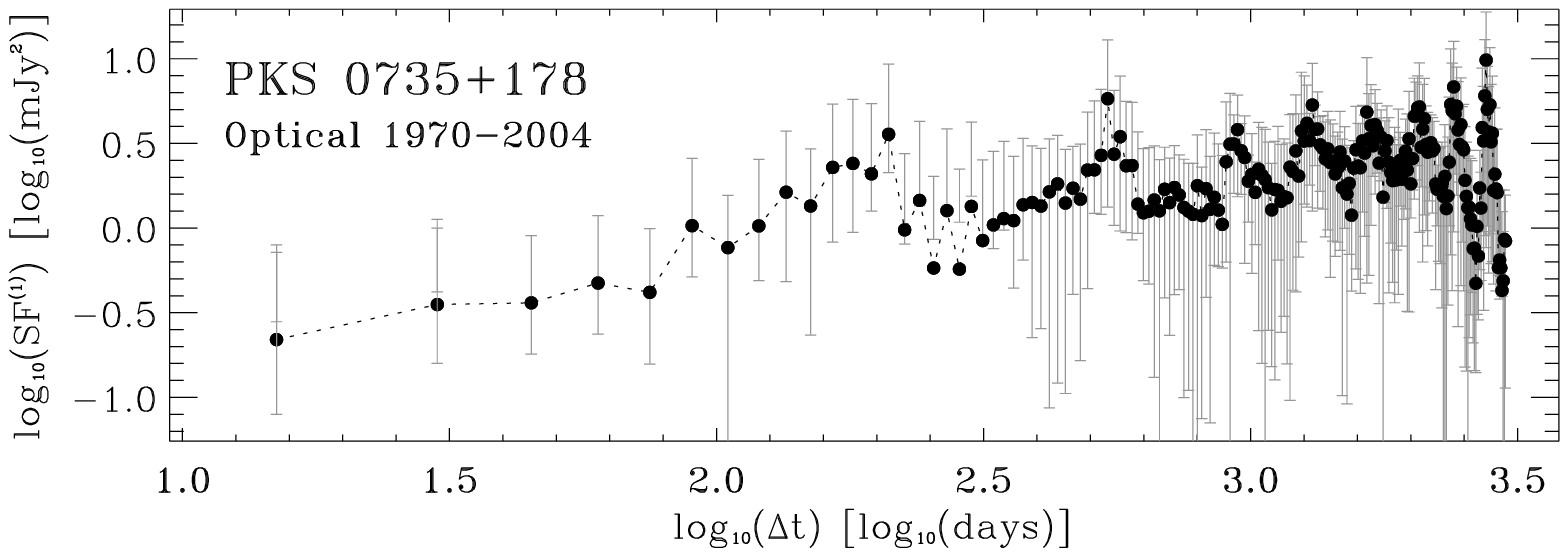}}} \\[-3mm]
\hspace{-4mm}{\resizebox{8.6cm}{!}{\includegraphics{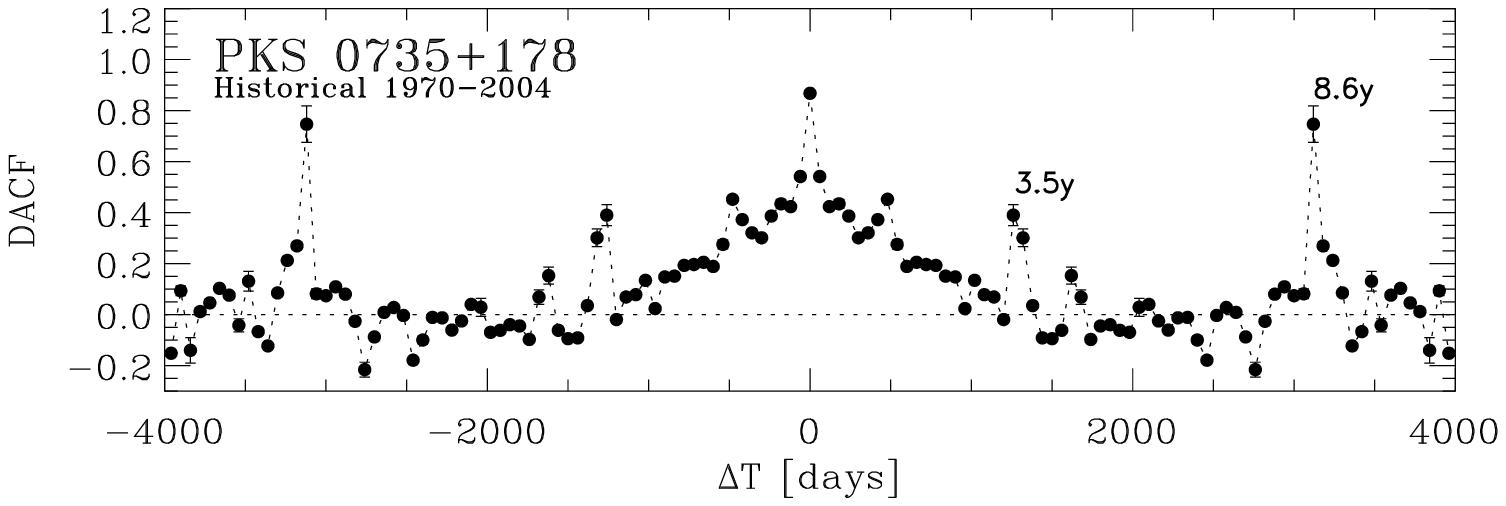}}}
\hspace{-4mm}{\resizebox{8.6cm}{!}{\includegraphics{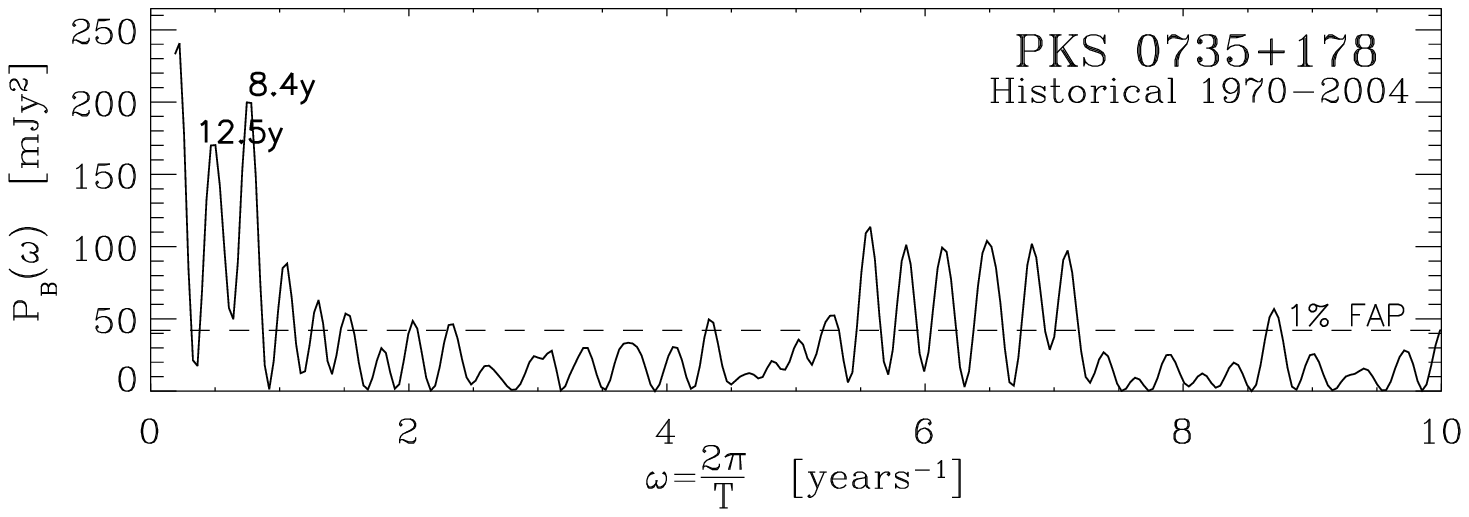}}} \\[-3mm]
\hspace{-4mm}{\resizebox{8.6cm}{!}{\includegraphics{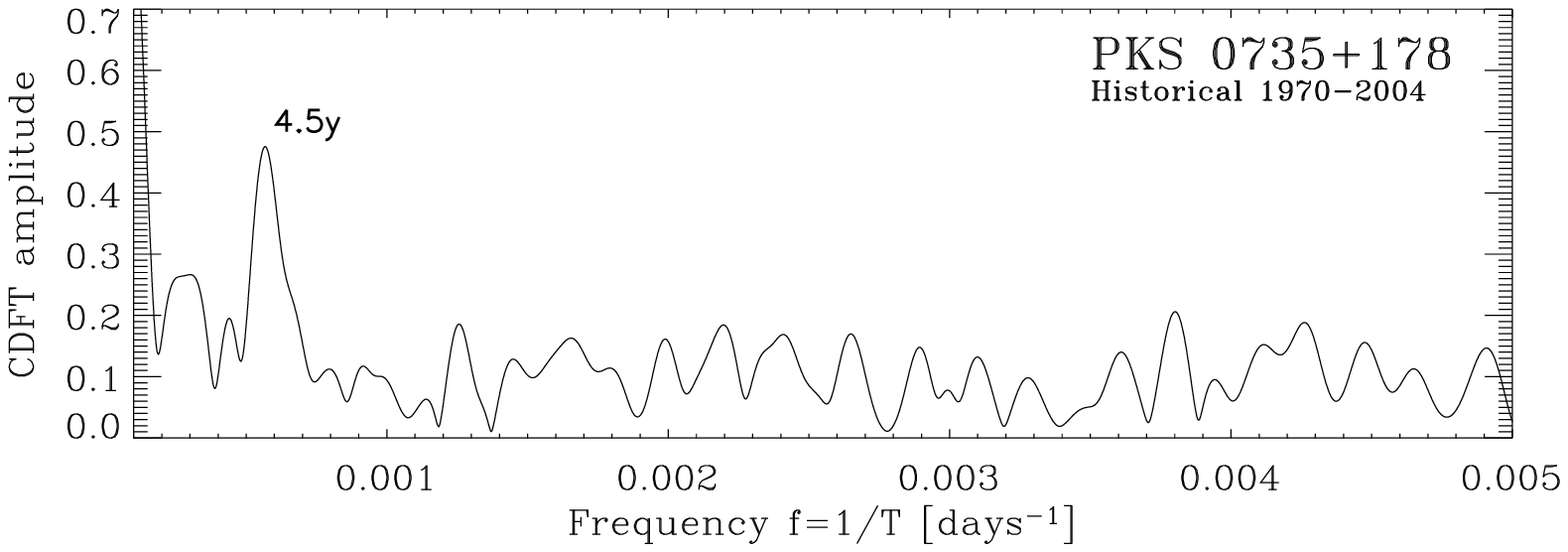}}}
\hspace{-4mm}{\resizebox{8.6cm}{!}{\includegraphics{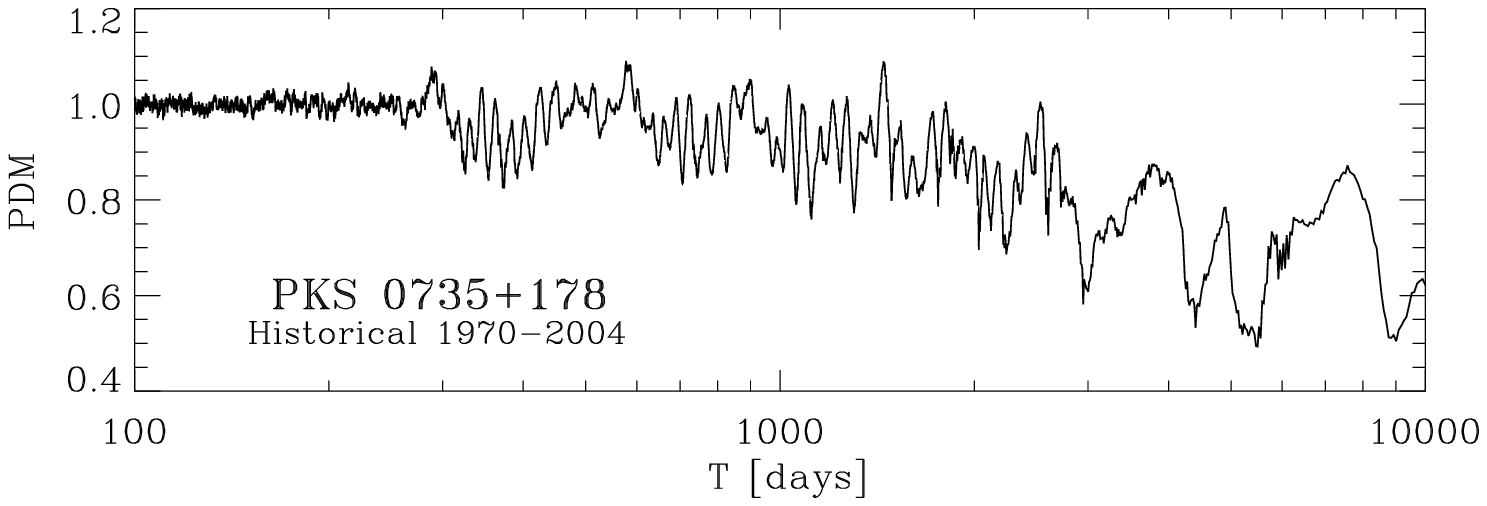}}}\\[-3mm]
\hspace{-4mm}{\resizebox{8.6cm}{!}{\includegraphics{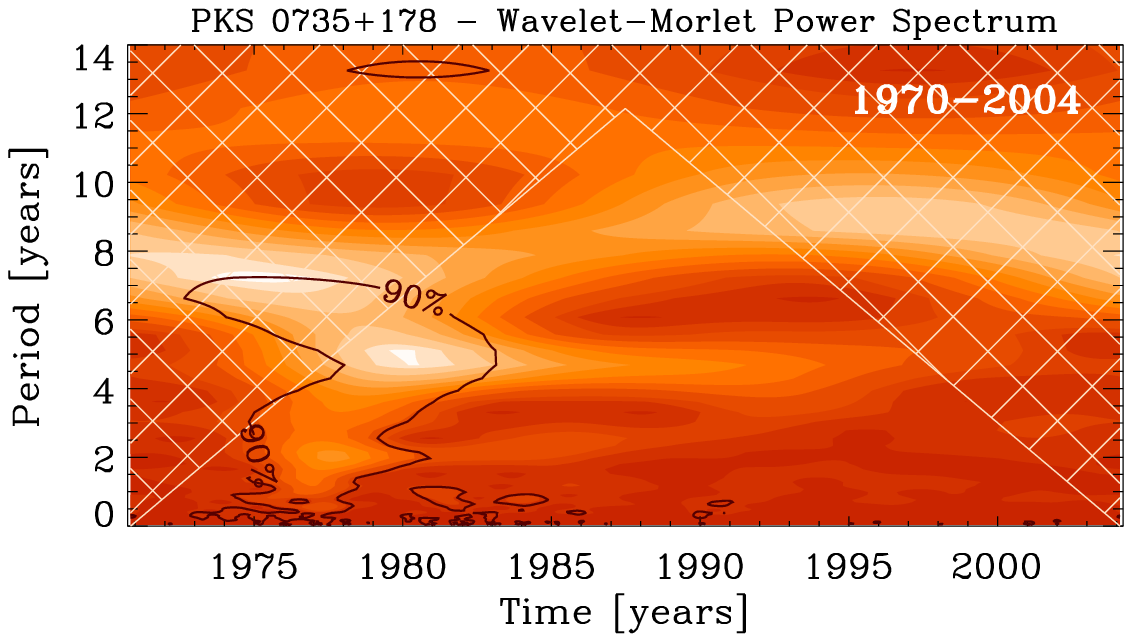}}}
\hspace{-4mm}{\resizebox{8.6cm}{!}{\includegraphics{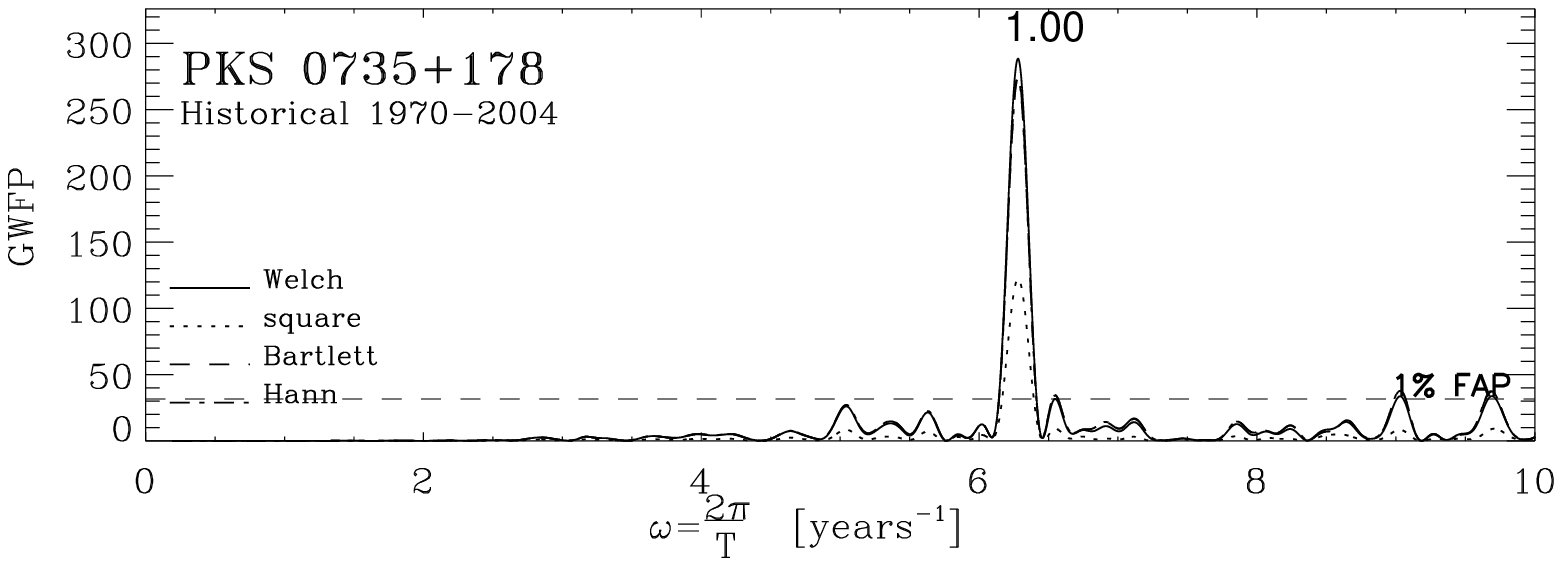}}}
\vspace{-6mm}
\end{tabular}
\end{center}
\caption{Panels from left to right and following below: the whole
historical (1970--2004) flux light curve of PKS 0735+178 in
$B$-band and plots from the related time-series analysis. SF (data
bin: 1 day, SF bin: 15 days), DACF (data bin: 1 day, DACF bin: 60
days), LSP, CDFT, PDM, Morlet-CWT scalogram, and GWFP (gap
threshold 0.3 years). Issues and results from these diagrams are
described in the text.} \label{fig:statisticalplot19702004}
\end{figure*}
%

The following methods, optimized or adapted for unevenly sampled
timeseries, are applied: the first-order Structure Function (SF),
the Discrete Auto Correlation Function (DACF), the Lomb-Scargle
Periodogram (LSP), the Discrete Fourier Transform in the ``Clean''
implementation (CDFT), the Phase Dispersion Minimization (PDM),
the scalogram of the Continuous Wavelet Transform (CWT) using
different waveforms, and the periodogram of the synthetic light
curve constructed on the empty gaps, using different window
functions (Gaps Window Function Periodogram GWFP).
\par The SF is equivalent to the power spectral density function
(PSD) of the signal calculated in the time domain instead of
frequency space, which makes it less dependent on sampling
problems, like windowing and alias \citep[see, e.g.
][]{rutman78,simonetti85,smith93}. The first order SF represents a
measure of the mean squared of the flux differences
$(F_{i}-F_{i+\Delta t})$  of $N$ pairs with the same time
separation $\Delta t$:
\begin{equation}\label{eq:structurefunction}
 {\rm SF}^{(1)}(\Delta t)=\texttt{}\frac{1}{N}\sum_{i=1}^{N}\left(F_{i}-F_{i+\Delta
 t}\right)^{2}.
\end{equation}
The general definition involves an ensemble average. Deep drops in
the SF shape means a small variance and provides the signature of
a possible characteristic time scales, but a wiggling pattern and
fake breaks (that can indicate false time scales) are common when
the sampling is not sufficient. Typically, the SF increases with
$\Delta t$ in a log-log representation, showing an intermediate
steep curve, whose slope $b$ is related to the power law index $a$
of the PSD by the relation $a=1+b$ (a typical PSD has indeed a
power-law dependence $P(f)\propto 1/f^{a}$ on the signal frequency
$f=1/t$). The maximum correlation timescale is reached when the SF
is constant for longer lags \citep{hughes92,lainela93}.
\par The DACF allows to study the level of auto-correlation in
unevenly sampled data sets \citep[see, e.g.
][]{edelson88,hufnagel92} without any interpolation or addition of
artificial data points. The pairs $(F_{i},F_{j})$ of a discrete
datasets are first combined in unbinned discrete correlations
\begin{equation}\label{eq:udacf}
  {\rm UDACF}_{ij}=\frac{(F_{i}-<F>)(F_{j}-<F>)}{\sigma_{F}\sigma_{F}},
\end{equation}
where $<F>$ is the average values of the sample and $\sigma_{F}$,
the standard deviation. Each of these correlations is associated
with the pairwise lag $\Delta t_{ij} = t_{j} - t_{i}$ and every
value represents information about real points. The DACF is
obtained by binning the ${\rm UDACF}_{ij}$ for each time lag
$\Delta t$, and averaging over the number $M$ of pairs whose time
lag $\Delta t_{ij}$ is inside $\Delta t$, i.e.: ${\rm DACF}(\Delta
t)=1/M\sum_{ij} {\rm UDACF}_{ij}$. The choice of the bin size is
governed by a trade-off between the desired accuracy in the mean
calculation and the desired resolution in the description of the
correlation curve. A preliminary time binning of data usually
leads to better results. The number of the real points per time
bin can vary greatly in the DACF, but data bins with an equal
population can be built together with Montecarlo estimations for
peaks and uncertainties, as done in the Fisher $z$-transformed
DACF method \citep[ZDACF,][]{alexander97}.
\par The LSP is a technique analogous to the Fourier analysis for
discrete unevenly sampled data trains, useful to detect the
strength of harmonic components with a certain angular frequency
$\omega=2\pi f$ \citep[see, e.g.
][]{lomb76,scargle82,horne86,papadakis93}.
\par In the CDFT method, first a ``dirty'' discrete Fourier transform
(DFT) for unequally spaced data is calculated and then an
interactive ``cleaning'' of the dirty DFT is performed \citep[see,
e.g. ][]{hoegbom74,roberts87,foster95}. The CDFT method is a
complex and one-dimensional version of a deconvolution algorithm
widely used in 2-dimensional image reconstruction. This technique
provide a simple way to understand and remove false peak artifacts
introduced by empty gaps. This method is effective especially in
describing and recognizing multiperiodic signals. A standard
(uncleaned) DFT method was implemented previously by
\citet{deeming75}.
\par The PDM method \citep[see, e.g. ][]{lafler65,jurkevich71,stellingwerf78}
try to minimize the variance of data at a constant phase with
respect to the mean value of the light curve. If a trial period is
close to a real period, the scattering of data against the derived
mean in the light curve constructed on such phase (the light curve
folded on such period) is small. The PDM method has no preference
for a particular periodical shape, it incorporates all the data
directly into the test statistic and it is well suited for small
and randomly spaced samples. A value is statistically significant
when the PDM drops towards zero.
\par The Wavelet method is used to transform a signal into another representation
able to showing the information in a more useful shape \citep[see,
e.g. ][]{daubechies92,foster96,percival02}. Wavelet transforms
(WT) permits a local decomposition of the scaling behavior in time
for each quantity (in contrast to the usual methods based on the
Fourier analysis), allowing the signal features and the frequency
of their ``scales'' to be determined simultaneously. Hence it is a
useful tool especially to detect typical timescales and identify
signals with exotic spectral features, transient information
content and non-stationary properties. WT are defined following
the Fourier theory, but wavelets can be formally described as
localized, oscillatory functions whose properties are more
attractive than sine and cosine functions.
\par WT is computed at different times
in the signal, using mother wavelets (orthogonal base functions
localized in both time and pulse spaces) of different frequency
and convolved on each occasion. In this way the power spectrum
(i.e. the modulus of the transform value) on a two dimensional
location-frequency plane is obtained (the so called wavelet
``scalogram''). A continuous WT of a one-dimensional (1D)
time-series is computed as a complex array at different times, the
real component being the amplitude and the imaginary component
providing the phase. The square of the transformed modulus gives
the wavelet power spectrum in function of both time and harmonic
frequencies. In our analysis we chose the Morlet complex-valued
waveform \citep[see, e.g. ][]{farge92}, composed of a plane wave
modulated by a Gaussian envelope of unit width:
\begin{equation}\label{eq:morlet}
\psi_{0}(\tau)=\frac{1}{\sqrt[4]{\pi}}e^{i \omega_{0}
\tau}e^{-\tau^{2}/2}
\end{equation}
where $\tau$ is the non-dimensional time parameter and
$\omega_{0}$ the non-dimensional frequency. Such continuous WT is
convoluted on the discrete sequence of the time series $\{F_i\}$
with scaled and translated versions of $\psi_{0}(\tau)$. It is
considerably faster to calculate such continuous WT in Fourier
space: the convolution theorem allows to do all the N convolutions
for a given scale simultaneously and efficiently in Fourier space
N being the number of points in the time series) using a standard
DFT \citep{kaiser94,percival02}.
\par
\begin{figure*}[htp!!!]
\begin{center}
\begin{tabular}{c}
\hspace{-3mm}{\resizebox{8.8cm}{!}{\includegraphics{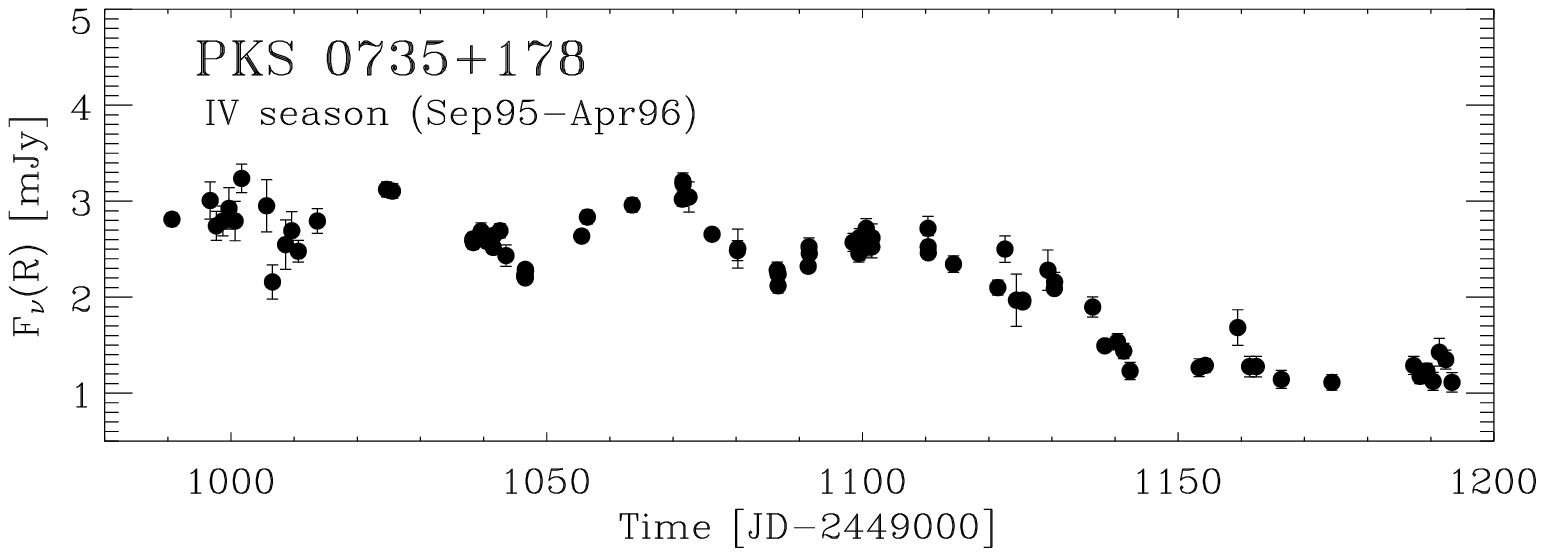}}}
\hspace{-3mm}{\resizebox{8.6cm}{!}{\includegraphics{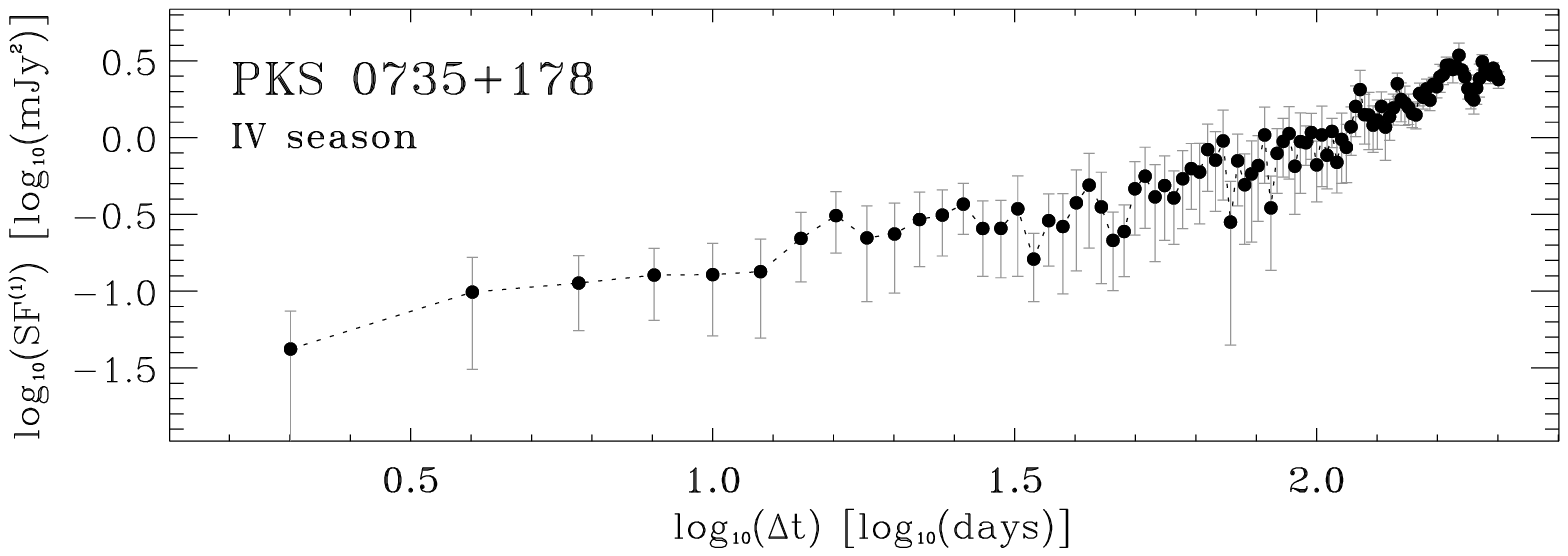}}} \\[-3mm]
\hspace{-3mm}{\resizebox{8.6cm}{!}{\includegraphics{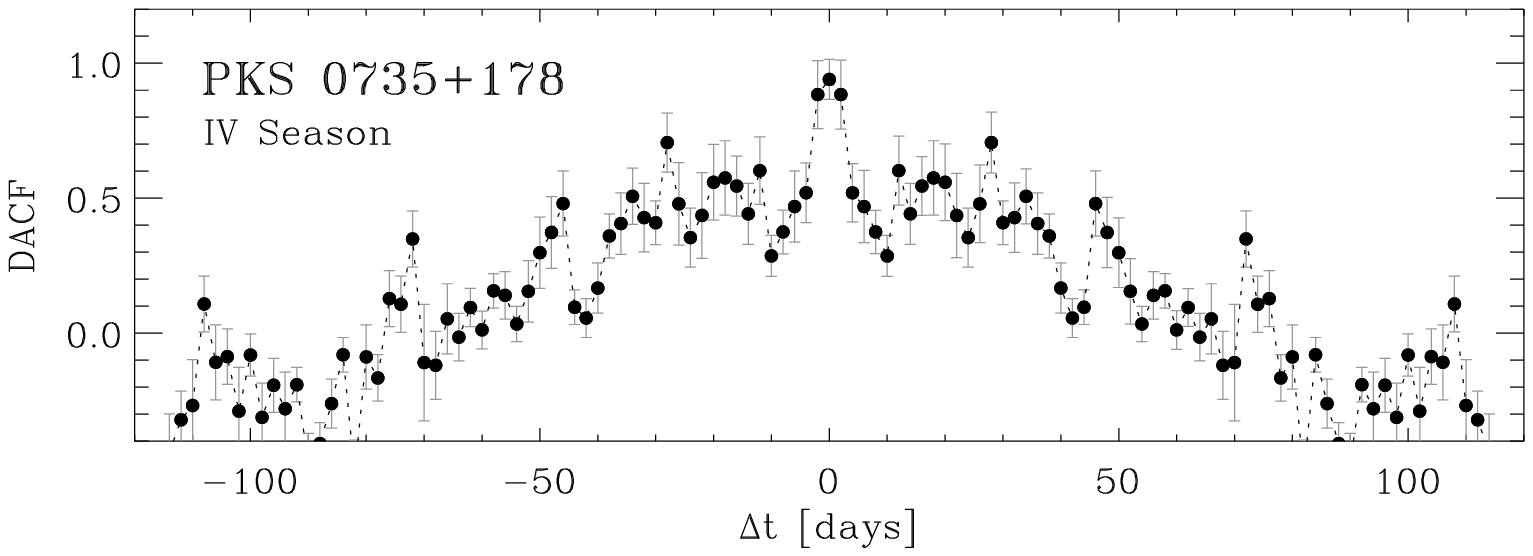}}} 
\hspace{-3mm}{\resizebox{8.6cm}{!}{\includegraphics{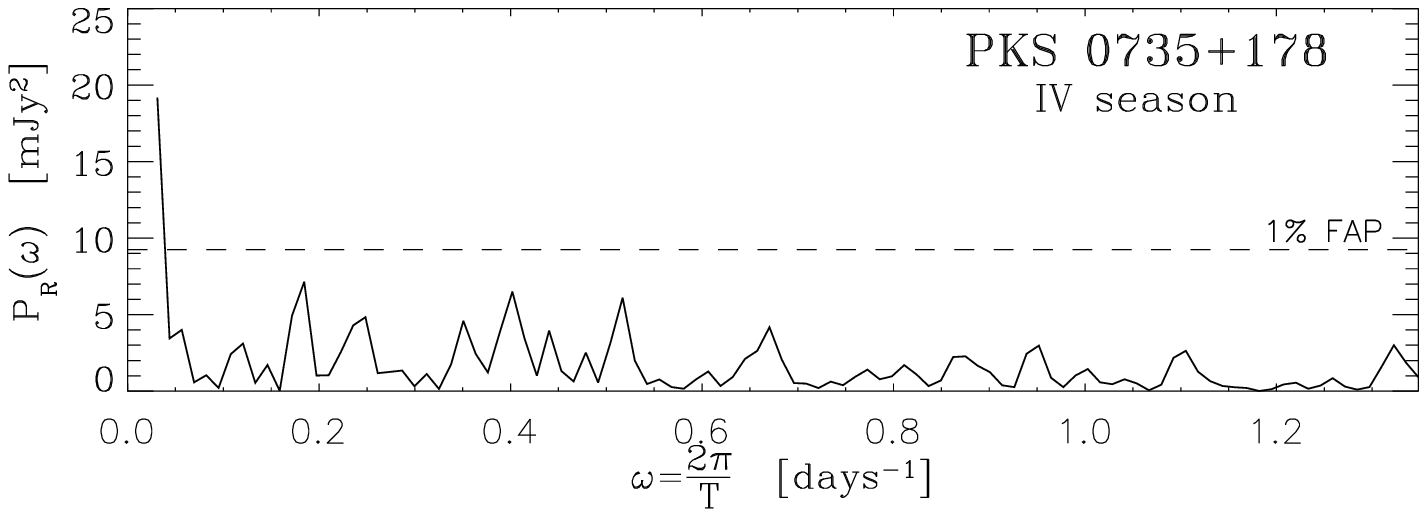}}} \\[-3mm]
\hspace{-3mm}{\resizebox{8.6cm}{!}{\includegraphics{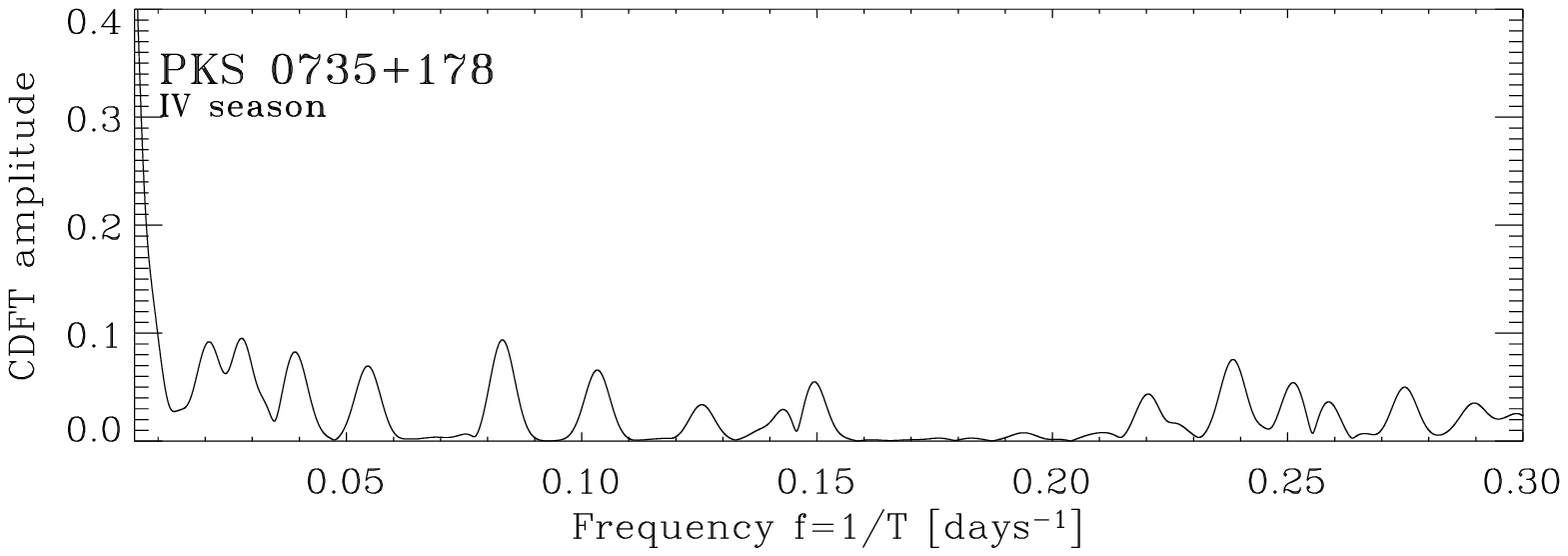}}} 
\hspace{-3mm}{\resizebox{8.6cm}{!}{\includegraphics{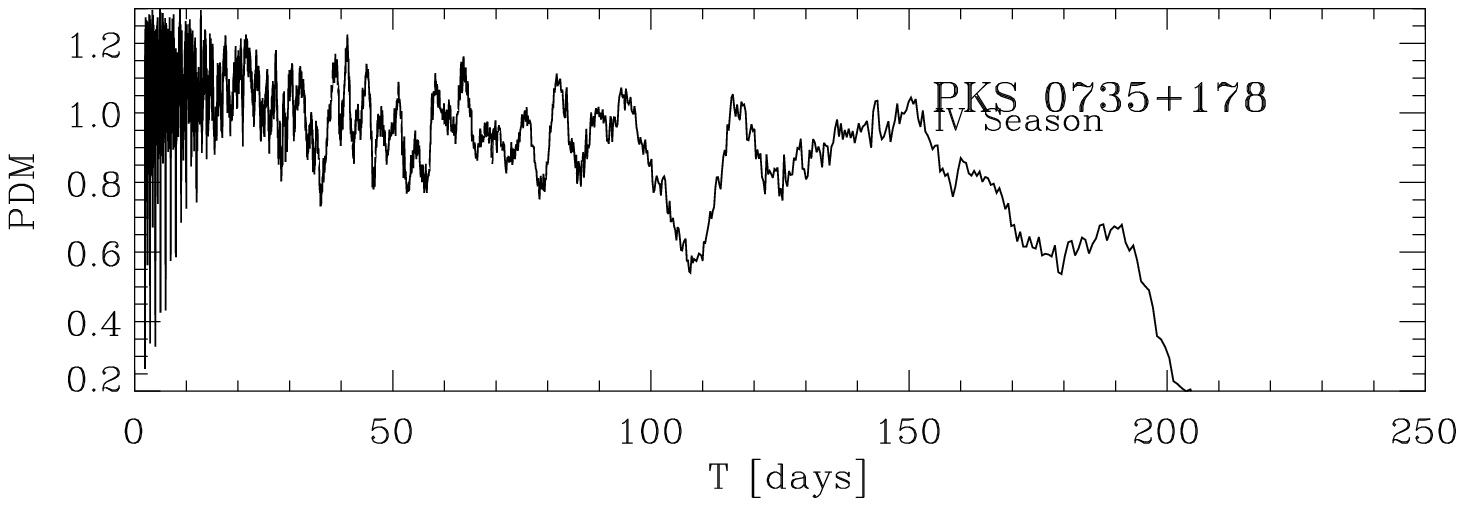}}}\\[-3mm]
\hspace{-4mm}{\resizebox{8.6cm}{!}{\includegraphics{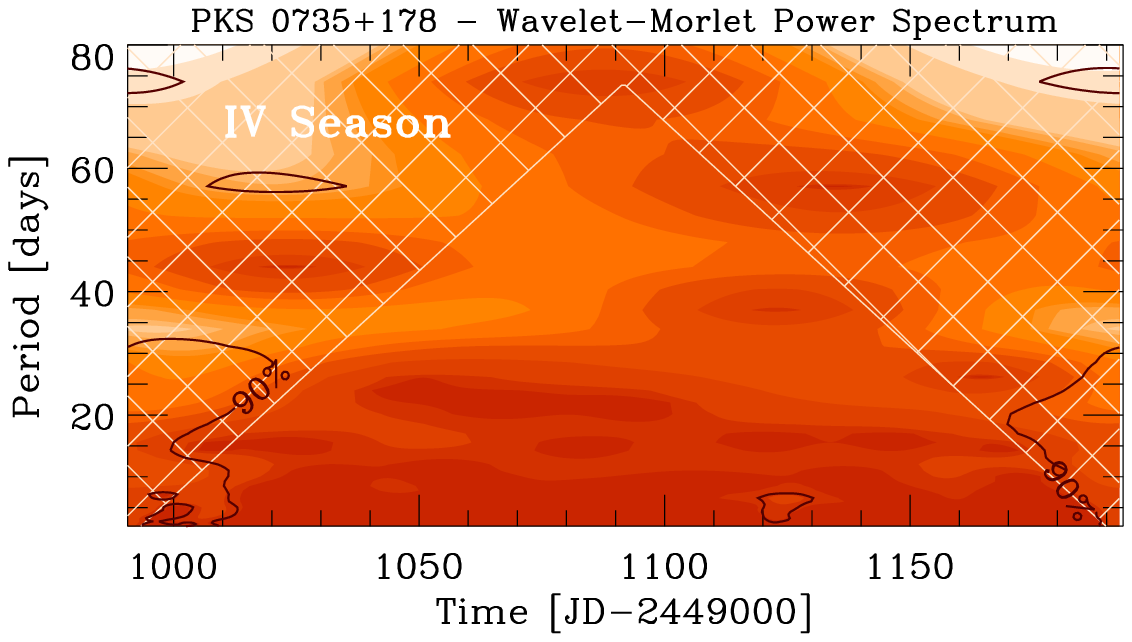}}}
\hspace{-3mm}{\resizebox{8.6cm}{!}{\includegraphics{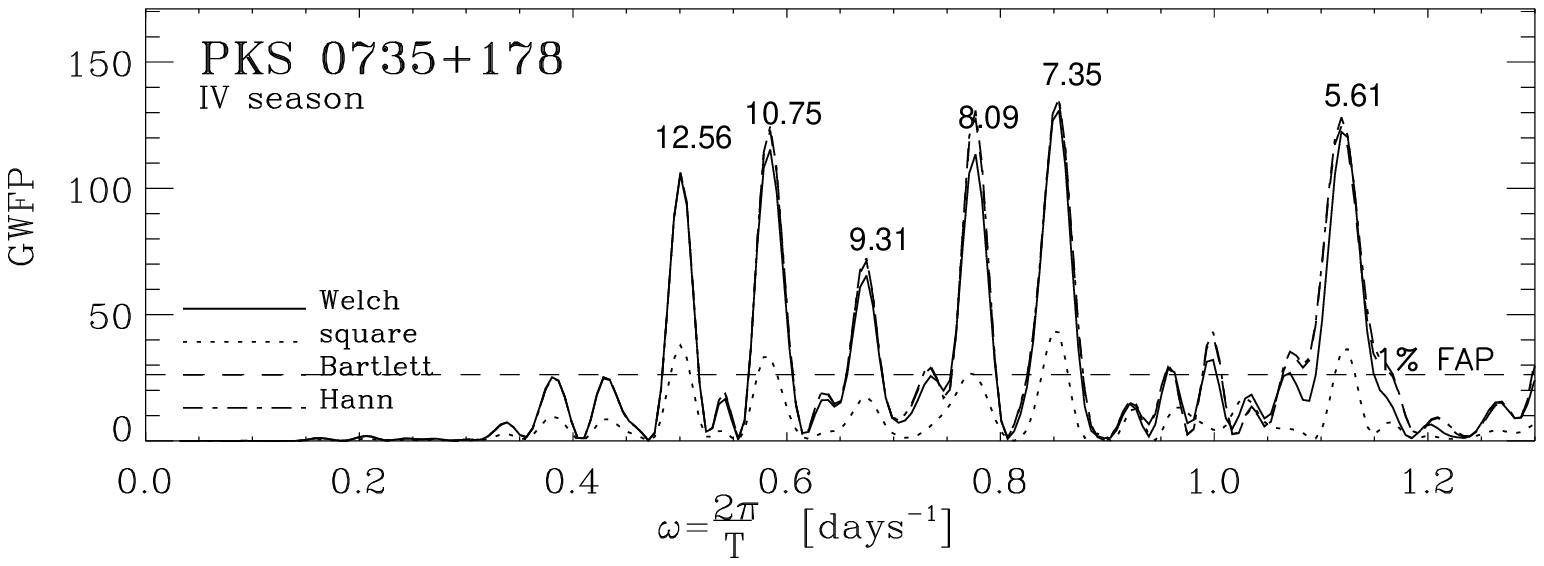}}}
\vspace{-5mm}
\end{tabular}
\end{center}
\caption{Panels from left to right and following below: the
$R$-band light curve of PKS 0735+178 in the IV observing season
(Sept. 1995, Apr. 1996) obtained by our monitoring programme, and
related functions produced by the time-series analysis.
SF (data bin: 1 day, SF bin: 2 days), DACF (data bin: 1 day, DACF
bin: 2 days), LSP, CDFT, PDM, Morlet-CWT scalogram, and GWFP (gap
threshold 5 days). Issues and results from these diagrams are
described in the text. General results of the analysis of all the
seasons from our dataset are summarized in
Table\ref{tab:timescalestable}.}
 \label{fig:statisticalplotIVseason}
\end{figure*}

\begin{figure*}[htp!!!]
\begin{center}
\begin{tabular}{c}
\hspace{-3mm}{\resizebox{8.8cm}{!}{\includegraphics{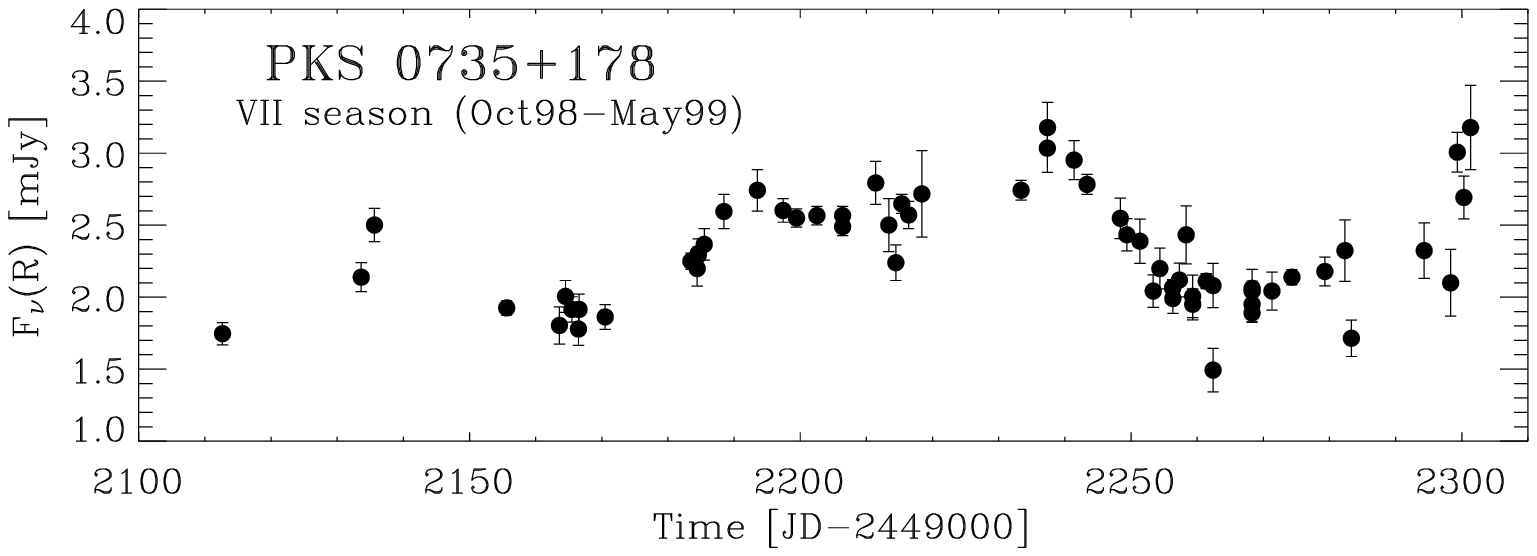}}}
\hspace{-3mm}{\resizebox{8.6cm}{!}{\includegraphics{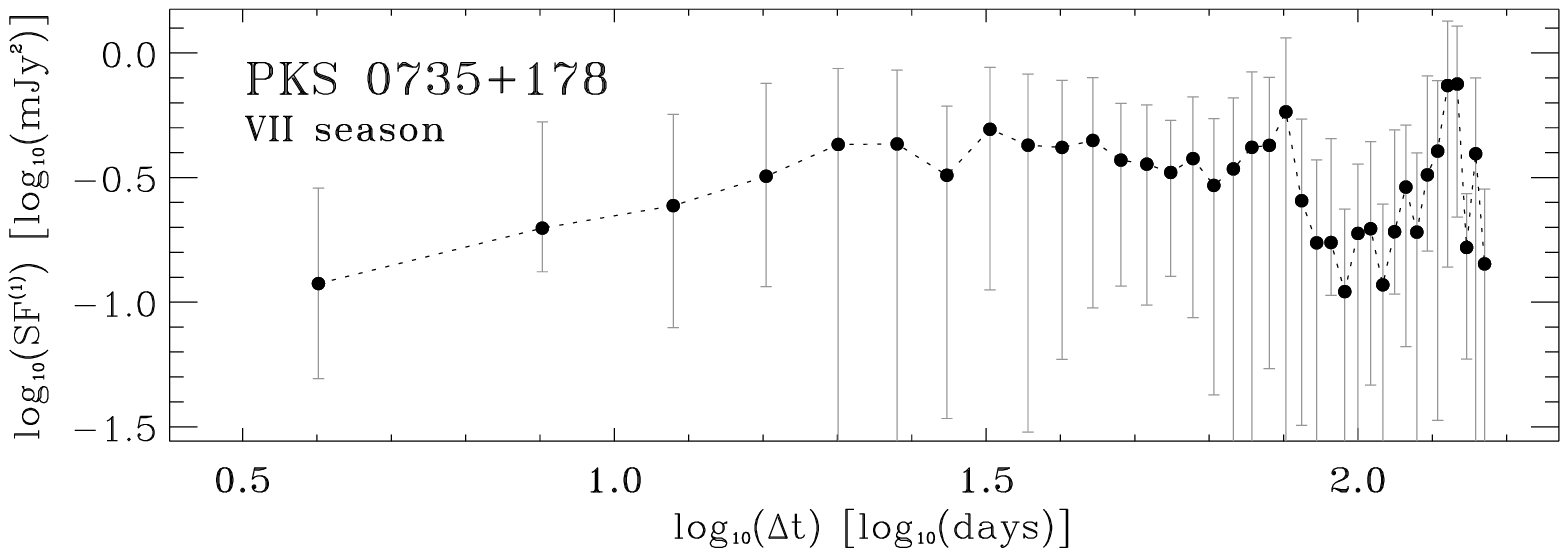}}} \\[-3mm]
\hspace{-3mm}{\resizebox{8.6cm}{!}{\includegraphics{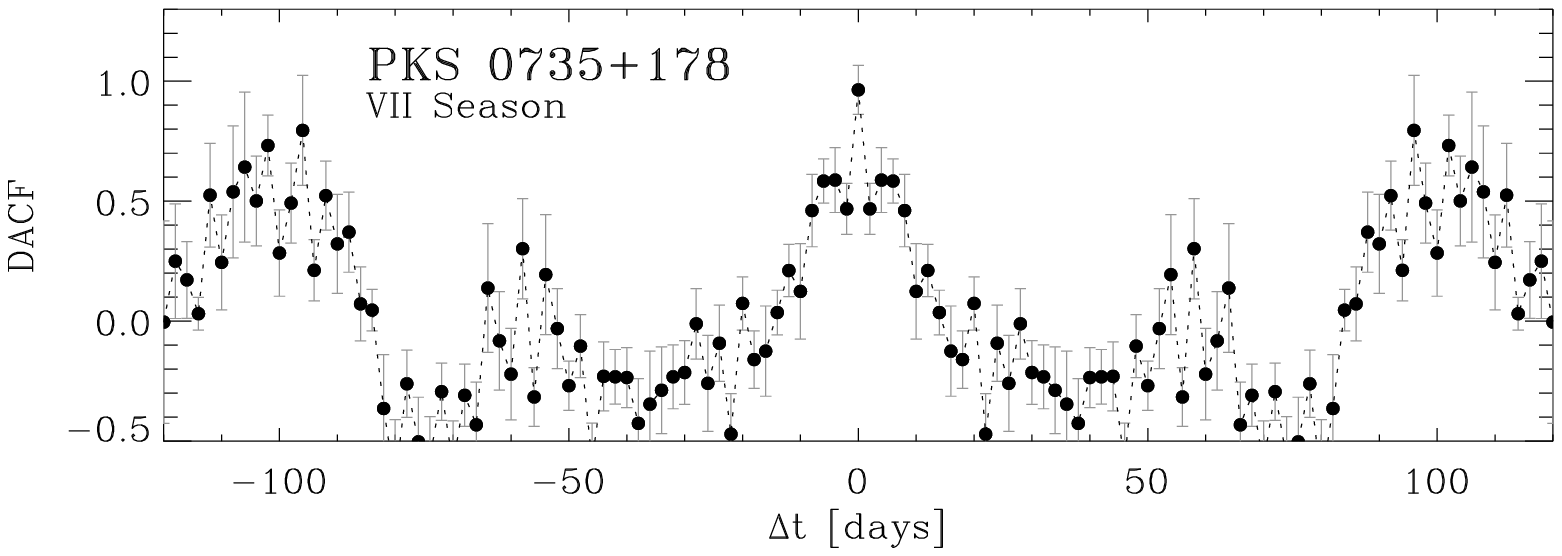}}}
\hspace{-3mm}{\resizebox{8.6cm}{!}{\includegraphics{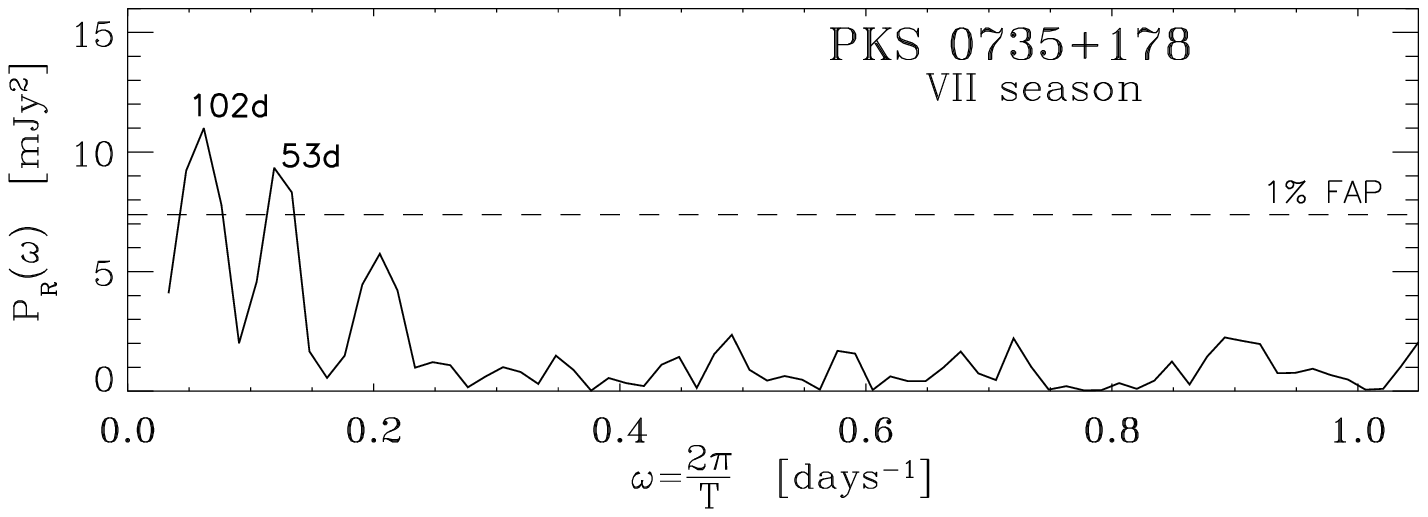}}} \\[-3mm]
\hspace{-3mm}{\resizebox{8.6cm}{!}{\includegraphics{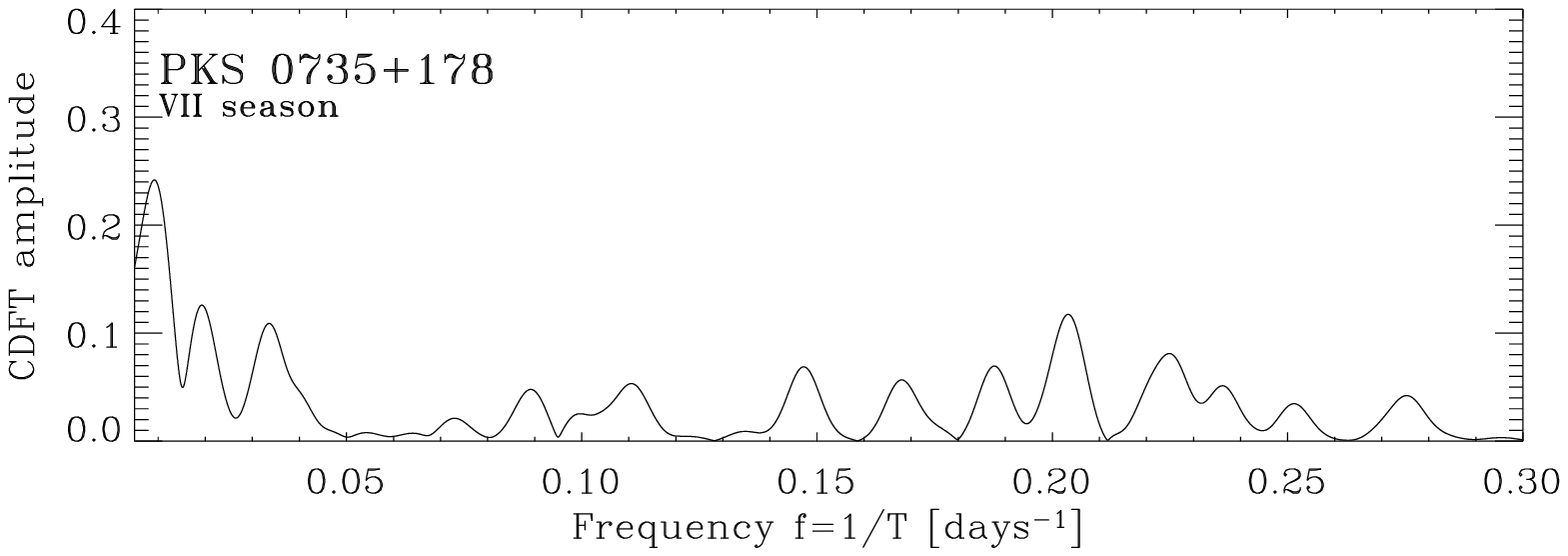}}}
\hspace{-3mm}{\resizebox{8.6cm}{!}{\includegraphics{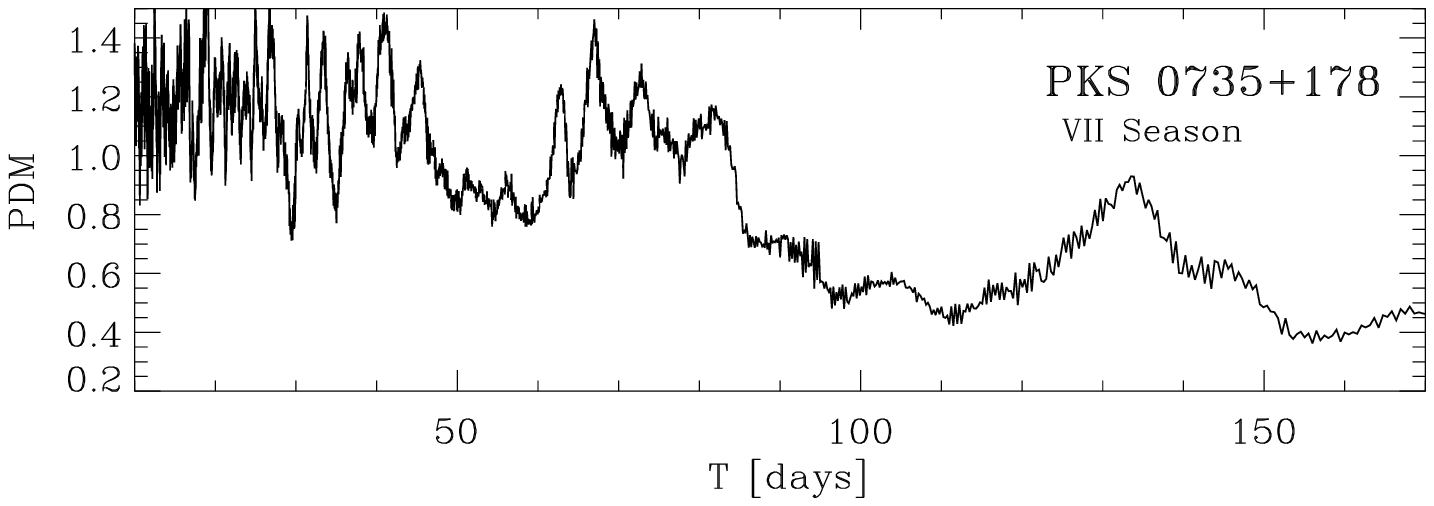}}}\\[-3mm]
\hspace{-4mm}{\resizebox{8.6cm}{!}{\includegraphics{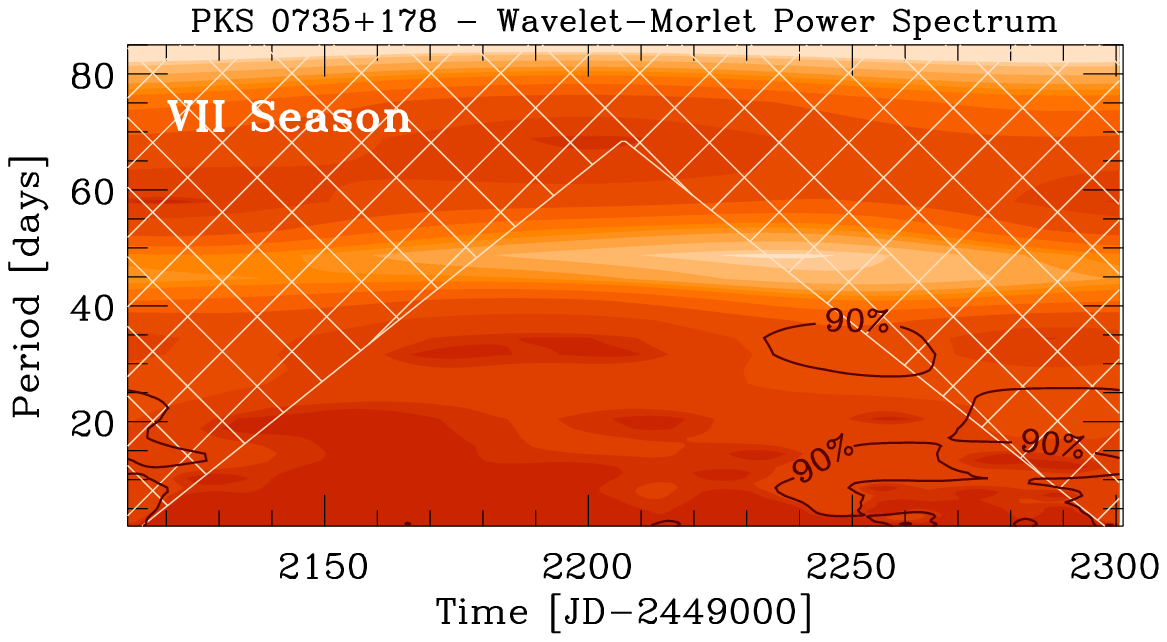}}}
\hspace{-3mm}{\resizebox{8.6cm}{!}{\includegraphics{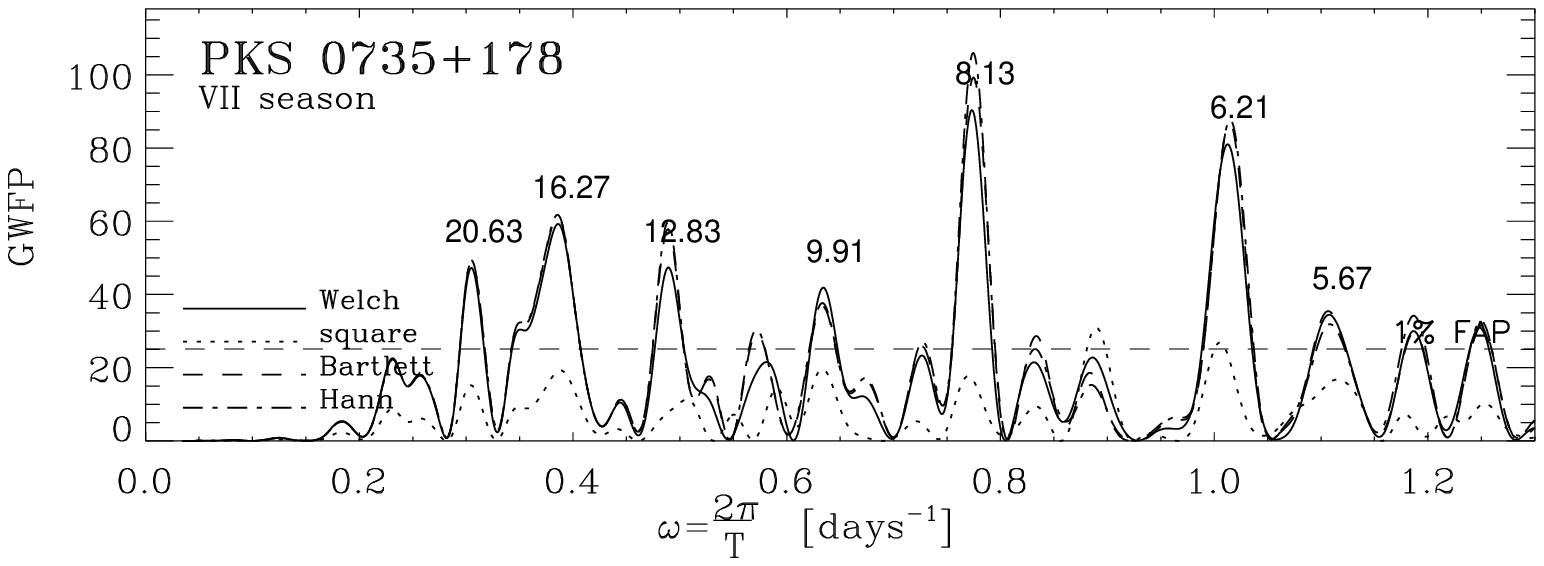}}}
\vspace{-5mm}
\end{tabular}
\end{center}
\caption{Panels from left to right and following below: the
$R$-band light curve of PKS 0735+178 in the VII observing season
(Oct. 1998, May 1999) obtained by our monitoring programme, and
related functions produced by the time-series analysis.
SF (data bin: 1 day, SF bin: 4 days), DACF (data bin: 1 day, DACF
bin: 2 days), LSP, CDFT, PDM, Morlet-CWT scalogram, and GWFP (gap
threshold 5 days). Issues and results from these diagrams are
described in the text. General results of the analysis of all the
seasons from our dataset are summarized in
Table\ref{tab:timescalestable}.}
 \label{fig:statisticalplotVIIseason}
\end{figure*}

\begin{figure*}[htp!!!]
\begin{center}
\begin{tabular}{c}
\hspace{-3mm}{\resizebox{8.8cm}{!}{\includegraphics{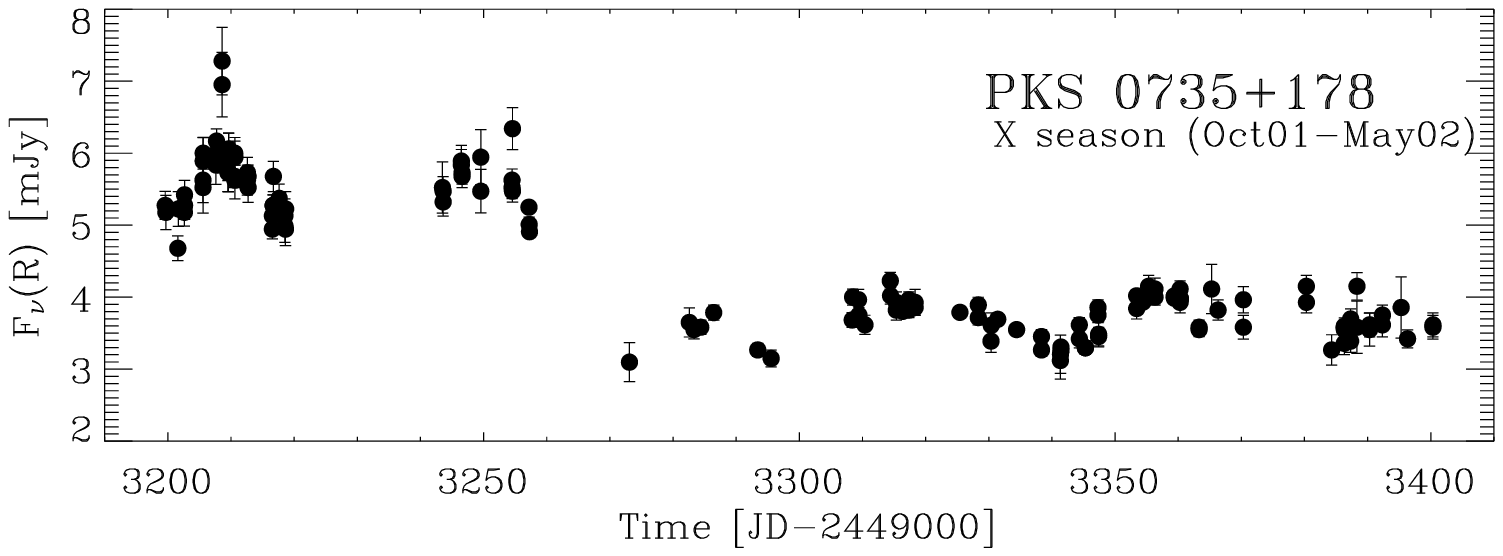}}}
\hspace{-3mm}{\resizebox{8.6cm}{!}{\includegraphics{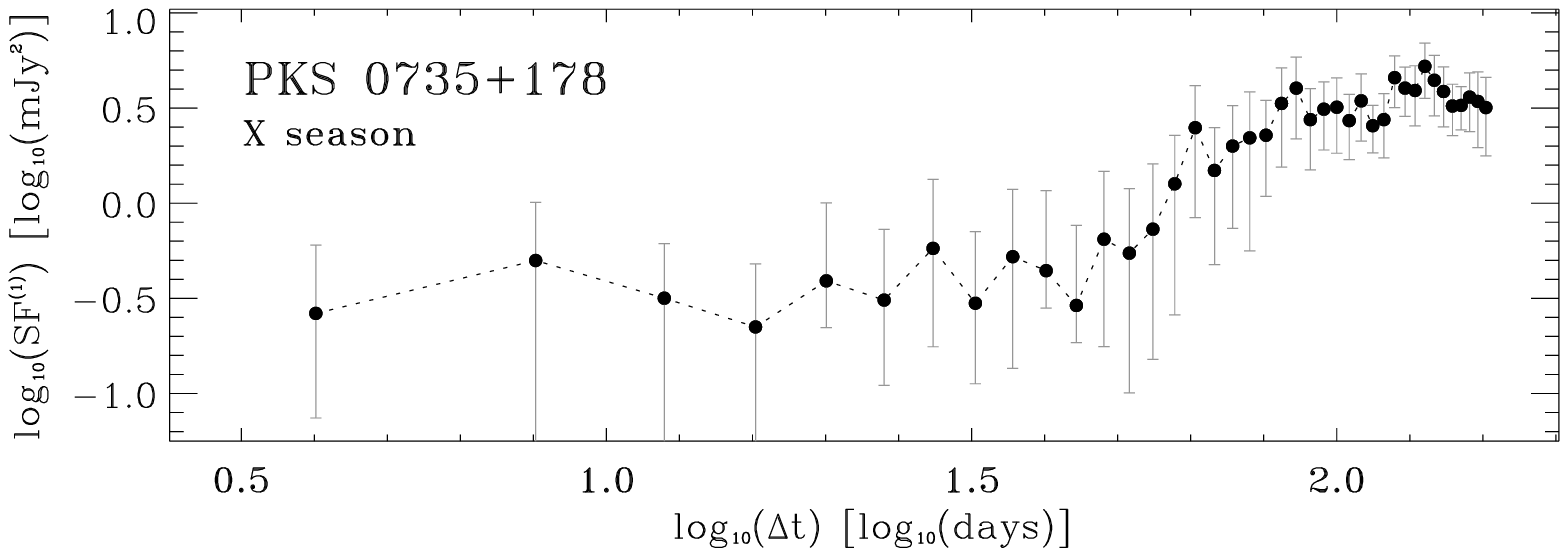}}}   \\[-3mm]
\hspace{-3mm}{\resizebox{8.6cm}{!}{\includegraphics{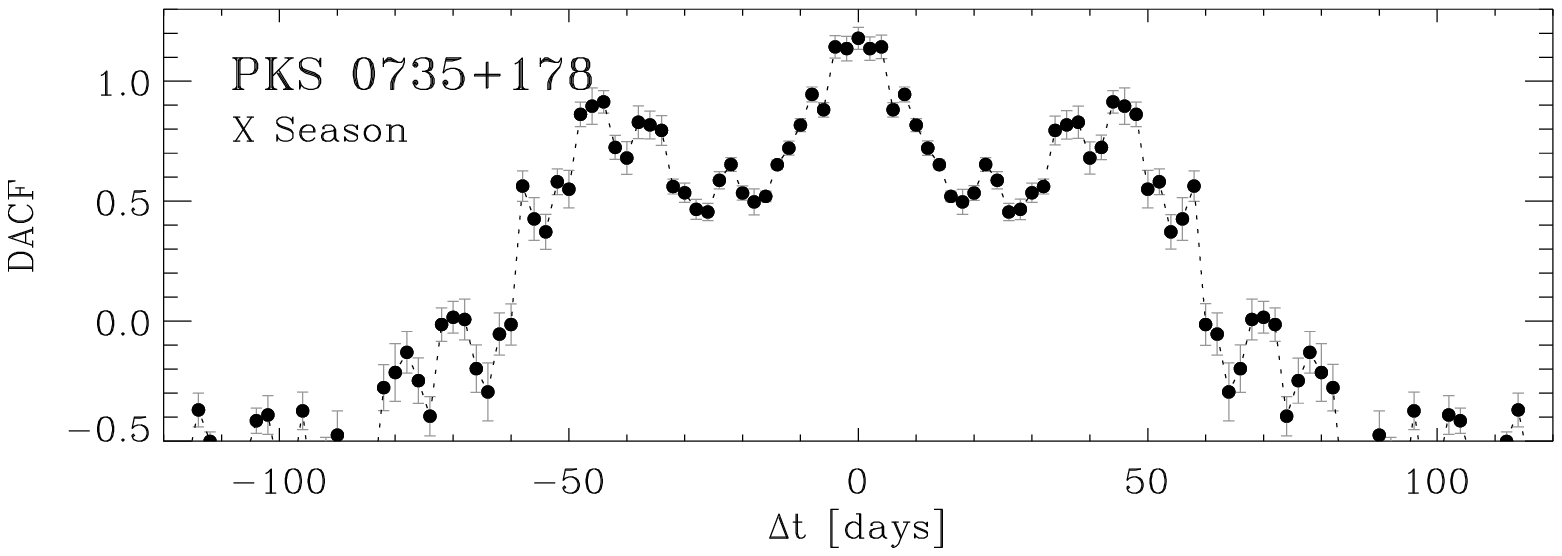}}}
\hspace{-3mm}{\resizebox{8.6cm}{!}{\includegraphics{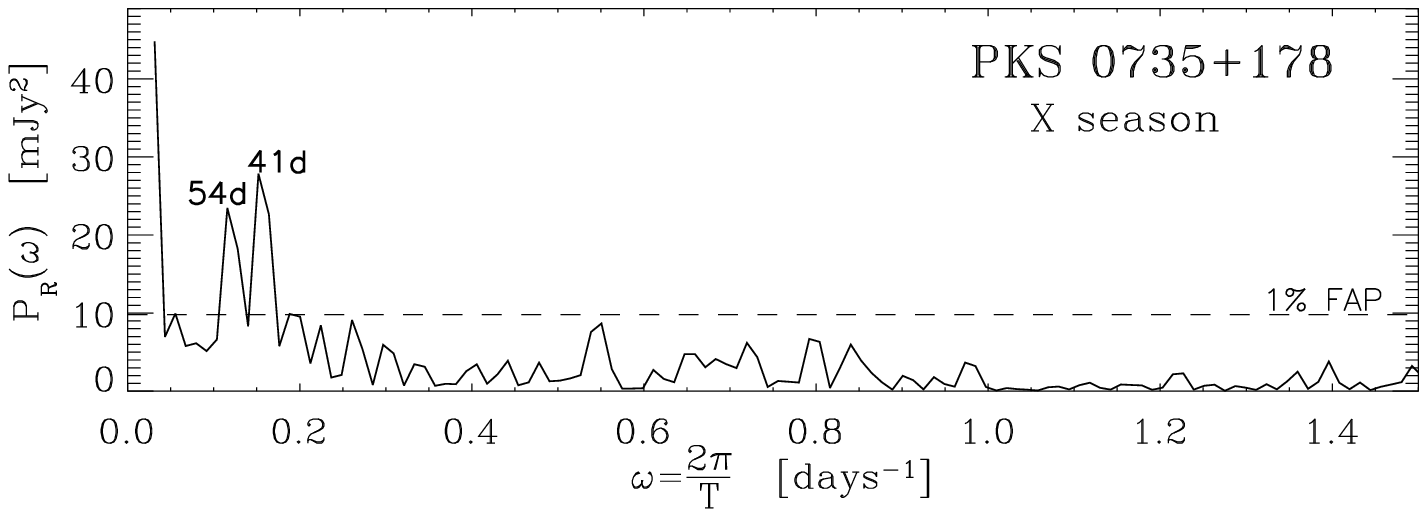}}}   \\[-3mm]
\hspace{-3mm}{\resizebox{8.6cm}{!}{\includegraphics{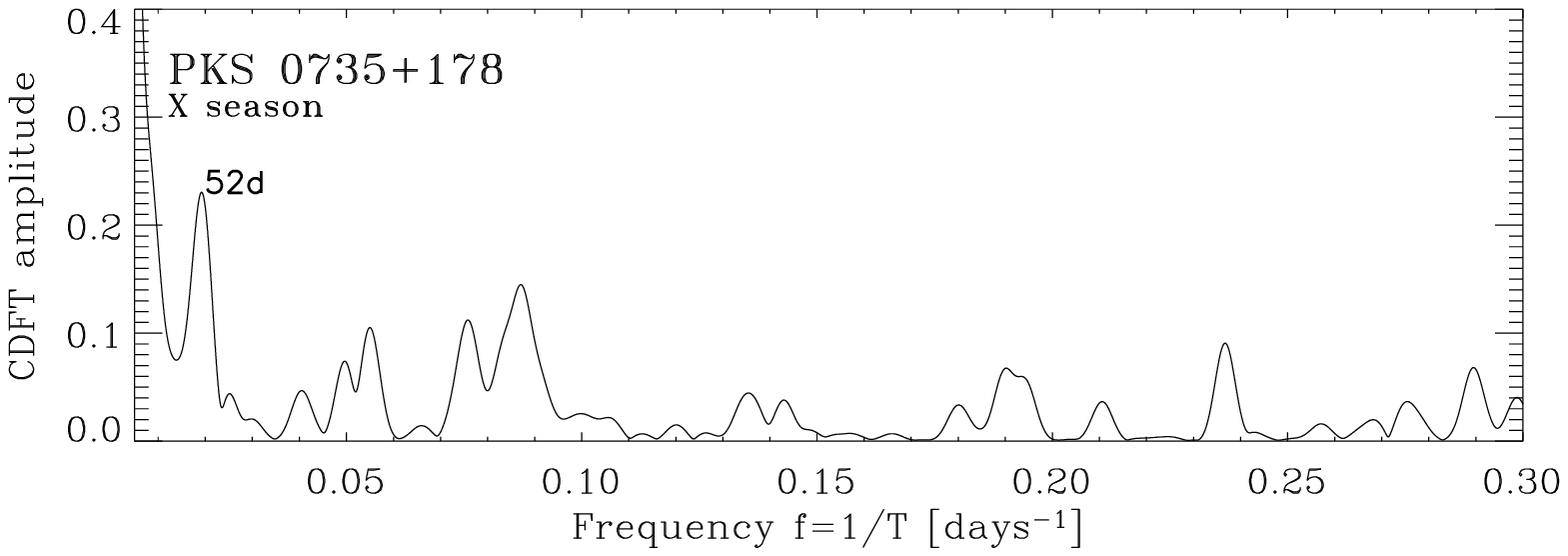}}}
\hspace{-3mm}{\resizebox{8.6cm}{!}{\includegraphics{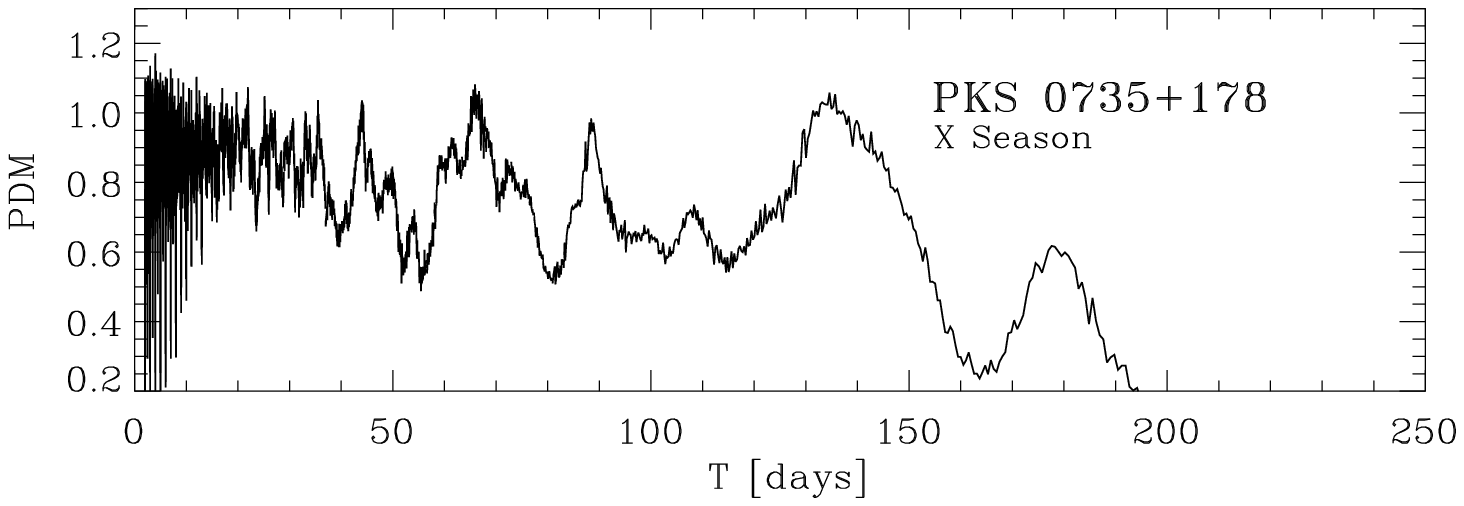}}}  \\[-3mm]
\hspace{-4mm}{\resizebox{8.6cm}{!}{\includegraphics{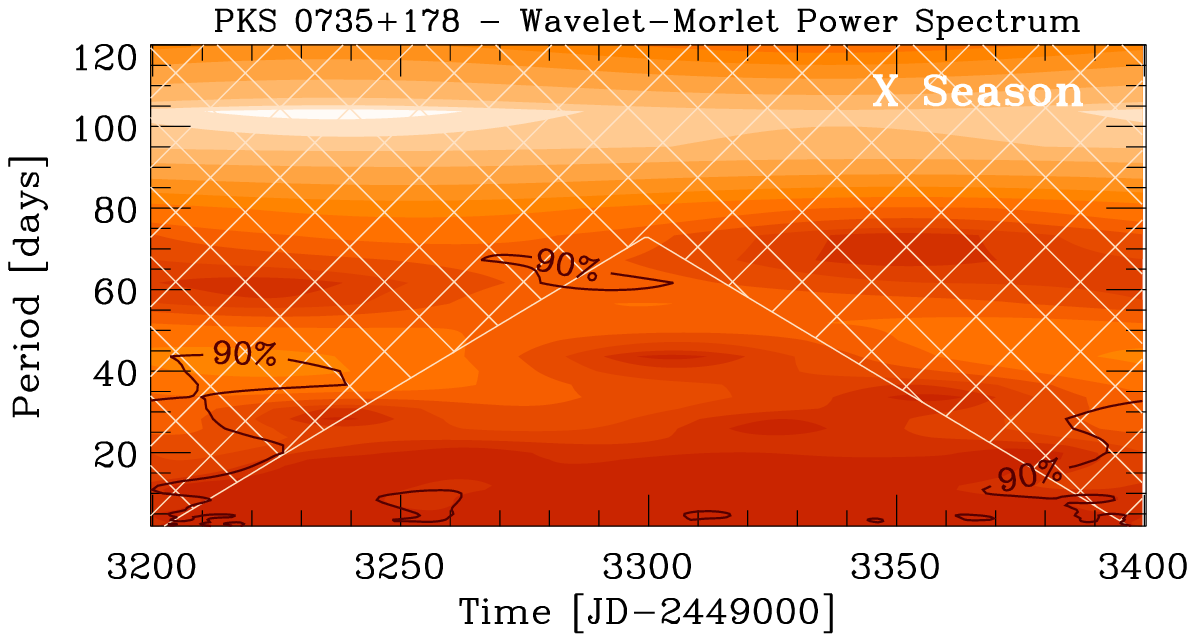}}}
\hspace{-3mm}{\resizebox{8.6cm}{!}{\includegraphics{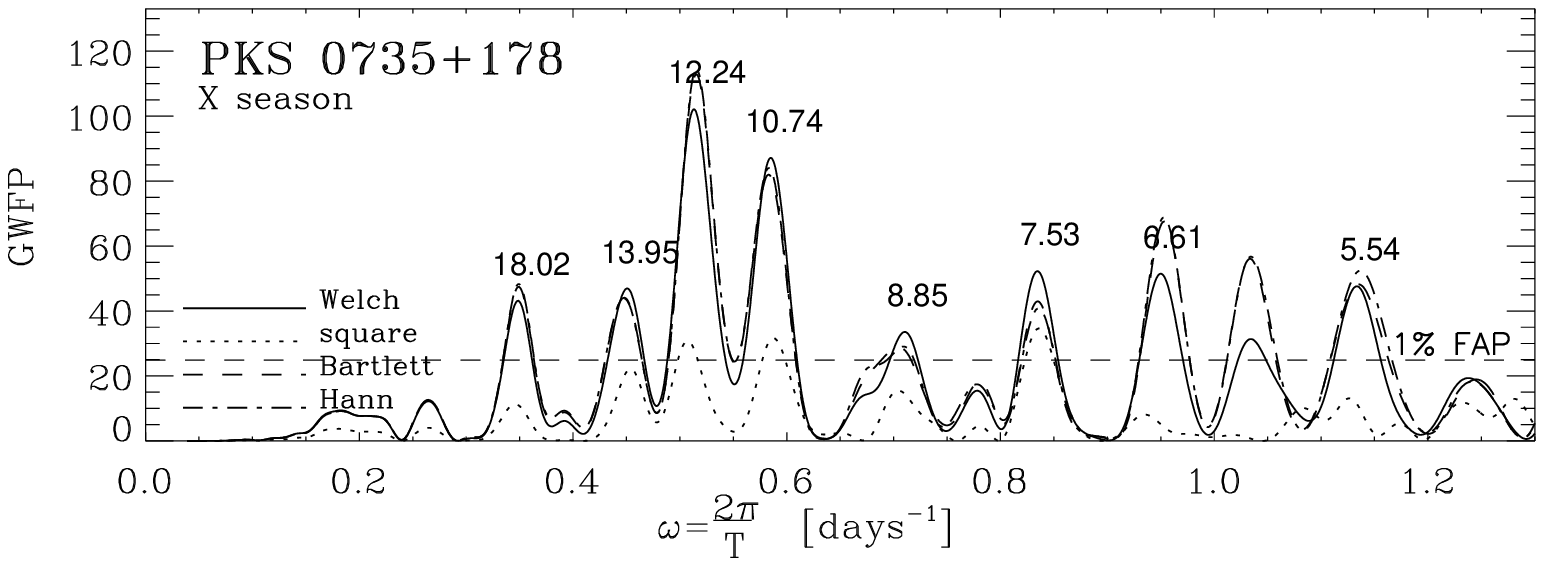}}}
\vspace{-5mm}
\end{tabular}
\end{center}
\caption{Panels from left to right and following below: the
$R$-band light curve of PKS 0735+178 in  the X observing season
(Oct. 2001, May 2002) obtained by our monitoring programme, and
related functions produced by the time-series analysis. SF (data
bin: 1 day, SF bin: 4 days), DACF (data bin: 1 day, DACF bin: 2
days), LSP, CDFT, PDM, Morlet-CWT scalogram, and GWFP (gap
threshold 5 days). Issues and results from these diagrams are
described in the text. General results of the analysis of all the
seasons from our dataset are summarized in
Table\ref{tab:timescalestable}.}
 \label{fig:statisticalplotXseason}
\end{figure*}

 The temporal analysis of the PKS 0735+178 optical dataset
is performed using all the methods mentioned above separately on
each light curve with results summarized in Table
\ref{tab:timescalestable}. In particular the diagrams from the
time-series analysis of 5 light curves (the whole historical
1906-2004 and best sampled 1970-2004 $B$-band series, the seasonal
IV, VII and X $R$-band light curves from our dataset) are reported
in Fig. \ref{fig:statisticalplot19062004},
Fig.\ref{fig:statisticalplot19702004},Fig.\ref{fig:statisticalplotIVseason}
Fig.\ref{fig:statisticalplotVIIseason}, and
Fig.\ref{fig:statisticalplotXseason}. In the analysis of mid-term
scales, the problem of spurious artifacts given by seasonal gaps
with no data (solar conjunction with the source) was avoided
studying each seasonal light curve separately. The GWFP of the
best sampled 1970-2004 light curve
(Fig.\ref{fig:statisticalplot19702004} last panel, bottom, right)
shows indeed only one powerful peak placed at exactly 1-year
scale. This represents the periodical yearly recurrence of the
seasonal gaps.
%
%
%
%
%
%
%
%

\par The whole 1906-2004 light curve (Fig.
\ref{fig:statisticalplot19062004}) is patently affected by
substantial differences in data sampling, by void gaps, by a poor
sampling earlier than 1970 and a long empty interval (1958-1970).
Nevertheless several signal features and characteristic timescales
are pointed out by the different techniques (using both binned and
unbinned series): about 8.6y, 12-13y, 25y and 34y (see the summary
reported in Table\ref{tab:timescalestable}). The 13.7y timescale
(pointed out by nice features in the LSP and CWT functions) is the
same value claimed as the major component of a multi-periodical
trend by \citet{qian04} and \citet{fan97}, on the other hand the
8.6y timescale (suggested by the DACF and LSP) is probably to be
ascribed mainly by the temporal behaviour after 1970
(Fig.\ref{fig:statisticalplot19702004}), and it is reported by
\citet{qian04} too. Longer duration scales (e.g. 34y) are
difficult to be set out with confidence. Fake signal features, due
to recurrences and temporal patterns given by the empty gaps,
occurred at scales shorter than 8 years only (see the GWFP plot),
therefore minor and fainter hallmarks in the SF, LSP, CWT
corresponding to such shorter scales are neglected.
\par
\begin{figure}[t!!]
\centering
\begin{tabular}{l}
\hspace{-0.8cm}
{\resizebox{8.5cm}{!}{\includegraphics{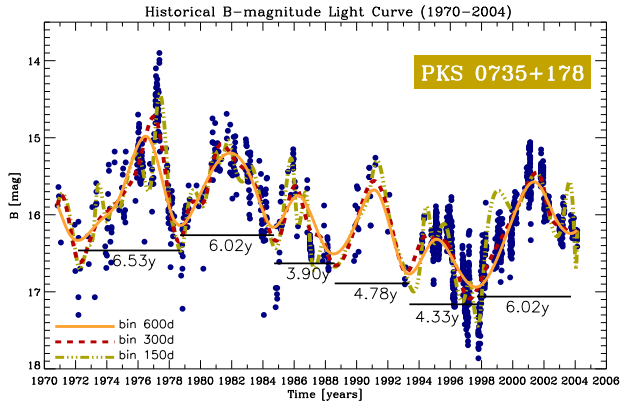}}}
\vspace{-0.2cm}
\end{tabular}
\caption{Cubic spline interpolations of the best sampled 1970-2004
historical light curve of PKS 0735+178. The continuous (orange)
spline curve is obtained with a data binning of 600 days, while
the other dotted/broken spline lines are obtained with binnings of
150 and 300 days. The interpolations show 6 main peaks already
visible with the 600-days bin. With this binning and considering
the separation between the troughs of the pseudo-sinusoidal curve,
we have cycles of about 6.0-6.5 years and 3.9-4.8 years. This raw
visual interpretation and spline envelope could be linked to one
of the fundamental components (such as 4.4-4.8 years) found with
the time series analysis (see Tab. \ref{tab:timescalestable}).
Hence both the hypotheses of a multi-component recurrent trend, or
a pseudo-periodicity (with period drifts) and modulating this
light curve can be plausible.}
\label{fig:splines19702004}
\end{figure}

 The best sampled portion of the historical light curve
(1970--2004, Fig.\ref{fig:statisticalplot19702004}) spans 33 years
and has a quite fair continuous, regular and long-term coverage:
867 nights with 1 datapoint at least, an average number of data
points per night of 1.8, an average gaps among data of 7.8 days
and a maximum gap of 1.6 years (Table\ref{tab:timescalestable}).
The more relevant characteristic timescales suggested by the
different temporal analysis methods are around 4.5y, 8.6y, and
12.5y (other few values such 3.5y, 7.4y, 11.8y years could be
traced back to the previous mentioned, if we consider the
finite-resolution accuracy of these methods). The 4.5, 8.6, 12.5
years scales might be considered multiples harmonic signatures of
a fundamental (coherent, absolute, or transient periodical, or
again with drifting duration) component of about 4 years (see also
the spline visual envelope in Fig.\ref{fig:splines19702004}). A
characteristic timescales of about 4.8 years was previously
claimed also by \citet{webb88} and \citet{smith87}, while the 8.6
years scale was recently suggested by \citet{qian04}. On the other
hand no evidence for the periodical signature of 14.2 years
previously claimed by \citet{fan97} are observed, but only weak
hints for scales in the 11.6-13.5 years range are found out. The
power spectral density in the $1/f^{a}$ regime, shows a slope
index $a$ between 1.5 and 2 (i.e. next to a pure shot-noise
behaviour). The GWFP show a very powerful fake signature at 1.0
years as expected, produced by the recurrent 1-year gap between
subsequent observing seasons. This artifact due to sampling it is
not completely neglected by the methods used (see e.g. the
residual peaks around $\omega \sim 6$ i.e. $\sim 2\pi$ in the LSP
plot), therefore it is very useful to develop and make use of the
GWFP technique in conjunction with the other time series methods.
No other (longer) fake features due to the irregular sampling are
pointed out by the GWFP, therefore all the other characteristic
timescales claimed in this light curve can be considered due to
real variability. In the CWT scalogram a significant and localized
pulse in power is visible (with a scale around 4.8 years and
located in correspondence of the big 1977 outburst). Another
localized bump gives a scale of about 7.4y. An elongated and lower
intensity band in the CWT scalogram, corresponding to the last
epochs (about 1994-2004) indicates again a coherent time scale
between the values 8.2-8.9 years. Results are quite similar using
different suitable CWT mother functions as the Mexican-hat and
Paul waveforms in this light curve.
\par As example, in the time series analysis of the IV observing season of our $R$-band dataset (Sept. 1995, Apr.
1996, Fig.\ref{fig:statisticalplotIVseason}) a main modulating
trend quite monotonically decreasing is observed. This is stated
for example by the shape of the SF in the logarithmic plot (second
panel, top right of Fig.\ref{fig:statisticalplotIVseason}) being
quite linearly monotonic and without any turnover flattening
(given by the reaching of a maximum correlation timescale). In
this general trend 4 or 5 moderate and secondary oscillations can
be visually identified (possibly related to the unique and weak
signal of a characteristic timescale between 28-34 days, as hinted
by the DACF, PDM, and CWT). Other characteristic scales are not
displayed, and the artifact noise given by the irregular gaps, is
important only at timescales below 13 days (see the GWFP last
panel, bottom right). The power spectral density in the $1/f^{a}$
regime, shows a power index $a=1.97\pm 0.25$, i.e. a temporal
variability mode like the shot noise (brown noise) signal.
\par In the $R$-band light curve during the VII observing season
(Oct. 1998, May 1999, Fig. \ref{fig:statisticalplotVIIseason}) a
brightening stage (between about $JD=2451170$ and $JD=2451265$,
i.e. 95 days long) looking as produced by two blended main flares
of about 50 days duration, might be supposed by a visual
inference. Such values (about 95 and 50 days) are pointed out
indeed by the SF, DACF, LSP, PDM and CWT functions
(Table\ref{tab:timescalestable}) as characteristic timescales
(scales where is more power in the signal). The SF and PDM methods
suggests also a possible timescale about 30 days. Fake features
given by the irregular gaps are important only at timescales below
21 days (GWFP last panel, bottom right of Fig.
\ref{fig:statisticalplotVIIseason}). The power spectral density
function shows a power index $a=1.64\pm 0.09$, i.e. a temporal
fluctuation mode placed at halfway between the flickering and the
shot noise behaviour.
\par The plots from the analysis of the X observing season
data (Oct. 2001, May 2002, Fig.\ref{fig:statisticalplotXseason})
obtained by our monitoring programme, show an high state with two
relevant flares, followed by a low and rather oscillating phase
(after date $JD=2452273$). Unfortunately there was a long
observing gap (about 25 days between $JD=2452218$ and
$JD=2452243$) during the more active phase of this season, that is
also the brightest optical state of the source recorded in our
1994-2004 database. Characteristic timescales of about 52-55 days
are suggested by the LSP, CDFT, and PDM, while a 41 days scale is
also found by the DACF and LSP (Table\ref{tab:timescalestable}).
Fake features given by the irregular gaps are influential only at
timescales below 18 days (GWFP last panel, bottom right,
Fig.\ref{fig:statisticalplotXseason}). The power spectral density
function shows a power index $a=1.46\pm 0.17$, i.e. again a
temporal fluctuation mode placed at halfway between the flickering
and the shot noise behaviour.
\par The summary of the temporal statistics and analysis results
is showed in Table\ref{tab:timescalestable}. About statistics the
following information is reported: the observing interval and its
duration (98.1 years in total, and between 144 days and 203 days
regarding our observing seasons); the number of the effective
observing nights with one data point at least (867 data points in
the 1970-2004 light curve, and between 20 and 62 in our dataset
observing seasons); the average number of data points per
observing night in the interval (spanning between 1 and 2.3,
implying no data clustering); the average separation between 2
successive data points (3 days on average on all our 10 observing
seasons); the maximum separation (maximum empty gap) between 2
successive data points (no more than 25 days in the worst sampled
season). In time intervals where the SF slope can be recognized in
the log-log representation, we calculated its power index $b$
trough a linear regression. About the time series analysis the
following quantities are reported (when possible): characteristic
timescales calculated by deep drops in the SF, the power law index
$a=1+b$ of the PSD in the $1/f^{a}$ regime calculated by the SF,
characteristic timescales inferred from the SF turnover to the
plateau produced by times longer than the maximum correlation lag,
timescales estimated from power peaks in the DACF/ZDACF, in the
LSP and in the CDFT, timescales indicated by deep drops in the PDM
and by peaks in CWT scalogram.
\par The criteria adopted in order to possibly avoid to quote fake features and
artifacts due to the irregular sampling and gaps, or edge effects,
in Table\ref{tab:timescalestable} are the following: 1) only
timescales shorter than 1/2 or 1/3 of the interval duration were
considered; 2) only the more relevant signatures in each method
are preliminary considered; 3) among these we discarded the
features that are not indicated by more than one method on the
same light curve portion, when the signature is not particularly
strong; 4) if some timescales, among the remaining, is still
matched by the synthetic GWFP, these are discarded too. Most of
the applied methods take well into account the power of artifacts
given by the irregular dataset and recurrences of gaps, as showed
by
Fig.\ref{fig:statisticalplot19062004},\ref{fig:statisticalplot19702004},
\ref{fig:statisticalplotIVseason},\ref{fig:statisticalplotVIIseason},
and \ref{fig:statisticalplotXseason} (see the comparison of the
first 7 panels with the last GWFP panel). However residual
spurious power can be still present in the functions (see e.g. the
spurious peaks of the LSP around $\omega \sim 6$ i.e. $T=1$ year,
in Fig.\ref{fig:statisticalplot19702004} as mentioned).
\par On long timescales the main temporal components we
found can be grouped in three main ranges of values: between 4.4
and 4.8 years, between 8.2 and 8.6 years and between 10.8 and 13.2
years
(Fig.\ref{fig:statisticalplot19061958},Fig.\ref{fig:statisticalplot19062004},
Fig.\ref{fig:statisticalplot19702004} and
Table\ref{tab:timescalestable}). These components could modulate
the long term optical light curve of PKS 0735+178 with roughly
cyclical oscillations. A characteristic timescale of about 4.8
years was previously claimed also by \citet{webb88} and
\citet{smith95} while a 8.6 years scale was recently suggested by
\citet{qian04}. On the other hand we did not find any strong
evidence of further and longer-duration ($> 14$ years) timescales
reported in literature \citep[e.g. ][ and
Sect.\ref{par:0735optpropintro}]{fan97,qian04}, but only weak
hints of a 25 years and a 34 years signature.
\par Some parts of the light curve could contribute with
different typical scales to the overall series, while data are
treated in varying ways, with timescales suppressed or enhanced as
the relative weight of different segments of the light curve
changes. The variability scale of 8.2-8.6 years could be an
important finding and a real signature of a dominant and possibly
quasi-periodical component anyway, because it was found in both
the whole 1906-2004 series and its best-sampled portion.
%
%
%
%
Moreover some uncertainty and statistical dispersion in the values
found by different methods is expected, especially when the
sampling is irregular, and here the dispersion range is small (0.4
years). In addition fast and/or isolated flares randomly occurred
and uncorrelated to any general trend, can provide loud
contributes to the power spectrum, disproving any periodicity
hypothesis based on a long but under-sampled historical light
curve. The better sampled portion of the PKS 0735+178 light curve
did not disprove the characteristic timescale mentioned above,
therefore it is reasonable to suppose a possible dominant period
around 8.5 years. This hypotheses is open to future investigations
based on prolonged monitoring observations. In this view the
shorter 4.4-4.8 years scale found might be a submultiple of the
previous component. Hence this would be the real period of the
fundamental harmonics (pointed out only by the best sampled
1970-2004 portion because of the sufficient sampling to detect
it). A support corroboration of this conjecture is also provided
by the spline interpolation reported in
Fig.\ref{fig:splines19702004}: 6 major maxima and cycles are
outlined between 1970 and 2004, with a duration between 3.9-4.8
years and 6.0-6.5 years. This crude visual interpretation could be
linked to the possible fundamental (and possible
duration-drifting) modulating component of 4.5 years. Finally the
group of longer-duration scales found in the broader range
10.8-13.2 years are detected in each piece of the historical light
curve (see Tab. \ref{tab:timescalestable}), but these values are
probably too much scattered to mask a strict periodicity
signature.
\par On intervals shorter than 200 days, monitored by our observations
in 10 subsequent seasons (from the III starting in Oct.1994 to the
XII ending in Feb.2004) and analyzed deeply with the methods
mentioned above, there is no evidence for one single and pure
periodical features, but there are signatures of several
characteristic timescales of mid duration, commonly found in
different observing seasons. These ``recurrent'' and ``common''
timescales are distributed in few groups of values: 18 days, 24-25
days, 27-28 days, between 40 and 42 days, between 50 and 56 days,
65-66 days, between 76 and 79 days, and 95-96 days. In particular
the timescales of 27-28 days are found in 3 observing seasons
(might be related to the synodical month interference), timescales
between 50 and 56 days are observed in 6 seasonal light curves,
and timescales between 76 and 79 days are detected in 4 seasons.
About these timescales, several hypotheses can be proposed. 1)
These temporal signatures could be the result of a rough
multi-periodical behaviour given by the superimposition of few
harmonic components (spanning from about 2 dozen of days to about
100 days). 2) They could be produced by pseudo-periodical cycles,
with a drift of the period duration around a fundamental value of
27 days for example (50-56 and 76-79 days could be though as rough
multiples in this case). 3) Such characteristic timescales could
be produced by different periodical stages of transitory nature,
with a sort of time-localized periodicity surviving only for
limited epochs. 4) Again they could be the result of a variability
mode endowed of few typical duty-cycles showing similar and
recurrent peak shapes and durations, but occurring at random
times. An improved and continuous monitoring, with an higher
precision photometry and a better sampling, will provide more
significant statistical results about the optical variability of
PKS 0735+178 at these timescales.
%
%
%
%
\section{Summary and conclusions}\label{par:summaryconclusions}
Blazars are one of the most exciting class of AGN, and the primary
know extragalactic sources emitting high energy gamma-rays.
Variability monitoring is an important effort in the study of
blazars for several reasons (even if well sampled light curves are
a very big challenge to be obtained at optical wavelengths
normally).
%
%
%
%
%
%
\pagebreak \noindent
\begin{landscape}
%
\begin{table}[htp!!!]
\hspace{+0.5cm}%
\caption[]{Summary of temporal statistics and characteristic
timescales found in the optical light curves of PKS 0735+178 (when
possible) using 7 different methods. Data sets investigated: 3
historical $B$-band flux light curves (the complete 1906-2004
curve, the best sampled 1970-2004 part, and the 1906-1958
portion), and 10 separated $R$-band flux light curves obtained in
each observing season of our monitoring program (from the III
season started in Oct.1994 to the XII season ended in Feb.2004).
The following data are reported in the table columns (from left to
right) for each light curve: (1) optical band and name of the
light curve interval; (2) duration of the light curve; (3) number
of the effective observing nights N$_{on}$ with one data point at
least; (4) average number of data points per observing night
$<n>$; (5) average separation between two successive data points;
(6) maximum separation (maximum empty gap) between two successive
data points; (7) power-law index $a$ of the power spectral density
($PSD$) in the $1/f^{a}$ regime ($f=1/T$), calculated in the
time-domain through the first order structure function $SF$; (8)
characteristic timescales $T_{dr}$ calculated by deep drops in the
$SF$; (9) characteristic timescales $T_{to}$ inferred from the
$SF$ turnover to a long-lag plateau in the log-log representation;
(10) timescales $T_{pe}$ estimated from peaks in the discrete
auto-correlation function ($DACF$); (11) timescales $T_{pe}$
derived by peaks in the Lomb-Scargle periodogram $LSP$; (12)
timescales $T_{pe}$ derived by peaks in the `clean''
implementation of the discrete Fourier transform ($CDFT$); (13)
timescales $T_{dr}$ estimated from deep drops in the phase
dispersion minimization function $PDM$; (14) timescales pointed
out by peaks in the 2-D contour plot of the wavelet scalogram (the
two-dimensional energy density function provided by the continuous
wavelet transform $CWT$ computed using a Morlet mother waveform).
The errors in the timescales are difficult to be estimated in
general, and the significance of a scale is high when it is
relevant according to a method, and when several methods point out
similar values. Columns from (2) to (5) show that the sampling is
fairly regular without high data clustering. The criteria adopted
in order to avoid possibly the quotation of fake features and
artifacts due to the irregular sampling in this table, and
results displayed here, are described in the text of the paper.} \label{tab:timescalestable} %
\vspace{-0.5cm} \hspace{-0.2cm}
%
\includegraphics[width=6.7cm,angle=-90]{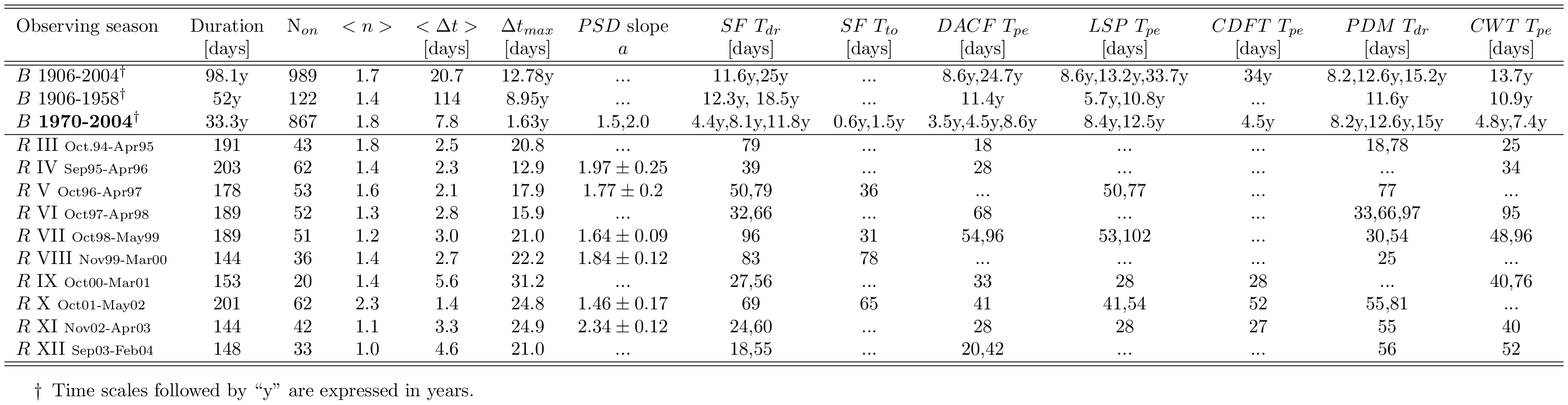}
\end{table}
%
\end{landscape}
\pagebreak \noindent
1) Short-term observations and multiwavelength (MW) snapshots
obtained during broad but limited-duration campaigns cannot
resolve all the puzzling questions about blazars, while long-term
monitoring allows to investigate the behaviour and evolution of
the emitted flux on different scales. 2) The knowledge about time
variability is crucial like the spectral variability in constraint
emission models, and the observed behaviour on long scales can be
cross-correlated with the MW observations usually available on
short timescales. 3) Radio-optical monitoring could be considered
a farsighted effort too: it enable to construct long-term records
of variability for several sources, useful for future researches.
4) Even if most of blazars seems to exhibit an irregular,
uncorrelated and unpredictable temporal behaviour, their optical
light curve shapes appear to be not trivial (sometimes signatures
of long-term memory, temporal self similarity and intermittence
are displayed), whereas in few known cases
periodical/quasi-periodical components cannot be ruled out. 5) In
addition, flare triggers and target of opportunity alerts for
space observatories and large-size telescopes are usually based on
a regular and constant monitoring. 6) Time series analysis of
sparse data sets (like blazar light curves), is a challenging,
interdisciplinary subject, being developed and applied on a wide
variety of present-day research topics outside astrophysics. 7)
Our fairly novel investigation and results on mid-term optical
timescales (days, weeks), could be also compared to the analysis
of blazar gamma-ray light curves that will be provided, at the
same scales, by the forthcoming Gamma-ray Large Area Space
Telescope (GLAST). In fact this high-energy space observatory will
be a large field-of-view and all-sky monitor for flares and
variability, allowing to record flux variations on over timescales
$> 1 $ day on hundreds of $\gamma$-ray blazar-like sources. 8)
Moreover worldwide international collaborations and the
participation of amateur and schools/universities optical
telescopes are now possible (thanks to the development of CCD
photometry and automation technology) meaning a valuable link for
education and public outreach.
\par With this in mind, during a long-term and painstaking
optical monitoring programme, we have obtained, collected and
analyzed the largest amount of optical data in 4 colors ever
published on the prominent blazar PKS 0735+178, thanks to the
collaboration of 3 professional observatories (Perugia, Torino and
Tuorla) and 1 amateur facility (Sabadell).  Furthermore a new
$VRI$ photometric calibration of 7 comparison stars in the field
of this blazar is presented (Tab. \ref{tab:ourcompstars}), joint
with the reconstruction and analysis of the whole historical light
curve (spanning now from 1906 to 2004). These optical data are
rather unique with respect to continuity, sampling and duration
for this source, and the associated data analysis enough
comprehensive, despite of natural difficulties (weather/seeing
conditions, seasonal gaps, technical problems or limited
manpower). About 500 nights of observations, collected in more
than 10 years (period 1993-2004), and providing 1332 new $BVRI$
final data points on PKS 0735+178 are reported and investigated,
aiming to a quantitative statistical description of the data set,
a characterization of the multi-band behaviour, and an
investigation of variability over 3 decades in time, for the first
time in this blazar.
\par During the last 10 years, PKS 0735+178 continued to show rapid
and large--amplitude optical variations typical of blazars, even
if the source remained in a rather low or intermediate brightness
state (mag $R>14$), showing a mild flaring activity. However
starting from the end of 1997 the source showed a clear increase
of the average brightness until 2001, when an active phase
occurred. In this decennium typical variations of about 2 mag in
less than half-year are observed joint with a general wiggling
pattern produced by a superimposition or succession of flares, and
modulated by a slower (possibly achromatic and oscillating)
long-term trend. The quiescent and mild-activity reported from
1994 to the second half of 2000 in the optical band, was also
pointed out recently in radio bands, as a period of quiescent flux
activity and highly twisted jet geometry \citep{agudo06}. In the
whole $\sim$100-years history of PKS 0735+178 five brightest
outbursts and active phases were observed: the last occurred in
the period Feb.2001-Oct.2001, and the brightest outburst was
observed in May 1977 (when the source reached its historical
optical maximum $B \simeq 13.9$).
\par The analysis of the continuum optical spectrum of this blazar
suggests, as expected, a correlation between the fluxes in $B$ and
$R$ bands, while the long-term variability of the spectral index
$\alpha$ appear to be essentially achromatic and independent by
the wavelength (Fig. \ref{fig:alphavsflux}). High-amplitude and
isolated flares can imply correlated spectral changes (usually a
flattening, i.e. bluer when brighter), but our data showed usually
a rather erratic evolution of $\alpha$ as a function of the flux,
and few or weak hints of non-thermal signatures (see e.g. Fig.
\ref{fig:3loops}). At these mid-term timescales and without an
increased sampling, it is reasonable to expect that the
superimposition of pure synchrotron optical fluctuations and
emission peaks cannot be easily disentangled from slower
variability patterns produced by different mechanisms.
\par A summary of the quantitative temporal analysis
performed in each single observing season of our $R$-band light
curves is reported in Tab.\ref{tab:timescalestable}. Intervals
$<200$ days are investigated discovering quite common
characteristic scales of variability falling especially into value
ranges of 27-28 days, 50-56 days and 76-79 days. These signatures
are the stronger contribution to the power spectrum, possibly
produced by correlated flares with typical duty cycles emitted by
some charge/discharge-like mechanisms. On the contrary hand these
typical timescales might represent the effects of single or few
powerful random events that are completely uncorrelated and
infrequent. On other words, even if a characteristic and intrinsic
timescale is found, this does not mean necessarily a discovery of
a dominant modulation or periodicity. In fact such evidence could
well be produced by events of random or transitory nature, or be
misrepresented by a combination of different underlying components
affected by an insufficient observing sampling.
\par Moreover the shot--noise (Brownian/brown noise) behaviour,
pointed-out by the values of the power-law index $a$ of the PSD
(computed in each season and falling between 1.46 and 2.34, see
Tab.\ref{tab:timescalestable}), reflects the nature of the
variations and can be linked to the findings cited above. Red and
brown noise are termed usually as $1/f^{a}$ (power-law decline)
fluctuations, meaning that the occurrence of a specific variation
is inversely proportional to its strength. Brownian variability
can be produced by a sequence of random pulses endowed of
long-term memory, where independent/discrete events and parallel
relaxation processes (like shocks and knots, electron density
fluctuations, magnetic field turbulence or plasma instabilities)
might generate a succession of mild optical flares and
oscillations with similar duty-cycles, on mid-term timescales. We
remark that fluctuation in the statistical moments and parameters
like the PSD and variance may be intrinsic in red/brown noise
processes \citep[see, e.g.][]{vio05}, even if our analysis is
essentially phenomenological and model independent. In addition
when we analyzed separately different segments of the light curve
and applied methods like the wavelets, we took already into
account non-stationarity problems.
\par About the historical behaviour of PKS 0735+178,
we found 3 main characteristic timescales having an extension of
about 4.5 years, about 8.5 years (possibly signatures of the same
4.5 years fundamental component) and between about 11 and 13 years
(Table\ref{tab:timescalestable}). These scales could be the result
of an oscillating and achromatic trend, modulating (with a
pseudo-periodic or multi-component course) the long-term
variability. Pseudo-periodicity can imply drifts in duration and
modulation, while a multi-component trend has several different
scales contributing in the composite modulation. In particular in
the first hypothesis the majority of the the long-term
characteristic timescales found might be multiple signatures of a
base component slightly drifting and varying around the value of
4.5 years. The visual inference based for example on the light
curves of Fig. \ref{fig:storica} and Fig.
\ref{fig:splines19702004}, can supports this statistical finding.
A rather ``humped'' or ``multi-bumped'' cyclical activity is
evident, possibly meaning a bimodal course, defined by an
alternation of active and quiescent stages. The limited temporal
range of observations having a sufficient sampling (1970-2004)
does not allow to understand if this pseudo-cyclical activity is a
stable or transitory phenomena, and several hypotheses can explain
the observed trend. A quasi/pseudo-periodical behaviour (with a
fundamental component drifting/oscillating around a value of 4.5
years); a multi-component modulation (by possibly different
correlated mechanisms); a mere random or transitory occurrence; a
combination of the previous scenarios (for example a mixture of a
multi-component trend with a quasi-periodical component). The
achromatic behaviour reported in section \ref{par:specindexes}, is
in agreement with a dynamical model implying slower variations of
the base level flux and long-term modulations, resulting for by
variations in the beaming factor of the jet.
\par This kind of optical course could be better correlated to the
radio flux behaviour and the twisted (maybe precessing) jet of PKS
0735+178 observed in details by some years. This peculiar blazar
(both radio and X-ray selected, and also a gamma-ray EGRET source)
shows quite slow variations in the radio bands and a complex
morphology displaying several moving components. In previous
literature the optical and radio history of PKS 0735+178 have
already suggested a possible periodical activity. Periodicity in
blazars has been debated for more than 40 years and several models
were developed to explain this prospect \citep[for example
precessing or helical jets and supermassive binary black holes,
see
e.g.][]{lehto96,sillanpaa96,villata98,rieger00,valtaoja00,ostorero04}.
However only in very few (and well publicized) cases there is
still a sufficient evidence of cyclical outburst (like in OJ 287),
and usually there is a general scepticism about widespread
periodicity in quasars/blazars light curves till now. On the other
hand the search for supermassive binary black holes in
extragalactic sources, should become a major research topic in the
next years.
\par As final consideration we point out that our 10-year observations probably
mapped 2 distinct phases: a stage of low or intermediate optical
luminosity (1994-2000) and a phase of mild flaring activity
(2001). That dual and possibly cyclic scenario might well be
confirmed by the behaviour of the radio flux and structure during
the same years \citep{gabuzda94,gomez01,agudo06}. The optical flux
is believed to be mainly originated in the very inner regions of
the jet, even if the stronger variability could occur much farther
from the central engine than previously expected
\citep{marscher05}. Hence it is reasonable to conceive a
correlation between the optical and radio flux on long-term
scales, and during the most active events. An improved continuous
and longer optical monitoring of PKS 0735+178, and the comparison
of the optical flare events with the ejection and evolution of the
superluminal radio knots based on long-term data records, will
allow to shed light on the physics of the jet flow and flare
mechanisms in this interesting object.
%
%
\begin{acknowledgements}
First of all we wish to thank warmly all the collaborators,
observers, students and technicians who contributed, in several
occasions in past years, to observations, technical support, data
reduction, analysis, an discussions within the four teams.
Thinking probably we forget someone, not intentionally, we would
like to express our gratitude to the following people. For the
Perugia team: M. Bagaglia, M. Luciani, N. Marchili, S. Pascolini,
V. Picarelli, N. Rizzi, F. Roncella, G. Sciuto, C. Spogli. For the
Torino team S. Bosio, M. Cavallone, M. Chiaberge, S. Crapanzano,
G. De Francesco, G. Ghisellini, G. Latini, L. Ostorero, M. Puccio,
G. Sobrito. For the Tuorla team: R. Rekola. For the Sabadell team:
J.M. Coloma, R. Costa, S. Esteva, E. Forn\'{e}, R. Ramajo.
\par Dr. Bochen Qian is to be thanked for providing most of the
historical optical data. Dr. Jun-Hui Fan is to be thanked for
accounting general comments. Dr. Alex W. Fullerton is to be
thanked for providing the CLEAN code. A gratefully acknowledge is
for the anonymous referee because his detailed report helped to
improve and clarify the paper. The groups of the Perugia, Torino
and Tuorla Observatories belonging to the Research Training
Network ENIGMA, acknowledge the funding by the European
Community's Human Potential Programme under contract
HPRN-CT-2002-00321. The Perugia and Torino optical monitoring
programmes have been partly supported by the Italian Ministry for
Instruction, University and Research (MIUR) under grant
Cofin2001/028773. The Tuorla monitoring programme has been
partially supported by the Academy of Finland. This research has
made use also of: SIMBAD database (CDS, Strasbourg), NASA/IPAC NED
database (JPL CalTech and NASA), HEASARC database (LHEA NASA/GSFC
and SAO), and NASA's ADS digital library.
\end{acknowledgements}

\bibliographystyle{aa}

\begin{thebibliography}{}

\bibitem[Agudo et al.(2006)]{agudo06} Agudo, I., G{\' o}mez, J.~L., Gabuzda, D.~C., Marscher, A.~P.,
Jorstad, S.~G \& Alberdi, A. \ 2006, \aap, 453, 477

\bibitem[Alexander (1997)]{alexander97} Alexander, T. 1997, in Astronomical Time Series, Eds. Maoz,
Sternberg \& Leibowitz, Dordrecht: Kluwer, p. 163

\bibitem[Aller et al.(1999)]{aller99} Aller, M.~F., Aller, H.~D.,
Hughes, P.~A., \& Latimer, G.~E.\ 1999, \apj, 512, 601

\bibitem[Bai et al.(1999)]{bai99} Bai, J.~M., Xie, G.~Z., Li,
K.~H., Zhang, X., \& Liu, W.~W.\ 1999, \aaps, 136, 455

\bibitem[B\aa \aa th \& Zhang(1991)]{baath91} B\aa \aa th, L.~B.~\& Zhang,
F.~J.\ 1991, \aap, 243, 328

\bibitem[Bessell(2005)]{bessell05} Bessell, M.~S.\ 2005, \araa,
43, 293

\bibitem[Bessell(1990)]{bessell90} Bessell, M.~S.\ 1990, \pasp,
102, 1181

\bibitem[Bessell(1979)]{bessell79} Bessell, M.~S.\ 1979, \pasp,
91, 589

\bibitem[Blake(1970)]{blake70} Blake, G.~M.\ 1970, \aplett, 6,
201

\bibitem[Bregman et al.(1984)]{bregman84} Bregman, J.~N., Glassgold, A. E.,
Huggins, P. J., Aller, H. D., Aller, M. F., Hodge, P. E., Rieke, G. H.,
Lebofsky, M. J., Pollock, J. T., et al.\ 1984, \apj, 276, 454

\bibitem[Bregman, Glassgold, \& Huggins(1981)]{bregman81}
Bregman, J.~N., Glassgold, A.~E., \& Huggins, P.~J.\ 1981, \apj, 249, 13

\bibitem[Brown et al.(1989)]{brown89} Brown,
L.~M.~J., Robson, E.~I., Gear, W.~K., \& Smith, M.~G.\ 1989, \apj, 340, 150

\bibitem[Burbidge \& Hewitt(1987)]{burbidge87} Burbidge, G.~\&
Hewitt, A.\ 1987, \aj, 93, 1

\bibitem[B{\"o}ttcher et al.(2005)]{boettcher05} B{\"o}ttcher, M., Harvey, J.,
Joshi, M., Villata, M., Raiteri, C. M., Bramel, D., Mukherjee, R.,
Savolainen, T., Cui, W., Fossati, G., et al.\ 2005, \apj, 631, 169

\bibitem[B{\" o}ttcher \& Chiang(2002)]{boettcher02} B{\" o}ttcher,
M.~\& Chiang, J.\ 2002, \apj, 581, 127

\bibitem[Carswell et al.(1974)]{carswell74} Carswell, R.~F.,
Strittmatter, P.~A., Williams, R.~E., Kinman, T.~D., \& Serkowski,
K.\ 1974, \apjl, 190, L101

\bibitem[Ciaramella et al.(2004)]{ciaramella04} Ciaramella, A.,
Bongardo, C., Aller, H. D., Aller, M. F., De Zotti, G.,
L\"{a}hteenmaki, A., Longo, G., Milano, L., Tagliaferri, R.,
Ter\"{a}sranta, H., Tornikoski, M., \& Urpo, S. \ 2004, \aap, 419,
485

\bibitem[Ciprini et al. (2004)]{ciprini04} Ciprini, S., Tosti, G., Ter{\"
a}sranta, H., \& Aller, H.~D.\ 2004, \mnras, 348, 1379

\bibitem[Clements et al.(1995)]{clements95} Clements, S.~D., Smith, A.~G.,
Aller, H.~D., \& Aller, M.~F.\ 1995, \aj, 110, 529

\bibitem[Cotton et al.(1980)]{cotton80} Cotton, W. D., Wittels, J. J.,
Shapiro, I. I., Marcaide, J., Owen, F. N., Spangler, S. R., Rius, A.,
Angulo, C., Clark, T. A., Knight, C. A. 1980, \apjl, 238, L123

\bibitem[Daubechies(1992)]{daubechies92} Daubechies, I., 1992, Ten Lectures on
Wavelets, Philadelphia: Soc. for Industrial \& Applied Math.
(SIAM)

\bibitem[Deeming(1975)]{deeming75} Deeming, T.~J.\ 1975, \apss,
36, 137

\bibitem[Della Ceca et al.(1990)]{dellaceca90} Della Ceca, R.,
Palumbo, G.~G.~C., Persic, M., Boldt, E.~A., Marshall, E.~E., \& de Zotti,
G.\ 1990, \apjs, 72, 471

\bibitem[Edelson \& Krolik(1988)]{edelson88} Edelson, R.~A.~\&
Krolik, J.~H.\ 1988, \apj, 333, 646

\bibitem[Elvis et al.(1992)]{elvis92} Elvis, M., Plummer, D., Schachter, J.,
\& Fabbiano, G.\ 1992, \apjs, 80, 257

\bibitem[Falomo \& Ulrich(2000)]{falomo00} Falomo, R.~\& Ulrich,
M.-H.\ 2000, \aap, 357, 91

\bibitem[Fan \& Lin(2000)]{fan00} Fan, J.~H.~\& Lin, R.~G.\ 2000, \apj, 537, 101

\bibitem[Fan \& Lin(1999)]{fan99} Fan, J.~H.~\& Lin, R.~G.\
1999, \apjs, 121, 131

\bibitem[Fan et al.(1997)]{fan97} Fan, J.~H., Xie, G.~Z.,
Lin, R.~G., Qin, Y.~P., Li, K.~H., \& Zhang, X.\ 1997, \aaps, 125,
525

\bibitem[Farge(1992)]{farge92} Farge, M., 1992, Annu. Rev. Fluid Mech.,
24, 395

\bibitem[Fiorucci \& Munari(2003)]{fiorucci03} Fiorucci, M., \&
Munari, U.\ 2003, \aap, 401, 781

\bibitem[Fiorucci \& Tosti(1996)]{fiorucci96} Fiorucci, M., \& Tosti, G. 1996, A\&AS,
 116, 403

\bibitem[Fiorucci, Ciprini, \& Tosti(2004)]{fiorucci04} Fiorucci,
M., Ciprini, S., \& Tosti, G.\ 2004, \aap, 419, 25

\bibitem[Fiorucci, Tosti, \& Rizzi(1998)]{fiorucci98} Fiorucci, M., Tosti, G., \& Rizzi, N. 1998,
PASP, 110, 105

\bibitem[Foster(1996)]{foster96} Foster, G.\ 1996, \aj, 112,
1709

\bibitem[Foster(1995)]{foster95} Foster, G.\ 1995, \aj, 109,
1889

\bibitem[Gabuzda, G{\' o}mez, \& Agudo(2001)]{gabuzda01} Gabuzda, D.~C., G{\' o}mez, J.~L.,
\& Agudo, I.\ 2001, \mnras, 328, 719

\bibitem[Gabuzda et al.(1994)]{gabuzda94} Gabuzda, D.~C., Wardle,
J.~F.~C., Roberts, D.~H., Aller, M.~F., \& Aller, H.~D.\ 1994, \apj, 435,
128

\bibitem[Gabuzda, Wardle, \& Roberts(1989)]{gabuzda89} Gabuzda,
D.~C., Wardle, J.~F.~C., \& Roberts, D.~H.\ 1989, \apj, 338, 743

\bibitem[Georganopoulos \& Marscher (1998)]{georganopoulos98} Georganopoulos, M., \& Marscher, A. P. 1998, \apjl, 506, L11

\bibitem[Ghisellini et al.(1997)]{ghisellini97} Ghisellini, G., Villata, M., Raiteri, C.
M., Bosio, S., de Francesco, G., Latini, G., Maesano, M., Massaro,
E., Montagni, F., Nesci, R. al.\ 1997, \aap, 327, 61

\bibitem[Ghosh et al.(2000)]{ghosh00} Ghosh, K.~K., Ramsey,
B.~D., Sadun, A.~C., \& Soundararajaperumal, S.\ 2000, \apjs, 127,
11

\bibitem[G{\' o}mez et al.(2001)]{gomez01} G{\' o}mez, J.~L., Guirado, J.~C., Agudo, I.,
Marscher, A.~P., Alberdi, A., Marcaide, J.~M., \& Gabuzda, D.~C.\ 2001, \mnras, 328, 873

\bibitem[G{\' o}mez et al.(1999)]{gomez99} G{\' o}mez, J., Marscher, A.~P., Alberdi, A.,
\& Gabuzda, D.~C.\ 1999, \apj, 519, 642

\bibitem[Gu et al.(2006)]{gu06} Gu, M.~F., Lee, C.-U., Pak,
S., Yim, H.~S., \& Fletcher, A.~B.\ 2006, \aap, 450, 39

\bibitem[Hanski, Takalo, \& Valtaoja(2002)]{hanski02} Hanski,
M.~T., Takalo, L.~O., \& Valtaoja, E.\ 2002, \aap, 394, 17

\bibitem[Hartman et al.(1999)]{hartman99} Hartman, R.~C.,
Bertsch, D. L., Bloom, S. D., Chen, A. W., Deines-Jones, P.,
Esposito, J. A., Fichtel, C. E., Friedlander, D. P., Hunter, S.
D., McDonald, L. M. et al. 1999, \apjs, 123, 79

\bibitem[Homan et al.(2002)]{homan02} Homan, D.~C., Ojha, R., Wardle, J.~F.~C.,
Roberts, D.~H., Aller, M.~F., Aller, H.~D., \& Hughes, P.~A.\ 2002, \apj, 568, 99

\bibitem[Horne \& Baliunas(1986)]{horne86} Horne, J. H., \& Baliunas, S. L. 1986, \apj, 302, 757

\bibitem[H{\" o}gbom(1974)]{hoegbom74} H{\" o}gbom, J.~A.\ 1974,
\aaps, 15, 417

\bibitem[Hufnagel \& Bregman(1992)]{hufnagel92} Hufnagel, B.~R.~\&
Bregman, J.~N.\ 1992, \apj, 386, 473

\bibitem[Hughes, Aller \& Aller(1992)]{hughes92} Hughes, P. A., Aller, H. D., \& Aller, M. F. 1992, \apj, 396, 469

\bibitem[Hutchings, Johnson, \& Pyke(1988)]{hutchings88} Hutchings,
J.~B., Johnson, I., \& Pyke, R.\ 1988, \apjs, 66, 361

\bibitem[Jurkevich(1971)]{jurkevich71} Jurkevich, I.\ 1971, \apss,
13, 154

\bibitem[Kaiser(1994)]{kaiser94} Kaiser, G., 1994, A Friendly Guide to
Wavelets, Boston: Birkhauser editions

\bibitem[Katajainen et al.(2000)]{katajainen00} Katajainen, S., Takalo, L. O., Sillanp\"{a}\"{a}, A.,
et. al 2000, A\&AS, 143, 357

\bibitem[Kataoka et al.(2000)]{kataoka00} Kataoka, J., Takahashi,
T., Makino, F., Inoue, S., Madejski, G.~M., Tashiro, M., Urry,
C.~M., \& Kubo, H.\ 2000, \apj, 528, 243

\bibitem[Kellermann et al.(2004)]{kellermann04} Kellermann, K.~I.,
Lister, M. L., Homan, D. C., Vermeulen, R. C., Cohen, M. H., Ros,
E., Kadler, M., Zensus, J. A., \& Kovalev, Y. 2004, \apj, 609, 539

\bibitem[Kellermann et al.(1998)]{kellermann98} Kellermann, K.~I., Vermeulen, R.~C.,
Zensus, J.~A., \& Cohen, M.~H.\ 1998, \aj, 115, 1295

\bibitem[Kidger(1989)]{kidger89} Kidger, M.~R.\ 1989, \aap, 226, 9

\bibitem[Kirk, Rieger, \& Mastichiadis(1998)]{kirk98} Kirk,
J.~G., Rieger, F.~M., \& Mastichiadis, A.\ 1998, \aap, 333, 452

\bibitem[Kubo et al.(1998)]{kubo98} Kubo, H., Takahashi, T.,
Madejski, G., Tashiro, M., Makino, F., Inoue, S., \& Takahara, F.\
1998, \apj, 504, 693

\bibitem[K\"{u}hr et al.(1981)]{kuhr81} K\"{u}hr, H., Witzel, A., Pauliny-Toth, I.
I. K., \& Nauber, U. 1981, A\&A, 45, 367

\bibitem[Lafler \& Kinman(1965)]{lafler65} Lafler, J., \&
Kinman, T.~D.\ 1965, \apjs, 11, 216

\bibitem[Lainela \& Valtaoja(1993)]{lainela93} Lainela, M.~\&
Valtaoja, E.\ 1993, \apj, 416, 485

\bibitem[L{\"a}hteenm{\"a}ki \& Valtaoja(1999)]{lahteenmaki99}
L{\"a}hteenm{\"a}ki, A., \& Valtaoja, E.\ 1999, \apj, 521, 493

\bibitem[Lehto \& Valtonen(1996)]{lehto96} Lehto, H.~J.~\&
Valtonen, M.~J.\ 1996, \apj, 460, 207

\bibitem[Lin \& Fan(1998)]{lin98} Lin, R.~G.~\& Fan, J.~H.\
1998, \apss, 259, 67

\bibitem[Lomb(1976)]{lomb76} Lomb, N.~R.\ 1976, \apss, 39, 447

\bibitem[Lu(1972)]{lu72} Lu, P.~K.\ 1972, \aj, 77, 829

\bibitem[Madejski \& Schwartz(1988)]{madejski88} Madejski,
G.~M.~\& Schwartz, D.~A.\ 1988, \apj, 330, 776

\bibitem[Marscher(2005)]{marscher05} Marscher, A.~P.\ 2005,
Mem. Soc. Astron. Italiana, 76, 168

\bibitem[Marscher(1980)]{marscher80} Marscher, A.\ 1980, \nat,
288, 12

\bibitem[Marscher(1977)]{marscher77} Marscher, A.~P.\ 1977, \aj,
82, 781

\bibitem[Massaro \& Trevese(1996)]{massaro96} Massaro, E.~\&
Trevese, D.\ 1996, \aap, 312, 810

\bibitem[Massaro et al.(1995)]{massaro95} Massaro, E., Nesci, R.,
Perola, G.~C., Lorenzetti, D., \& Spinoglio, L.\ 1995, \aap, 299,
339

\bibitem[Mead et al.(1990)]{mead90} Mead, A.~R.~G., Ballard,
K.~R., Brand, P.~W.~J.~L., Hough, J.~H., Brindle, C., \& Bailey,
J.~A.\ 1990, \aaps, 83, 183

\bibitem[McGimsey et al.(1976)]{mcgimsey76} McGimsey, B.~Q.,
Williamon, R.~M., \& Miller, H.~R.\ 1976, \aj, 81, 750

\bibitem[Nilsson et al.(2006)]{nilsson06} Nilsson, K. et al. 2006,
in preparation

\bibitem[Nolan et al.(2003)]{nolan03} Nolan, P.~L., Tompkins,
W.~F., Grenier, I.~A., \& Michelson, P.~F.\ 2003, \apj, 597, 615

\bibitem[Ojha et al.(2004)]{ojha04} Ojha, R., Homan, D.~C.,
Roberts, D.~H., Wardle, J.~F.~C., Aller, M.~F., Aller, H.~D., \&
Hughes, P.~A.\ 2004, \apjs, 150, 187

\bibitem[Ostorero, Villata, \& Raiteri(2004)]{ostorero04}
Ostorero, L., Villata, M., \& Raiteri, C.~M.\ 2004, \aap, 419, 913

\bibitem[Papadakis \& Lawrence(1993)]{papadakis93} Papadakis,
I.~E.~\& Lawrence, A.\ 1993, \mnras, 261, 612

\bibitem[Percival \& Walden (2002)]{percival02} Percival,  D.B., \& Walden,
A. T. 2002, Wavelet Methods for Time Series Analysis, Cambridge:
Cambridge University Press

\bibitem[Perlman \& Stocke(1994)]{perlman94} Perlman, E.~S.~\&
Stocke, J.~T.\ 1994, \aj, 108, 56

\bibitem[Pollock et al.(1979)]{pollock79} Pollock, J.~T., Pica,
A.~J., Smith, A.~G., Leacock, R.~J., Edwards, P.~L., \& Scott,
R.~L.\ 1979, \aj, 84, 1658

\bibitem[Pursimo et al.(2002)]{pursimo02} Pursimo, T., Nilsson,
K., Takalo, L.~O., Sillanp{\" a}{\" a}, A., Heidt, J., \&
Pietil{\" a}, H.\ 2002, \aap, 381, 810

\bibitem[Pursimo et al.(1999)]{pursimo99} Pursimo, T., Nilsson, K.,
Sillanp\"{a}\"{a}, A., Takalo, L. O., \& Heidt, J. \ 1999, ASP
Conf.~Ser.~159: BL Lac Phenomenon, 385

\bibitem[Qian \& Tao(2004)]{qian04} Qian, B.~\& Tao, J.\ 2004,
\pasp, 116, 161

\bibitem[Raiteri et al.(2003)]{raiteri03}  Raiteri, C. M., Villata, M., Tosti, G., Nesci,
R., Massaro, E., Aller, M. F., Aller, H. D., Ter{\"a}sranta, H.,
Kurtanidze, O. M., Nikolashvili, M. G.,  et al.\ 2003, \aap, 402,
151

\bibitem[Ravasio et al. (2004)]{ravasio04} Ravasio, M., Tagliaferri, G.,
Ghisellini, G., \& Tavecchio, F.\ 2004, \aap, 424, 841

\bibitem[Rector \& Stocke(2001)]{rector01} Rector, T.~A.~\&
Stocke, J.~T.\ 2001, \aj, 122, 565

\bibitem[Rieger \& Mannheim(2000)]{rieger00} Rieger, F.~M.~\&
Mannheim, K.\ 2000, \aap, 359, 948

\bibitem[Roberts, Lehar, \& Dreher(1987)]{roberts87} Roberts,
D.~H., Lehar, J., \& Dreher, J.~W.\ 1987, \aj, 93, 968

\bibitem[Rutman(1978)]{rutman78} Rutman, J. 1978, Proc. IEEE, 66, 1048

\bibitem[Sagar et al.(2004)]{sagar04} Sagar, R., Stalin, C.~S.,
Gopal-Krishna, G., \& Wiita, P.~J.\ 2004, \mnras, 348, 176

\bibitem[Scargle(1982)]{scargle82} Scargle, J.~D.\ 1982, \apj,
263, 835

\bibitem[Scarpa et al.(2000)]{scarpa00} Scarpa, R., Urry, C.~M.,
Falomo, R., Pesce, J.~E., \& Treves, A.\ 2000, \apj, 532, 740

\bibitem[Schlegel, Finkbeiner, \& Davis(1998)]{schlegel98} Schlegel, D. J.,
Finkbeiner, D. P., Davis, M. 1998, \apj, 500, 525

\bibitem[Sillanp\"{a}\"{a} et al.(1996)]{sillanpaa96} Sillanp\"{a}\"{a}, A.,
Takalo, L. O., Pursimo, T., Nilsson, K., Heinamaki, P.,
Katajainen, S., Pietila, H., Hanski, M., Rekola, R., Kidger, M. et
al.\ 1996, \aap, 315, L13

\bibitem[Simonetti, Cordes, \& Heeschen(1985)]{simonetti85}
Simonetti, J.~H., Cordes, J.~M., \& Heeschen, D.~S.\ 1985, \apj,
296, 46

\bibitem[Sitko \& Sitko(1991)]{sitko91} Sitko, M.~L.~\& Sitko,
A.~K.\ 1991, \pasp, 103, 160

\bibitem[Smith \& Nair(1995)]{smith95} Smith, A.~G.~\& Nair,
A.~D.\ 1995, \pasp, 107, 863

\bibitem[Smith et al.(1993)]{smith93}
Smith, A.~G., Nair, A.~D., Leacock, R.~J., \& Clements, S.~D.\
1993, \aj, 105, 437

\bibitem[Smith, Leacock, \& Webb(1988)]{smith88} Smith, A.~G.,
Leacock, R.~J., \& Webb, J.~R.\ 1988, in LNP Vol.~307: Active
Galactic Nuclei, ed. by H. R. Miller \& P. J. Wiita, Berlin:
Springer-Verlag, 158

\bibitem[Smith et al.(1987)]{smith87}
Smith, P.~S., Balonek, T.~J., Elston, R., \& Heckert, P.~A.\ 1987,
\apjs, 64, 459

\bibitem[Smith et al.(1985)]{smith85} Smith, P.~S., Balonek,
T.~J., Heckert, P.~A., Elston, R., \& Schmidt, G.~D.\ 1985, \aj,
90, 1184

\bibitem[Stellingwerf(1978)]{stellingwerf78} Stellingwerf, R.~F.\
1978, \apj, 224, 953

\bibitem[Stickel, Fried, \& Kuehr(1993)]{stickel93} Stickel, M.,
Fried, J.~W., \& Kuehr, H.\ 1993, \aaps, 98, 393

\bibitem[Takalo(1991)]{takalo91} Takalo, L. O. 1991, A\&AS, 90, 161

\bibitem[Takalo et al.(1992)]{takalo92}Takalo, L. O., Sillanp\"{a}\"{a}, A., Nilsson, K., Kidger, M., de Diego,
J. A., \& Piirola, V. 1992, A\&AS, 94, 37

\bibitem[Ter{\" a}sranta et al.(2004)]{terasranta04} Ter{\"
a}sranta, H., Achren, J., Hanski, M., Heikkil\"{a}, J.,
Holopainen, J., Joutsamo, O., Juhola, M., Karlamaa, K.,
Katajainen, S., Kein\"{a}nen, P., et. al\ 2004, \aap, 427, 769

\bibitem[Ter\"{a}sranta et al.(1992)]{terasranta92} Ter\"{a}sranta, H.,
Tornikoski, M., Valtaoja, E., Urpo, S., Nesterov, N., Lainela, M., Kotilainen, J.,
Wiren, S., Laine, S., Nilsson, K., \& Valtonen, L. \ 1992, \aaps, 94, 121

\bibitem[Tommasi et al.(2001)]{tommasi01} Tommasi, L., Palazzi, E., Pian, E., Piirola, V., Poretti,
E., Scaltriti, F., Sillanp\"{a}\"{a}, A., Takalo, L., Treves, A.
2001, \aap, 376, 51

\bibitem[Tornikoski et al.(1994)]{tornikoski94} Tornikoski, M.,
Valtaoja, E., Ter\"{a}sranta, H., Smith, A.~G., Nair, A.~D., Clements, S.~D.,
\& Leacock, R.~J.\ 1994, \aap, 289, 673

\bibitem[Tosti, Pascolini, \& Fiorucci(1996)]{tosti96} Tosti, G., Pascolini, S.,
\& Fiorucci, M. 1996, PASP, 108, 706

\bibitem[Valtaoja et al.(2000)]{valtaoja00} Valtaoja, E., Ter{\"
a}sranta, H., Tornikoski, M., Sillanp{\" a}{\" a}, A., Aller,
M.~F., Aller, H.~D., \& Hughes, P.~A.\ 2000, \apj, 531, 744

\bibitem[Valtaoja et al.(1993)]{valtaoja93} Valtaoja, L., Karttunen, H., Valtaoja, E.,
Shakhovskoy, N. M., \& Efimov, Y. S. 1993, A\&A, 273, 393

\bibitem[Valtaoja et al.(1991)]{valtaoja91} Valtaoja, L.,
Sillanp\"{a}\"{a}, A., Valtaoja, E., Shakhovskoi, N.~M., \&
Efimov, I.~S.\ 1991, \aj, 101, 78

\bibitem[Veron \& Veron(1975)]{veron75} Veron, P., \& Veron,
M.~P.\ 1975, \aap, 39, 281

\bibitem[Villata et al.(2004)]{villata04} Villata, M., Raiteri, C. M., Kurtanidze, O. M.,
Nikolashvili, M. G., Ibrahimov, M. A., Papadakis, I. E., Tosti,
G., Hroch, F., Takalo, L. O., Sillanp\"{a}\"{a}, A. et al.\ 2004,
\aap, 421, 103

\bibitem[Villata et al.(2002)]{villata02} Villata, M., Raiteri, C. M., Kurtanidze, O. M.,
Nikolashvili, M. G., Ibrahimov, M. A., Papadakis, I. E.,
Tsinganos, K., Sadakane, K., Okada, N., Takalo, L. O. et al.\
2002, \aap, 390, 407

\bibitem[Villata et al.(1998)]{villata98} Villata, M., Raiteri,
C.~M., Sillanpaa, A., \& Takalo, L.~O.\ 1998, \mnras, 293, L13

\bibitem[Vio et al.(2005)]{vio05} Vio, R., Kristensen, N.~R.,
Madsen, H., \& Wamsteker, W.\ 2005, \aap, 435, 773

\bibitem[Voges et al.(1999)]{voges99} Voges, W., Aschenbach, B.,
Boller, Th., Br\"{a}uninger, H., Briel, U., Burkert, W., Dennerl, K., Englhauser, J., Gruber, R., Haberl,
et al.\ 1999, \aap, 349, 389

\bibitem[Wagner(1996)]{wagner96} Wagner, S.~J.\ 1996, \aaps,
120, 495

\bibitem[Webb et al.(1988)]{webb88} Webb, J. R., Smith, A. G., Leacock, R. J., Fitzgibbons,
G. L., Gombola, P. P., \& Shepherd, D. W. 1988, AJ, 95, 374

\bibitem[Wing(1973)]{wing73} Wing, R.~F.\ 1973, \aj, 78, 684

\bibitem[Xie et al.(1992)]{xie92} Xie, G.~Z., Li, K.~H., Liu,
F.~K., Lu, R.~W., Wu, J.~X., Fan, J.~H., Zhu, Y.~Y., \& Cheng,
F.~Z.\ 1992, \apjs, 80, 683

\bibitem[Zhang et al.(2004)]{zhang04} Zhang, X., Zhang, L.,
Zhao, G., Xie, Z.-H., Wu, L., \& Zheng, Y.-G.\ 2004, \aj, 128,
1929

\bibitem[Zhang et al.(2002)]{zhang02} Zhang, L.~Z., Fan, J.-H.,
\& Cheng, K.-S.\ 2002, \pasj, 54, 159

\bibitem[Zekl et al.(1981)]{zekl81} Zekl, H., Klare, G., \& Appenzeller, I. 1981, A\&A, 103, 342

\end{thebibliography}

\end{document}